\documentclass[preprintnumbers,nofootinbib,amsmath,amssymb]{revtex4}

\def\ga{\mathrel{\raise.3ex\hbox{$>$\kern-.75em\lower1ex\hbox{$\sim$}}}}
\def\la{\mathrel{\raise.3ex\hbox{$<$\kern-.75em\lower1ex\hbox{$\sim$}}}}

\newcommand\beq{\begin{equation}}
\newcommand\eeq{\end{equation}}
\newcommand\beqar{\begin{eqnarray}}
\newcommand\eeqar{\end{eqnarray}}
\usepackage{graphicx}
\usepackage{epsfig}
\usepackage{epsf}
\usepackage{rotating}
\usepackage{verbatim}

 \usepackage{multirow}

\begin{document}

\preprint{ArXiv:0909.3524} \preprint{UMN--TH--2815/09} \vskip 0.2in
\title{

Ghost instabilities of cosmological models with vector fields \\

nonminimally coupled to the curvature
}

\author{Burak Himmetoglu$^{(1)}$, Carlo R. Contaldi$^{(2)}$ and Marco Peloso$^{(1)}$}
\address{$^{(1)}${\it School of Physics and Astronomy, University of Minnesota, Minneapolis, MN 55455, USA}}
\address{$^{(2)}${\it Theoretical Physics, Blackett Laboratory, Imperial College, London, SW7 2BZ, UK}}

\begin{abstract}

  We prove that many cosmological models characterized by vectors
  nonminimally coupled to the curvature (such as the Turner-Widrow
  mechanism for the production of magnetic fields during inflation,
  and models of vector inflation or vector curvaton) contain
  ghosts. The ghosts are associated with the longitudinal vector
  polarization present in these models, and are found from studying
  the sign of the eigenvalues of the kinetic matrix for the physical
  perturbations. Ghosts introduce two main problems: (1) they make the
  theories ill-defined at the quantum level in the high energy/sub
  horizon regime (and create serious problems for finding a well
  behaved UV completion); (2) they create an instability already at
  the linearized level. This happens because the eigenvalue
  corresponding to the ghost crosses zero during the cosmological
  evolution. At this point the linearized equations for the
  perturbations become singular (we show that this happens for all the
  models mentioned above). We explicitly solve the equations in the
  simplest cases of a vector without vev in a FRW geometry, and of a
  vector with vev plus a cosmological constant, and we show that
  indeed the solutions of the linearized equations diverge when these
  equations become singular.

\end{abstract}

\maketitle

\section{Introduction}

Although the WMAP measurements of the Cosmic Microwave Background
(CMB) strongly support the inflationary paradigm \cite{WMAP}, several
studies pointed out some peculiar features in the data that seem at
odd with the simplest inflationary predictions. These so called
`anomalies' include the low power in the quadrupole
moment~\cite{cobe,wmap1,lowl}, the alignment of the lowest
multipoles~\cite{axis}, a $\sim 5^\circ$ cold spot with suppressed
power \cite{cold}, an asymmetry in power between the northern and
southern ecliptic hemispheres~\cite{asym}, and broken rotational
invariance~\cite{Groeneboom:2008fz}.~\footnote{Ref. \cite{ArmendarizPicon:2008yr}
  found an upper limit on the anisotropy, which is compatible with the
  result of \cite{Groeneboom:2008fz}.} The significance of some of
these effects has increased in the latest
studies~\cite{Groeneboom:2008fz,Hoftuft:2009rq}, based on the WMAP-5
years data. These observations have motivated a number of studies both
on data analysis (how to construct estimators that can assess the
degree of violation of statistical isotropy, see
eg.~\cite{estimators}) and on theory (how to construct models that
reproduce these features). One can for instance attempt to ascribe the
departure from statistical isotropy to initial conditions at the onset
of inflation, either on the background evolution
\cite{gcp,uzan,Dimastrogiovanni:2008ua,gkp}, or on a super-horizon
isocurvature mode \cite{erickcek}.~\footnote{The super-horizon mode
  breaks the translational, and therefore can give rise to the
  asymmetry claimed in~\cite{asym}; the Bianchi-I backgrounds
  considered in \cite{gcp,uzan,Dimastrogiovanni:2008ua,gkp} have
  instead planar symmetry, and have the correct structure to explain
  the violation of rotational invariance observed
  by~\cite{Groeneboom:2008fz}. General expressions for the correlation
  $\langle a_{\ell m} \, a_{\ell' m'}^* \rangle$ between different
  multipoles were given in \cite{gcp} for generic breaking of
  rotational invariance, and in \cite{trans} for generic breaking of
  translational invariance.}

Ref. \cite{string} associates the super-horizon mode with a long cosmic string. In general, however, in these models the breaking of statistical isotropy is built-in as an initial or ``boundary'' condition, rather than being predicted from first principles (e.g., from a given lagrangian). A second major problem that the above models suffer is that inflation rapidly removes any initial background anisotropy~\cite{wald}, and blows to unobservably large scales any perturbation that was present at its onset. Therefore, these proposals require a minimal (and, therefore, tuned) amount of inflation. To improve over these two problems, one may consider introducing in the inflationary model some nonminimal ingredients that contrast the rapid isotropization caused by the inflaton. This has been realized through the addition of quadratic curvature invariants to the gravity action~\cite{barrow}, with the use of the Kalb-Ramond 
axion~\cite{nemanja}, or of vector fields~\cite{ford}. \footnote{While the models studied here have been proposed in the context of primordial inflation, vector fields with nonvanishing spatial vacuum expectation value (vev) have been also employed as sources of the late time time acceleration~\cite{dark}. See also \cite{dark2} for a different model of anisotropic dark energy.}

The present work continues a series of previous papers in which we
studied the stability of some of the models with vector fields. In
Ref. \cite{hcp1}, we presented a general discussion valid for three
groups of models, characterized by (i) a potential $V \left( A^2
\right)$ for the vector \cite{ford}, (ii) a fixed spatial norm of the
vector, enforced by a lagrange multiplier \cite{acw}, or (iii) a
nonminimal coupling of the vector to the scalar curvature
\cite{Turner:1987bw,mukhvect,soda1,Dimopoulos}. We showed that all
these models have ghost instabilities. Although few explicit
computations were given in the Letter \cite{hcp1}, we explained the
physical reasons behind the instability: the terms that characterize
the above models break the U(1) symmetry that would be otherwise
associated to the vector field. This introduces an additional
polarization (the longitudinal vector mode) that, in these models,
turns out to be a ghost. In Ref. \cite{hcp2} we provided the explicit
computations for the case (ii).

In the present work, we present  explicit computations for the models in the group  (iii). The first of such models \cite{Turner:1987bw}  is a well known mechanism for the generation of magnetic fields. Among other things, it was pointed out there that, for the specific action
\begin{equation}
S = \int d^4 x \, \sqrt{-g} \left[ - \frac{1}{4} F^2 + \frac{1}{12} R \, A^2 \right]
\end{equation}
the equations for the vector field (more precisely, those governing the vev, and the transverse polarizations) are the same as those of a scalar field minimally coupled to the curvature. This fact was also exploited in a series of recent papers, motivated by the WMAP anomalies. The work \cite{mukhvect} studies inflation driven by $N$ vector fields. The simplest realization of   \cite{mukhvect} is characterized by three mutually orthogonal vectors with equal vev: this provides a homogeneous and isotropic (Friedmann-Robertson-Walker, FRW) evolution; however, one can envisage the more complicated situation in which a large number $N$ of fields is present, with random orientation. This provides a nearly isotropic expansion, with  a naturally small,  ${\rm O } \left( 1 / \sqrt{N} \right)$, anisotropy. Ref. \cite{soda1} provides a simpler version of the same idea, in which a single vector field plus an inflaton field (in a sense, replacing the average effect of the vectors) are present. Finally, Refs. \cite{Dimopoulos} used the vector as a curvaton field, in order to produce a nearly scale invariant spectrum of primordial perturbations. 

The presence of ghosts has not emerged in previous stability studies
of such models, due their partial nature. The original work
\cite{Turner:1987bw} studied the evolution of modes of the magnetic
field, associated to the transverse photon polarizations, but did not
discuss the role of the additional longitudinal mode. The results of
\cite{Dimopoulos} are based on the $\delta N$ formalism, which
computes the classical evolution of super-horizon modes, assuming that
the quantum theory is under control (as it happens in the case of
scalar field inflation, for which it was
developed). \footnote{Moreover, we show below that also the classical
  evolution for the longitudinal mode leads to a divergence at
  super-horizon, when the total mass of the vector vanishes.} 
 Ref.~\cite{mukhvect2} studied the gravity waves (GW)
in the model of vector inflation \cite{mukhvect}, assuming that the
coupling of these modes with the other perturbations - which, for this
model, is present already at the linearized level, see
eq.~(\ref{coupledTT}) below - can be disregarded. Finally,
Ref. \cite{mukhvect3} studied the linearized equations of motion for
vector inflation either in the short wavelength, or in the long
wavelength regime. \footnote{In inflationary models, the wavelength of
  a perturbation grows nearly exponentially, while the Hubble rate $H$
  is nearly constant. Therefore, at sufficiently early times, the
  wavelength of any mode is smaller than the horizon scale $H^{-1}
  \,$, while the opposite is true at sufficiently late times. The
  moment at which the two scales are equal is dubbed ``horizon
  crossing''; we remark that perturbations of different size cross the
  horizon at different times (larger modes, exit the horizon earlier
  during inflation).} The study of the linearized equations alone does
not allow to see whether a perturbation is a positive or negative
energy mode, and for this reason the presence of ghosts does not
appear in this analysis.~\footnote{Cf. the case of a scalar field,
  with lagrangian ${\cal L } = \pm \left( \dot{\phi}^2 - m^2 \phi^2
  \right) /2$ (dot denotes derivative with respect to time, and we
  ignore spatial dimensions); the equation of motion is $\ddot{\phi} +
  m^2 \, \phi = 0$ for either overall sign of the lagrangian, so that
  no instability appears from the equation. However, the minus sign in
  the kinetic term corresponds to a negative energy field (a ghost).}
Moreover, as we show below, the linearized equations of motion for the
perturbations become singular close to horizon crossing, in a regime
where neither the long nor the short wavelength analyses apply. In
Section \ref{vector-inflation}, we provide a more detailed discussion
of some claims made in \cite{mukhvect3}.

It is important to stress that the ghost instability takes place also
for those of the above models which have a FRW background. The
presence of a ghost in these models is indeed due to the specific sign
of the effective mass term $M^2 = - R / 6 \,$ for the vector induced
by the coupling to the curvature. Indeed, as we showed in \cite{hcp1},
a negative mass squared for a vector results in a ghost, and not
simply in a tachyon as in the scalar case. We stress that a massless
vector has only the two transverse polarizations; it is therefore not
surprising that the mass term controls the nature of the longitudinal
mode. Another example on how the mass term controls the nature of a
mode is given by the more complicated case of the graviton mass $m_1^2
h^\mu_\mu h^\nu_\nu - m_2^2 h_{\mu\nu} h^{\mu\nu}$. It is well known
that, unless $m_1 = m_2$ (namely, the Fierz-Pauli mass term
\cite{Fierz:1939ix}) the linearized spectrum of perturbations has a
ghost.

The most obvious problem associated to a ghost is the instability of
the vacuum. If a ghost is coupled to a normal field (and, in all the
above theories, there are at least gravitational couplings), the
vacuum will decay in ghost-nonghost excitations, with a rate which is
UV divergent (since the final state quanta can have arbitrarily large
momentum without violating energy conservation). To avoid the
associated instantaneous vacuum decay, theories with ghosts are
thought to be consistent only as effective theories, valid below some
energy scale $\Lambda$. It has been shown in \cite{cline} that, a
theory which has a ghost today coupled gravitationally to positive
energy fields is phenomenologically viable only if $\Lambda \la {\rm
  MeV} \,$, otherwise we would see signatures of the vacuum decay in
the diffuse $\gamma-$ray background. A stronger coupling would result
in a tighter bound on $\Lambda$. All the above models discuss physics
at much greater energy scales than MeV, so, a necessary (but, as we
will see, not sufficient) condition for them to be consistent is that
the ghost  vanishes at late times. The easiest way to achieve this
is to add a positive contribution $m^2 > 0$ to the mass term, $M^2 = -
R/6 + m^2$, and to require that $m^2$ is lower than $-R/6$ during
inflation, but greater than it today. In fact, this is already built
in in the models of vector inflation \cite{mukhvect,soda1}, for which
$m^2$ controls the slow roll evolution of the vev of the vector. The
same can be also done in the case of vector curvaton, where it is
found that the mass provides an ${\rm O } \left( m^2 / H^2 \right)$
departure from scale invariance ($H$ being the Hubble rate during
inflation) \cite{Dimopoulos}. For the case of photons, there are
strong upper limits on the allowed value for $m$ (see the limits on
the PDG Particle Listing \cite{Amsler:2008zzb}). These limits are
however less stringent than the present value of $R/6 = {\rm O} \left(
  10^{-33} \, {\rm eV} \right)^2 \,$.

Even if it is straightforward to eliminate the ghost(s) from the
present spectrum, this does not eliminate the instability of these
models during or after the inflationary stage. We will comment on
problems that arise at the nonlinear level in the Discussion Section
\ref{sec:discussion}. We stress here that the ghost instability of
these models manifests itself already at the linearized level. The
spectrum of the theory is obtained by computing the quadratic action
for the dynamical perturbations $Y_i$ around the given background. We
do so by following, and generalizing to the present case, the exact
same steps that are done in the standard computation of scalar field
inflation to obtain the canonical modes of the system \cite{mfb}. The
quadratic action is characterized by a kinetic $K_{ij} \, \dot{Y}_i^*
\, \dot{Y}_j$ term, a mass term, and a mixed term (see
eq.~(\ref{action-integrated})). The eigenvalues of $K$ control the
nature of the physical modes (i.e., a mode is a positive / negative
energy excitation if the corresponding eigenvalue is positive /
negative). The kinetic matrix depends on background quantities, and
thus on time. As a consequence, its eigenvalues depend on time, and
the nature of a mode can change during the background evolution. From
our computations, we find two different types of behavior, according
to whether the vev of the vector is or is not vanishing:

\begin{itemize}

\item $\langle \vec{A} \rangle = 0$: one eigenvalue of $K$ is negative during most of the sub-horizon regime; it changes sign at some moment close to horizon crossing, without passing through zero (it does so by diverging). It crosses zero later on, when the total mass $M^2$ vanishes.

\item $\langle \vec{A} \rangle \neq 0$: one eigenvalue of $K$ starts positive, but crosses zero at some point close to horizon crossing, and remains negative for some amount of time.

\end{itemize}

The different behaviors are due to the fact that, for a nonvanishing
vev, the perturbations of the vector are mixed with those of the
metric at the linearized level (i.e. in the quadratic action), and
this affects the spectrum of the theory. However, we see that the
mixing with gravity does not remove the ghost. The case of vanishing
vev applies to the computations of \cite{Turner:1987bw} and
\cite{Dimopoulos}.~\footnote{Notice, however, that for a vector
  curvaton the vev cannot be exactly zero if one wants to realize a
  violation of statistical isotropy of the perturbations. Indeed, to
  have a violation of statistical isotropy, one needs to ``single
  out'' different direction(s); this is precisely provided by the vev
  of the vector. We  follow the computations of
  \cite{Dimopoulos} which are performed under the assumption of zero
  or negligible vev. For a nonvanishing vev, the model of
  \cite{Dimopoulos} belong to the second class of models just
  mentioned, and the ghost instability manifests itself in the second
  way just mentioned.} The case of a vector with a vev applies to the
model of \cite{soda1}. The vector inflation model of \cite{mukhvect}
is more complicated, since it contains an arbitrary number $N$ of
vectors (and, hence, of ghosts). We study the simplest realization,
with three mutually orthogonal vectors. We find that, in this case,
the model has three ghosts. Two eigenvalues behave as in the $\langle
\vec{A} \rangle = 0$ case just mentioned, while the third eigenvalue
behaves as in the $\langle \vec{A} \rangle \neq 0$ case. Hence, it
appears that in this case only one linear combination of the ghosts is
affected by the coupling to gravity.

As we show below, the system of linearized equations for the
perturbations becomes singular when one of the eigenvalues of the
kinetic matrix crosses zero. Correspondingly, we expect that the
linearized solutions diverge at this moment. We explicitly solved the
equations in the case of a single vector with no vev, and in the case
of one vector with nonvanishing vev plus a cosmological constant. We
did not solve the equations for the case of vector inflation (due to
technical difficulties: the system contains $18$ gauge invariant
perturbations in its simplest realization). However we stress that,
also for this model, we explicitly proved that (i) there are ghosts
(which is by itself enough to pose serious doubts on any prediction obtained from
this model; see the Discussion Section), and (ii) the equations for
the perturbations become singular at some finite moments of time.

The plan of the paper is the following. In Section \ref{sec:review},
we review the basic mechanism for the models of generation of
primordial magnetic fields, vector inflation, and vector curvaton. In
Section \ref{sec:ghost}, we outline the computation of the quadratic
action for the physical modes of the system. We also show there why
vanishing eigenvalues of the kinetic matrix $K$ result in singular
linearized equations for the perturbations. In Section \ref{sec:novev}
we present the computations for the $\langle \vec{A} \rangle = 0$
case. We show that the theory has a ghost, and that the linearized
perturbations diverge where the total mass of the vector vanishes. In
Section \ref{sec:vev} we study the case of nonvanishing vev. We first
discuss the simplest possibility of a vector and a cosmological
constant. For this case, we both compute the quadratic action, and
solve the linearized equations of the perturbations. Once again we
verify that the modes diverge precisely when one eigenvalue of the
kinetic matrix $K$ crosses zero. We then study the case of one vector
and one inflaton \cite{soda1}, and the simplest realization of vector
inflation \cite{mukhvect}. For these models we compute the kinetic
term of the physical perturbations, and we study how the eigenvalues
evolve in time. Conclusions and discussions are given in Section
\ref{sec:discussion}.

\section{Review of some models with a $R \, A^2$ term}

\label{sec:review}

In this Section, we briefly review the reasons for introducing a
nonminimal coupling of a vector field to the curvature. We start from
the quadratic action of a vector field with a generic time-dependent
mass:
\begin{equation}
S = \int d^4 x \sqrt{-g} \left[ - \frac{1}{4} F^2 - \frac{1}{2} \, M \left( t \right) ^2  A^2 \right]
\label{acmass}
\end{equation}
leading the equations of motion
\begin{equation}
\frac{1}{\sqrt{-g}} \, \partial_\nu \left( \sqrt{-g} \, F^{\mu \nu} \right) + M^2 \, A^\mu = 0
\label{eqmass}
\end{equation}
Moreover, due to its antisymmetry, the field strength $F_{\mu \nu}$ satisfies the identity
\begin{equation}
\partial_{\mu}\, F_{\rho\nu} + \partial_{\nu}\, F_{\mu\rho} +
\partial_{\rho}\, F_{\nu\mu} = 0 \label{mx2}
\end{equation}

Turner and Widrow \cite{Turner:1987bw} suggested a mechanism for the
generation of primordial magnetic fields during inflation, starting
from the action (\ref{acmass}).~\footnote{More precisely,
  ref.~\cite{Turner:1987bw} studied a more general action with a
  quadratic term of the type $R_{\mu \nu} A^\mu A^\nu$ also included;
  we disregard this term in the present work.} In this mechanism, the
vector field $A_\mu$ is the electromagnetic field, and the mass term
is proportional to the scalar curvature $R$. Following
\cite{Turner:1987bw}, we use conformal time $\eta$, defined by the
line element $d s^2 = a^2 \left( \eta \right) \left( - d \eta^2 + d
  \vec{x}^2 \right)$ in the present discussion, so that the electric
and magnetic fields are ($\epsilon_{123} = + 1$)
\begin{equation}
F_{i0} = a^2 \, E_i \;\;\;,\;\;\; F_{ij} = a^2 \, \epsilon_{ijk} \, B_k
\end{equation}
The two equations (\ref{eqmass}) and (\ref{mx2}) can be easily combined into an equation for the magnetic field \cite{Turner:1987bw}:
\begin{equation}
\left( \partial_\eta^2 - \partial_{\vec{x}}^2 + a^2 \, M^2 \right) \left( a^2 \vec{B} \right) = 0
\end{equation}
From this equation, we find the well known result that, in the massless case ($M^2=0$), and in the large wavelength limit (i.e. negligible spatial gradient) , the amplitude of a magnetic field decreases as $\propto a^{-2}$. Correspondingly, its energy density decreases as  $\propto a^{-4}$. Consider instead the mass term \cite{Turner:1987bw}
\begin{equation}
M^2 = \xi \, R = 6 \, \xi \, \frac{a''}{a^3}
\end{equation}
where $\xi$ is a constant, and prime denotes derivative wrt $\eta
\,$. The equation of motion for the magnetic field then becomes
\begin{equation}
\left( \partial_\eta^2 -  \partial_{\vec{x}}^2 + 6 \, \xi \, \frac{a''}{a} \right) \left( a^2 \vec{B} \right) = 0
\end{equation}

During inflation, $a = - 1 / \left( H \, \eta \right)$ where the Hubble rate $H$ is nearly constant; inserting this into the equation of motion (and treating $H$ as constant), we find that, in the large wavelength regime,  the energy density of the magnetic field behaves as
\begin{equation}
\rho_B \propto \vec{B}^2 \propto a^{-5 + \sqrt{1-48 \, \xi}}
\end{equation}
Therefore, for $\xi < 0$ - corresponding to a negative $M^2$ in the action (\ref{acmass}) - $\rho_B$ is less affected by the expansion with respect to the massless case.

It is instructive to compare the behavior of the magnetic field with respect to that of a massless scalar field coupled to the curvature, characterized by the action
\begin{equation}
S = \int d^4 x \, \sqrt{-g} \left[ - \frac{1}{2} \partial_\mu \phi \partial^\mu \phi - \frac{1}{2} \, \xi_s \, R \, \phi^2 \right]
\end{equation}
This leads to the equation of motion
\begin{equation}
\left( \partial_\eta^2 - \partial_{\vec x}^2 +\left( 6 \xi_s - 1 \right) \frac{a''}{a} \right) \left( a \, \phi \right) = 0
\end{equation}
By comparing  this equation with the analogous expression for the vector field~\footnote{It is appropriate to compare the behavior of $a^2 \vec{B}$ with that of $a \, \phi$, since, due to the different structures of the kinetic terms, these are the canonically normalized fields in the two cases.}, it has been noted \cite{Turner:1987bw} that a vector field with $\xi = - 1/6$ behaves analogously to a scalar field minimally coupled to the curvature ($\xi_s = 0$). Conversely, the standard vector field $\xi =0$ is analogous to a conformally coupled scalar, $\xi_s = 1/6 \,$ (therefore, no magnetic field is produced by the inflationary expansion in the standard case).

This analogy has been recently exploited in \cite{mukhvect}, that proposed a mechanism of inflation driven by a combination of $N$ nonminimally coupled vector fields. The mass term of the vectors comprises of the coupling to the curvature plus a constant term:
\begin{equation}
S = \int d^4 x \, \sqrt{-g} \,  \sum_{a=1}^{N}\, \left[ -\frac{1}{4}\,
F^{(a)}_{\mu\nu}\, F^{(a)\, \mu\nu} - \frac{1}{2}\, \left( \xi \, R + m^2 \right)\, A_{\mu}^{(a)}\, A^{(a)\mu} \right]
\label{ac-gmv}
\end{equation}
The simplest case, and the one which has been most studied in  \cite{mukhvect}, is characterized by three mutually orthogonal vectors with equal vev:
\begin{equation}
\langle A^{(1)}_\mu \rangle = \left( 0 ,\, {\cal A} ,\, 0 ,\, 0 \right) \;\;\;,\;\;\;
\langle A^{(2)}_\mu \rangle = \left( 0 ,\, 0 ,\, {\cal A} ,\, 0 \right) \;\;\;,\;\;\;
\langle A^{(3)}_\mu \rangle = \left( 0 ,\, 0 ,\,  0 ,\, {\cal A} \right) \;\;\;,\;\;\;
{\cal A} \equiv  M_p \, a \left( t \right) \, B \left( t \right)
\label{3vev}
\end{equation}
These sources allow for a FRW background, controlled by the equations
of motion (we switch to physical time, and we denote by a dot derivative with respect to it)
\begin{eqnarray}
&& 
H^2 - \frac{\dot{B}^2}{2} - \frac{1}{2} \, m^2 \, B^2 = \frac{1+6 \, \xi}{2} \, H \, B \, \left( 2 \, \dot{B} + H \, B \right) \nonumber\\
&&
\ddot{B} +3 \, H \, \dot{B} + m^2 \, B = - \left( 1 + 6 \, \xi \right) B \left( \dot{H} + 2 \, H^2 \right)
\end{eqnarray}
For $\xi = -1/6$, and upon the identification $B = \phi / \left( \sqrt{3} M_p \right)$, we recover the same equations as those of chaotic inflation driven by a minimally coupled scalar field $\phi$.

Another compelling feature of the proposal of  \cite{mukhvect} is that it can naturally give a small violation of isotropy. Indeed, for a large number $N$ of vectors with random orientations and vev, one expects an almost isotropic expansion, with a deviation $\Delta H / H = {\rm O } \left( 1/ \sqrt{N} \right)$ between the expansion rates of the different directions. Ref. \cite{soda1} provides a slightly different mechanism in which a single nonminimally couple vector breaks the isotropy in one spatial direction, while a scalar field with greater energy density is responsible for the overall nearly isotropic expansion.

Vector fields with nonminimal coupling to the curvature have also been recently employed for the generation of a nearly scale invariant spectrum of perturbations \cite{Dimopoulos}. Consider the action (\ref{ac-gmv}) for a single field ($N=1$) and with $\xi = 1/6$. Following the discussion of  \cite{Dimopoulos}, we compute the evolution of the perturbations $\delta A_\mu$ assuming that the vev $\langle A_\mu \rangle$ can be neglected (see the remark we made about this in the Introduction). In this way, the perturbations of the vector do not mix with those of the metric at the linearized level. Moreover, with a negligible vector vev, we can study the evolution of $\delta A_\mu$ in an unperturbed FRW background. In the present discussion, we only consider  the transverse components of the perturbations, $\delta A_\mu = \left( 0 ,\, \delta \vec{A}^T \right) \,$, with $\partial_i 
\, \delta A_i^T = 0 \,$, since, as we will show in the following Section, there are serious problems with the longitudinal mode in these models. Due to the $\xi = - 1/6$ choice, the equation for this mode is identical to that of a minimally coupled curvaton scalar field:
\begin{equation}
\left[ \partial_\eta^2 - \partial_{\vec{x}}^2 + a^2 \left( m^2 - \frac{a''}{a^3} \right) \right] \delta A_i^T = 0
\label{eq-trans}
\end{equation}
We proceed as in the scalar curvaton case. For simplicity, we assume a dS background; 
in momentum space, the Fourier transform of $\delta A_i^T$ obeys to the equation
\begin{equation}
\left( \partial_\eta^2 + k^2 - \frac{2 \, \mu^2}{\eta^2} \right) \delta {\tilde A}_i^T = 0 \;\;\;,\;\;\;
\mu^2 \equiv 1 - \frac{m^2}{2 H^2}
\label{eq-at}
\end{equation}
Among the two solutions of this equation, we chose the one that reduces to the adiabatic vacuum at early times, when the mode is deeply inside the horizon:
\begin{equation}
\delta {\tilde A}_i^T = \frac{\sqrt{\pi}}{2} \, \sqrt{\vert \eta \vert} \, H_{\frac{1}{2} \sqrt{1 + 8 \mu^2}}^{(1)}  \left( 
\vert k \, \eta \vert  \right)
\label{norma-at}
\end{equation}
where $H^{(1)}$ is the Hankel function of the first kind. Indeed, up to an irrelevant phase, this solution reduces to ${\rm e}^{-i k \eta}/\sqrt{2 k}$ for $\vert k \eta \vert \gg 1 \,$. In the opposite late time / super horizon regime ($\vert k \eta \vert \ll 1 \,$), from the expansion of the Hankel function, we then find the power spectrum for the transverse modes  
\cite{Dimopoulos}
\begin{equation}
P_{Ti} \propto k^3 \, \vert \delta {\tilde A}_i^T \vert^2 \propto k^{3-\sqrt{1+ 8 \, \mu^2}} = k^{\frac{2 m^2}{3 H^2} + {\rm O} \left( \frac{m^4}{H^4} \right)} \label{spec-vec}
\end{equation}
Namely $m \ll H$ provides a small departure from scale invariance. We remark that this result follows from the solution (\ref{norma-at}). Indeed, eq. (\ref{eq-at}) has two solutions with two undetermined (and, in principle, $k-$dependent) integration constants. As it is customary, we chose the linear combinations of these solutions which reduces to the adiabatic vacuum in the early time sub-horizon regime. The phenomenological prediction (\ref{spec-vec}) crucially relies on the fact that the theory must be under control in this regime.

For all the models we have reviewed, the discussion presented in this Section disregards the role of the longitudinal polarization of the vector field(s). In the next Sections we show that, for all the models discussed, this mode turns out to be a ghost.

\section{Identification of ghosts, and their associated instability}
\label{sec:ghost}

In the next two Sections, we will see that the models described in the
previous Section have ghosts. We outline here the method we employ to
find the ghosts, and the instability associated with them. The ghosts
are among the physical excitations of the background geometries of the
various models. Therefore, we need to compute the spectrum of these
theories. To do so, we perturb the background solutions of a given
model, and we expand the action at quadratic order in the
perturbations. This is the free action for the perturbations
(interaction among these fields come from expanding the initial action
to higher orders), and the spectrum follows from the diagonalization
of this action.

There are two main issues in this computation. The first one is
associated to the gauge freedom that may be present in a theory. A
gauge theory is in a sense a redundant formulation of a physical
system, since different field configurations, related to each other by
a (nonsingular) gauge transformation, describe the same physics. In
the present context, the gauge freedom is the one associated with
general coordinate transformations. We can expand a gauge
transformation in a transformation acting on the background plus a
transformation acting on the perturbations (loosely speaking, we
decompose any gauge transformation into a ``big transformation'',
affecting the background, plus a ``small'' transformation, affecting
the perturbations on a given background). In any explicit computation,
one typically fixes the gauge freedom for the background and the
perturbations separately. Concerning the latter step, one can choose a
gauge that fixes the freedom completely (for instance, one can impose
that some perturbations vanish; one needs to show that this choice can
always be done, and that there is no residual freedom left);
equivalently, one can find a set of gauge invariant linear
combinations that do not change under the gauge transformation. Since
the theory is gauge invariant, it must be possible to write down the
equations of motion only in terms of the gauge invariant combinations,
and solve those expressions. This is the method that we adopt in our
explicit computations.

The second issue is that, even after removing the gauge redundancy,
the remaining perturbations do not all necessarily describe dynamical
degrees of freedom (e.g., physically propagating excitations). Modes
that do not correspond to dynamical excitations enter in the action
without time derivatives (up to boundary terms, which can be
disregarded in a theory without boundaries). Example of nondynamical
degrees of freedom are for instance the $\delta A_0$ component of a
vector, or the $\delta g_{\mu 0}$ components of the metric. In the
standard case (i.e., for ${\cal L} \supset F_{\mu \nu} F^{\mu \nu}$,
and ${\cal L} \supset R$), they enter in the action without time
derivatives. In general, the quadratic action for the perturbations
around a given background is formally of the type (in momentum space)
\begin{equation}
S = \int d^3k\, dt\, \left[ a_{ij}\, \dot{Y}_i^*\, \dot{Y}_j +
\left( b_{ij}\, N_i^*\, \dot{Y}_j + {\rm h.c.} \right) + c_{ij}\,
N_i^*\, N_j + \left( d_{ij}\, \dot{Y}_i^*\, Y_j + {\rm h.c.}
\right) + e_{ij}\, Y_{i}^*\, Y_{j} + \left( f_{ij}\, N_i^*\, Y_{j}
+ {\rm h.c.} \right) \right] \label{formal-act}
\end{equation}
where $Y_i$ are the dynamical modes, and $N_i$ the nondynamical ones. The coefficients $a_{ij} ,\, \dots ,\, f_{ij}$ depend on background quantities, and, for any given background solution, are functions of time.  From the action (\ref{formal-act}), we find the linearized equations
\begin{eqnarray}
&&
\frac{\delta S}{\delta Y_i^*} = 0 \,\,\, \Rightarrow \,\,\,
a_{ij} \, \ddot{Y}_j + \left[ \dot{a}_{ij} + d_{ij} - \left( d^\dagger \right)_{ij} \right] \dot{Y}_j
+ \left( b^\dagger  \right)_{ij} \, \dot{N}_j + \left[ \dot{d}_{ij} - e_{ij} \right] Y_j + \left[ 
\left( \dot{b}^\dagger \right)_{ij} - \left( f^\dagger \right)_{ij}   \right] N_j = 0 \label{eq-dyn}\\
&&
\frac{\delta S}{\delta N_i^*} = 0 \,\,\, \Rightarrow \,\,\,
c_{ij}\, N_j = -b_{ij}\, \dot{Y}_j - f_{ij}\, Y_j
\label{eq-nondyn}
\end{eqnarray}

We see explicitly that the equations of motion for the nondynamical modes are algebraic in them (they are constraint equations). As a consequence, the nondynamical modes do not introduce additional degrees of freedom, but are completely determined by the dynamical ones. We can obtain a system of differential equations for only the dynamical modes by inserting the solutions of eqs.~(\ref{eq-nondyn}) into equations (\ref{eq-dyn}):
\begin{equation}
K_{ij}\, \ddot{Y}_j + \left[ \dot{K}_{ij} + (\Lambda_{ij} - {\rm
h.c.}) \right]\, \dot{Y}_j + \left( \dot{\Lambda}_{ij} +
\Omega_{ij}^2 \right)\, Y_j = 0 \label{formal-eom}
\end{equation}
where
\begin{eqnarray}
&& K_{ij} \equiv a_{ij} - \left(b^{\dag}\right)_{ik}\,
\left(c^{-1}\right)_{km}\, b_{mj}
\nonumber\\
&& \Lambda_{ij} \equiv d_{ij} - \left(b^{\dag}\right)_{ik}\,
\left(c^{-1}\right)_{km}\, f_{mj}
\nonumber\\
&& \Omega_{ij}^2 \equiv - e_{ij} + \left(f^{\dag}\right)_{ik}\,
\left(c^{-1}\right)_{km}\, f_{mj} 
\label{klo}
\end{eqnarray}
To obtain instead the quadratic action for the dynamical modes, we insert the solutions of eqs.~(\ref{eq-nondyn}) into the action (\ref{formal-act}):
\begin{equation}
S \rightarrow \int d^3k\, dt\, \left[ \dot{Y}_i^*\, K_{ij}\,
\dot{Y}_j + \left( \dot{Y}_i^*\, \Lambda_{ij}\, Y_j +
{\rm h.c.} \right) - Y^*_i\, \Omega_{ij}^2\, Y_j \right] 
\label{action-integrated}
\end{equation}
The extremization of this action provides exactly the above equations (\ref{formal-eom}).

Eq.~(\ref{eq-nondyn}) and (\ref{formal-eom})  - or, equivalently, eqs.~(\ref{eq-dyn}) and (\ref{eq-nondyn}) - are the equations of motion for the perturbations. We can also obtain these equations in a different but equivalent form  by simply expanding at first order in the perturbations the general equations of motion of the model. In general, this last procedure is technically simpler than expanding the action at second order in the perturbations, and  extremizing it. However, while an action provides the equations of motion, in general the equations of motion do not fully determine the action. The quadratic action of the perturbations is necessary to quantize the perturbations, and provide their initial conditions. 

The procedure just outlined formalizes the steps that are done in the  standard case of scalar field inflation on a FRW background. Moreover, it extends it to an arbitrary number of dynamical and nondynamical fields. In the standard case, there are $5$ scalar modes in the perturbations of the metric and of the scalar field. Two of them are eliminated by gauge fixing (or, equivalently, by the use of gauge invariant variables). Only $1$ out of the $3$ remaining modes is dynamical. While one can show that the action can be written solely in terms of gauge invariant quantities without using the equations of motion, the constraint equations are used to eliminate the nondynamical modes (for instance, the constraint equation (10.39) is used to obtain the final action (10.59) in \cite{mfb}).  The final step in the computation is to normalize this final mode so that the kinetic term of this final action is canonical.~\footnote{ In the standard case, the canonically normalized scalar perturbation is the Mukhanov-Sasaki variable $v$ \cite{musa}. We explicitly verified that, in the standard case of scalar field inflation, the action (\ref{action-integrated}) obtained with the procedure described here coincides with the action (10.59) of \cite{mfb}.}

For a general problem, with more than one dynamical mode in the system,   one needs to diagonalize the kinetic matrix $K$. If one  eigenvalue of $K$ is negative, the corresponding eigenmode is a ghost (a field with negative energy). This signals an instability of the vacuum, which decays in ghost-nonghost excitations with a rate which diverges in the UV. A second type of instability takes place whenever, due to the cosmological background evolution, an eigenvalue of $K$ crosses zero, and $K$ is noninvertible.
Denote by $t_*$ one of the moments at which this happens. The equations of motion become singular for $t \rightarrow t_* \,$, since the second derivative of the modes diverge in this limit. We expect that, correspondingly, the solutions of the linearized system diverge for $t \rightarrow t_* \,$.~\footnote{To integrate the system (\ref{formal-eom}), one needs to invert the matrix $K$; in this way one obtains a system of equations of the form $\ddot{Y}_i = \dots$, where the right hand sides contain only zero or first time derivatives of the unknowns. This system can be then integrated numerically. The inversion of $K$
can be performed at all times $t < t_*$, and the resulting equations are regular until $t_*$; however, the expression for the second derivatives blows up for $t \rightarrow t_*$, and we expect that the solutions diverge at this moment. This is precisely what happens in the two examples below in which we explicitly solve the linearized equations. This is also what happened in our previous study \cite{hcp2} of the model \cite{acw} which presents an identical instability.} In the next two Sections, we show that both types of instability occur in the models we are studying.

\section{Ghost instability for $\langle A_\mu \rangle = 0$}
\label{sec:novev}

We start our analysis from the case in which the vector field has no vev. This is the case for the computations presented in \cite{Turner:1987bw,Dimopoulos} to study the evolution of the modes. On a technical level, the assumption of zero vev drastically simplifies the computation. Indeed, the actions we are studying (cf. eqs. (\ref{acmass}) and  (\ref{ac-gmv})), are quadratic in the vectors; therefore, for $\langle A_\mu \rangle = 0$, the perturbations of the vector(s) are decoupled from those of any other field at the linearized level.~\footnote{In the mechanism of \cite{Turner:1987bw}, the full action in not quadratic in $A_\mu$, since the photon is obviously coupled to the current of charged fields. This coupling is important from reheating on, when the conductivity of the charged particles become high, and affects the evolution of the magnetic field. However, the presence of these fields can be disregarded during the inflationary stage, as done in \cite{Turner:1987bw}.} Therefore, the action for the vector field already starts at quadratic order in the perturbations, and we can simply study the evolution of $\delta A_\mu$ in an unperturbed FRW background. The more involved case of $\langle A_\mu \rangle \neq 0$ is studied in the following Section.

Consider the action
\begin{equation}
S = \int d^4 x \sqrt{-g} \left[ - \frac{1}{4} F^2 - \frac{1}{2} \, M^2 \, A^2 \right]
\;\;\;,\;\;\; M^2 = - \frac{R}{6} + m^2
\label{acnovev}
\end{equation}
in the FRW background, $d s^2 = - d t^2 + a \left( t \right)^2 d {\vec x}^2 \,$. Parametrize the fluctuations of the vector field as $A_{\mu}=\left( \alpha_0, \, \partial_i\, \alpha + \alpha_i^T \right)$, where $i=1,2,3$, and where the modes $\alpha_i^T$ are transverse ($\partial_i \alpha_i^T= 0$). These two polarizations  are decoupled from the modes $\{ \alpha_0, \, \alpha \}$ (and also from each other) at the linearized level, and have been already studied in the previous Section (starting from eq.~(\ref{eq-trans})). As we saw, they are well behaved in all regimes. Here we study the nature and the evolution of the other two perturbations.

It is well known that a massive vector has three physical degrees of freedom. The two transverse ones are encoded in $\alpha_i^T$. Therefore, the two perturbations $\alpha_0$ and $\alpha$ encode only one physical mode, which is the longitudinal polarization of the massive vector. This can be immediately seen from the equations of motion for the two perturbations following from (\ref{acnovev}), which, in Fourier space, read~\footnote{The first equation is $\delta S / \delta A_0 \,$; the second equation has been obtained by combining $\delta S / \delta A_0 \,$ and $\delta S / \delta A_i \,$.}
\begin{eqnarray}
&&\alpha_0 = \frac{p^2}{p^2 + M^2} \dot{\alpha} \label{al0eq} \\
&&\ddot{\alpha} + \frac{\left( 3 p^2 + M^2 \right) H + p^2 \, \frac{\frac{d \, M^2}{d t}}{M^2}}{p^2 + M^2} \dot{\alpha} + \left( p^2 + M^2 \right) \alpha = 0 \label{aleq} 
\end{eqnarray}
where $p=k/a$ is the physical momentum of the perturbation, and, as usual, $H = \dot{a} / a$. While eq. (\ref{aleq}) is a second order differential equation, eq. (\ref{al0eq}) is an algebraic equation in $\alpha_0$. Therefore, $\alpha_0$ does not introduce additional degrees of freedom, but it is completely determined once $\alpha$ is known (compare these equations with the formal set of equations ({\ref{eq-dyn}) and (\ref{eq-nondyn})).

We can also see this from the quadratic action for the perturbations. Inserting the decomposition 
$A_{\mu}=\left( \alpha_0, \, \partial_i\, \alpha \right)$ in (\ref{acnovev}), and Fourier transforming the spatial coordinates, we find
\begin{equation}
S = \frac{1}{2} \int d t \, d^3 k \left[ p^2 \vert \dot{\alpha} \vert^2 - p^2 M^2 \vert \alpha \vert^2 - p^2 \left(
\alpha_0^* \, \dot{\alpha} + {\rm h.} \, {\rm c.} \right) + \left( p^2 +M^2 \right) \vert \alpha_0 \vert^2 \right]
\label{ac-aa0}
\end{equation}
The mode $\alpha_0$ enters in the action without time derivatives, confirming that it is a nondynamical mode. We integrate it out: we compute its equation of motion from this action (namely, eq. (\ref{aleq}) given above), we solve it, and we insert the solution back into (\ref{ac-aa0}). This leads to the action for the longitudinal mode:~\footnote{We remark that, in going from the action~(\ref{ac-aa0}) to the action~(\ref{ac-a}),   we are precisely following the general procedure that we outlined at a formal level in the previous Section.}
\begin{equation}
S = \frac{1}{2} \int d t \, d^3 k  \: p^2 \, M^2 \left( \frac{\vert \dot{\alpha} \vert^2}{p^2 + M^2} - \vert \alpha \vert^2 \right)
\label{ac-a}
\end{equation}
The extremization of this action reproduces the equation of motion (\ref{aleq}).~\footnote{This equation, in conformal time, was already given in~\cite{Dimopoulos}; this confirms our algebra. The action for the longitudinal mode could be also written starting from this equation, up to an overall factor. The procedure described here provides the complete action. We also note that the Ref. \cite{Dimopoulos} did not solved the equation at the moment in which the total mass of the vector vanishes; as we show here, the solution diverges at this moment.} We remark that the lagrangian in (\ref{ac-a}) is proportional to the $M^2$ term, and that, for $p^2 > \vert M^2 \vert$,  the longitudinal mode is a ghost (i.e. a field with negative energy, not simply a tachyon) whenever $M^2 < 0 \,$ \cite{hcp1,hcp2}.  In our previous work \cite{hcp2} we provided two additional proofs of the fact that $M^2 <0$ leads to a ghost. The first is based on the direct computation of the propagator (we find that one residue at the pole of the propagator has the ``wrong'' sign for $M^2 < 0$); the second on the Stuckelberg formalism (which is a convenient way to study the different polarizations of a massive vector or graviton; we rewrite $A_\mu = A_\mu^T + \partial_\mu \phi / \sqrt{\vert M^2 \vert}$, with $\partial^\mu \, A_\mu^T = 0$; it is then immediate to see that the field $\phi$ is a ghost for $M^2 < 0 \,$). 

The presence of a ghost signals the instability of the vacuum of the model, due to its (UV-divergent) decay in ghost-nonghost excitations. In addition, the equation of motion (\ref{aleq}) indicates that the system may be unstable already at the linearized level, whenever $\omega^2 \equiv p^2+M^2$ or $M^2$ vanish. Let us first discuss when this occurs. During inflation, $m^2 \ll R$ for the models we are studying. As a consequence, during inflation, $M^2 = m^2 - R / 6 \simeq - 2 H^2 < 0 $, while $\omega^2 = p^2 + M^2 \simeq p^2 - 2 H^2$ goes from positive to negative. We denote by $t_{\omega1}$ the moment at which $\omega^2$ vanishes; we note that this happens when the mode is close to horizon crossing, and that $\omega^2$ remains negative from $t_{\omega1}$ until  a time well after the end of inflation (since the mode is well outside the horizon, $p \ll H$, when inflation ends). After inflation, $R /6 = {\rm O} \left( H^2 \right)$ decreases as the universe expands, and it eventually drops below $m^2$. At this moment, which we denote by $t_M$, the mass term $M^2$ goes from negative to positive. Therefore, $\omega^2 = p^2 + M^2 > 0$ for all times $t \geq t_M \,$. Since we saw that $\omega^2 < 0$ at the end of inflation, there is a moment after the end of inflation, and before $t_M$, at which $\omega^2$ vanishes for the second time. We denote this moment by $t_{\omega2}$. In summary, $M^2$ vanishes at $t = t_M$, while $\omega^2$ vanishes at $t=t_{\omega1}, t_{\omega2}$. Denoting by $t_{\rm end}$ the moment at which inflation finishes, and by $t_0$ the present time, we have $t_{\omega1} < t_{\rm end} < t_{\omega2} < t_M < t_0 \,$.~\footnote{One may imagine that $m$ is so small, so that $t_M$ has not occurred yet. In this case the longitudinal polarization is still a ghost today; we disregard this possibility, due to the stringent limits on theories with ghosts found by \cite{cline} and discussed in the Introduction.}

To see whether the linearized system diverges when either  $\omega^2$ or $M^2$ vanish, we study the equation of motion (\ref{aleq}) for a finite interval of time close to $t_{\omega i}$ ($i=1,2$) or $t_M$. We assume that the equation of state $w$ of the source driving the expansion can be treated as constant in
this interval (which is certainly true, provided the interval is not too extended). We also assume that  $- 1 < w < 1/3$. This is a rather general assumption, since it includes the following cases: inflation (with $w \ga -1$), coherent oscillations of the inflaton after inflation (which, for a quadratic inflaton potential, give an average $w=0$), matter domination (if $m^2$ is sufficiently small, so that $M^2$ vanishes at this stage), and also radiation domination (for which the equation of state is slightly smaller than $1/3$, due to the masses of the particles in the thermal bath, or the thermal trace anomaly).~\footnote{The thermal trace anomaly is relevant for temperatures greater than the QCD phase transition, and gives $1/3 - w = {\rm O} \left( 10^{-3} \right)$ \cite{Cembranos:2009ds}. Mass thresholds give $1/3 - w = {\rm O} \left( 10^{-2} \right)$ when the temperature is close to the mass of a particle \cite{Coc:2006rt}.  For temperatures below any Standard Model particle, the departure from $w=1/3$, and $R=0$ can be neglected. However, as $R$ drops to zero, there will be a moment in which $M^2$ vanishes; we denote by $w$ the value of the equation of state at this moment.}

\bigskip  

{\bf Behavior of the linearized system for $p^2 + M^2 \rightarrow 0 \,$.} The scale factor evolves as
\begin{equation}
a = a_{\omega i} \left( \frac{t}{t_{\omega i}} \right)^{\frac{2}{3 \left( 1 + w \right)}}
\end{equation}
were we recall that $t_{\omega i}$ is either of the times at which $\omega^2 = 0$, and $a_{\omega i} $ is the value of the scale factor at that time. The mass and frequency squared are given by
\begin{eqnarray}
M^2  &=& m^2 - \frac{R}{6} = m^2 - \frac{2}{9 \, t^2} \, \frac{1 - 3 w}{\left( 1 + w \right)^2} 
\label{m20}\\
\omega^2 &=& p^2 + M^2 = p_{\omega i}^2 \, \left( \frac{t_{\omega i}}{t} \right)^{\frac{4}{3 \left( 1 + w \right)}} + m^2 - \frac{2}{9 \, t^2} \, \frac{1 - 3 w}{\left( 1 + w \right)^2}
\label{o20}
\end{eqnarray}
where $p_{\omega i}$ is the value of the physical momentum at $t_{\omega i}$. We then find
\begin{equation}
\omega \left( t_{\omega i} \right) = 0 \;\;\;\Rightarrow\;\;\; t_{\omega i} = \frac{\sqrt{2} \, \sqrt{1-3 \, w}}{3 \, \sqrt{m^2 + 
p_{\omega i}^2} \, \left( 1 + w \right)}
\end{equation}
We insert these expression for the momentum and the total mass  in eq.~(\ref{aleq}), and we Taylor expand  the resulting expression for $t \simeq t_{\omega i}$. We find:
\begin{equation}
\ddot{\alpha}  - \frac{\dot{\alpha}}{t - t_{\omega i} }  
+  C \, \left( t - t_{\omega i} \right) \, \alpha \approx 0
\label{eqappo0}
\end{equation}
where
\begin{equation}
C = \sqrt{\frac{2 \left( m^2 + p_{\omega i}^2 \right)}{1-3 w}} \, \left[ 3 m^2 \left( 1 + w \right) + p_{\omega i}^2 \left( 1 + 3 \, w \right) \right] 
\end{equation}
Eq. (\ref{eqappo0}) is solved by
\begin{equation}
\alpha \approx C_1 \, Ai' \left[ \left( - C \right)^{1/3} \left( t - t_{\omega i} \right) \right] +
C_2 \, Bi' \left[ \left( - C \right)^{1/3} \left( t - t_{\omega i} \right) \right] 
\end{equation}
where $C_{1,2}$ are two integration constants, and $Ai'$ and $Bi'$ the derivatives of the Airy functions $Ai$ and $Bi \,$, respectively. These two solutions are regular at $t = t_{\omega i}$, where they have the expansion series $Ai' ,\, Bi' = {\rm const.} + {\rm O} \left( t - t_{\omega i} \right)^2 \,$. Since the linearized term is absent, we find that $\alpha_0 \propto \dot{\alpha} / \left( t - t_{\omega i} \right)$ also remains finite as $t = t_{\omega i}$.

\bigskip  

{\bf Behavior of the linearized system for $ M^2 \rightarrow 0 \,$.} The scale factor evolves as
\begin{equation}
a = a_M \left( \frac{t}{t_M} \right)^{\frac{2}{3 \left( 1 + w \right)}}
\end{equation}
were we recall that $t_M$ is the times at which $M^2 = 0$, given by
\begin{equation}
t_M = \frac{\sqrt{2}}{3 \, m} \, \frac{\sqrt{1-3 w}}{1+w}
\end{equation}
(cf. eq. (\ref{m20})), and $a_M $ is the value of the scale factor at $t_M$. We Taylor expand eq. (\ref{aleq}) for $t \simeq t_M$:
\begin{equation}
\ddot{\alpha} + \frac{\dot{\alpha}}{t - t_M} + p_M^2 \, \alpha 
\approx 0
\end{equation}
where $p_M$ is the value of the physical momentum of the mode at $t_M \,$. This equation is integrated to give
\begin{equation}
\alpha \approx C_1\, J_0 \left( p_M \, (t_M - t) \right) +
C_2\, Y_0 \left( p_M\, (t_M - t) \right) \label{bessel}
\end{equation}
where $J_0, \, Y_0$ are, respectively, the Bessel functions of the first and
second kinds of order $0$, and $C_{1,2}$ are integration
constants. While the $J_0$ solution is regular at $t=t_M$,
the $Y_0$ solution has a logarithmic divergence. Correspondingly, the mode
$\alpha_0$ exhibits a linear divergence, as can be seen from equation (\ref{al0eq}). 

The only way to avoid the singularity is to arrange the initial conditions so that $C_2 = 0$. This must be done for every mode (namely, for any comoving momentum $k$) and for both the real and imaginary parts of the perturbations. We regard this as a completely unnatural assumption, since there is no reason why the initial conditions (set during inflation) should ``know'' about the singularity which is to occur later on when $M^2$ vanishes.

\begin{figure}[h]
\centerline{
\includegraphics[width=0.4\textwidth,angle=-90]{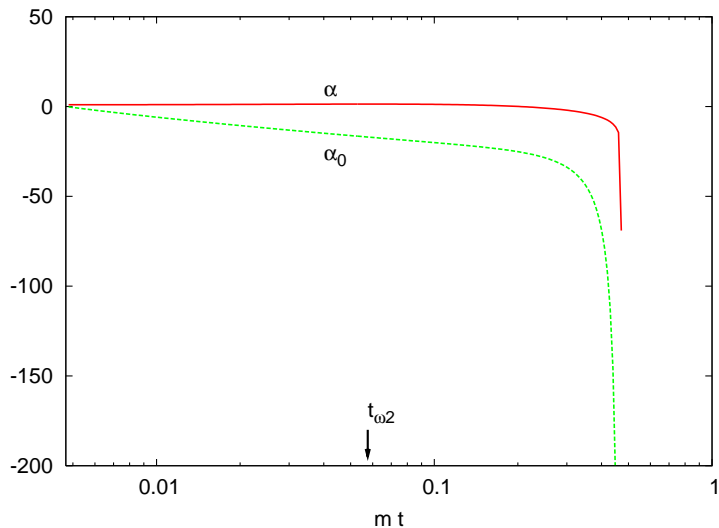}
} \caption{
Behavior of the linearized solutions representing the longitudinal mode. The system remains finite at the point $t_{\omega i}$, when the frequency vanishes, but it diverges at $t_M$, when the mass term $M^2$ of the vector vanishes. See the main text for the parameters chosen.
} \label{fig:novev}
\end{figure}

In conclusion, the solutions of the linearized system remain finite when $\omega^2 = p^2 + M^2 = 0$, while they diverge when $M^2 =0$.  To verify this behavior, we performed a numerical evolution of eq.~(\ref{aleq}) for the specific case of $w=0 \,$, $p_M =2 \,$. We then have $t_{\omega 2} \simeq 0.058 / m$ and $ t_M \simeq 0.47 / m \,$. We set the initial conditions $\alpha = 1,\, \dot{\alpha}=0$ at $t_M / 100 \,$. \footnote{The parameters of the evolution have no particular relevance, and have been chosen only for illustrative purposes. We have verified that other choices of parameters also confirm the behavior we have obtained analytically.} We plot in  Figure \ref{fig:novev} the resulting evolution of $\alpha$ and $\alpha_0$. We see that the system is regular at $t_{\omega 2}$, while it diverges at $t_M$ (we verified that $\alpha_0$ diverges linearly, and, correspondingly, $\alpha$ diverges logarithmically). 

It is worth noting that the solutions diverge when the kinetic term of the longitudinal mode vanishes, cf. eq.~(\ref{ac-a}). On the basis of what we wrote at the end of the previous Section, one should expect that this happens also for the cases in which the vector field(s) has a nonvanishing expectation value. We verify this explicitly in Section \ref{1vec-lambda}, for the simplest possible case of a vector field and a cosmological constant.

\section{Ghost instability for $\langle A_\mu \rangle \neq  0$}

\label{sec:vev}

As we mentioned, when the vector field(s) has a vev, its perturbations
mix with those of the metric already at the linearized level. The
computation then becomes significantly harder than that discussed in
the previous Section for the case of vanishing vev. Besides the
increased number of perturbations, the spatial vev of the vector(s)
breaks the isotropy of the background, so that, in general, one needs
to study the cosmological perturbations of a non FRW space.

We start by studying the case of a single vector field with a vev
along one spatial direction. The background has a $2$d isotropy in the
other two directions. In Subsection \ref{2dformalism} we review the
formalism for dealing with the coupled system of perturbations of the
vector and the metric in this context \cite{gcp,hcp2}. By exploiting
the symmetries of the background, the perturbations can be classified
in two distinct subsets, which are decoupled from each other at the
linearized level. We verified that one subset of perturbations, which
includes the transverse vector modes, is well behaved, and, for
brevity, we disregard it in the following discussions. The other
subset, which we discuss in details, includes the longitudinal
polarization of the vector.

In Subsection \ref{1vec-lambda} we apply this formalism to the
simplest case of a single vector field plus a cosmological
constant. In Subsection \ref{1vec-infla} we study the case in which
the cosmological constant is replaced by a scalar inflaton, which is
the model proposed in \cite{soda1}. Finally, in Subsection
\ref{vector-inflation} we study the model of vector inflation
\cite{mukhvect} (for which the formalism of Subsection
\ref{2dformalism} does not apply; for this reason, this Subsection is
a self-contained study).

\subsection{General formalism for a vector field in a $2$d isotropic background}
\label{2dformalism}

We review the general formalism for the study of perturbations \cite{gcp,hcp2}  for the case of a vector field with a nonvanishing vev along one spatial direction:
\begin{eqnarray}
\langle A_\mu \rangle &=& \left( 0 ,\, M_p \, a \left( t \right) \, B_1 \left( t \right) ,\, 0 ,\, 0 \right) \nonumber\\
ds^2 &=& -dt^2 + a^2 \left(t\right)\, dx^2 + b^2\left(t\right)\, \left( dy^2 + dz^2 \right)
\label{ansatz}
\end{eqnarray}
The background geometry is a Bianchi-I background with $2$d isotropy. The  complete set of  metric
and vector field perturbations can be written in a way which exploits the residual $2$d-isotropy of the system:
\begin{eqnarray}
\delta g_{\mu \nu} &=&
 \left( \begin{array}{ccc}  -2 \Phi & a\,
\partial_1\, \chi & b\left( \partial_i\, B + B_i \right) \\
 &  - 2 \, a^2 \, \Psi  & a\, b\, \partial_1\, \left(
 \partial_i\, \tilde{B} + \tilde{B}_i \right) \\
  & & b^2\, \left(  - 2\Sigma \, \delta_{ij} - 2\,
  \partial_i\, \partial_j\, E - \partial_i\, E_i - \partial_j\,
  E_i \right) \end{array} \right) \nonumber\\ \nonumber\\
 \delta A_\mu &=& \left( \alpha_0 ,\, \alpha_1 ,\, \partial_i \alpha + \alpha_i \right)
\label{defn-perts-2d}
\end{eqnarray}
where $i,j=2,3$ span the isotropic $2$d-plane. The perturbations $\{
\Phi, \, \chi, \, B, \, \Psi, \, \tilde{B}, \, \Sigma, \, E, \,
\alpha_0, \, \alpha_1, \, \alpha \}$ are $2$d scalar perturbations,
comprising one degree of freedom (d.o.f) each, and $\{ B_i, \,
\tilde{B}_i, \, E_i, \, \alpha_i \}$ are $2$d vector modes which
satisfy the transversality condition $(\partial_i\,
B_i=\dots=0)$. Therefore, the $2$d vector perturbations also comprise
one d.o.f. each. Contrary to the $3$d decomposition, there are no
tensor modes, since there are no degrees of freedom left in a
symmetric $2 \times 2$ transverse and traceless matrix. The vector and
scalar modes are decoupled at the linearized level (namely, in the
linearized system of equations for the perturbations, or,
equivalently, in their action, obtained by expanding the action for
the model at quadratic order in the perturbations).

We need to eliminate the redundancy associated with the freedom of
general coordinate transformations. We can do so by either choosing a
gauge that completely fixes this freedom, as done in \cite{gcp}, or by
combining the perturbations in gauge invariant modes \cite{hcp2}. We
choose this second method, since it provides a nontrivial check on our
algebra (since one needs to show that all the equations, and the
action, can be written solely in terms of these combinations). Among
different equivalent choices, we use the gauge invariant combinations
\cite{hcp2}
\begin{eqnarray}
\hat{\Phi} &=& M_p\, \left[ \Phi + \left( \frac{\Sigma}{H_b}
\right)^{\bullet} \right] \nonumber\\
\hat{\Psi} &=& M_p\, \left[ \Psi - \frac{H_a}{H_b}\, \Sigma +
\frac{b}{a}\, \partial_1^2\, \left( \tilde{B} + \frac{b}{a}\, E
\right) \right] \nonumber\\
\hat{B} &=& -\frac{M_p}{b}\, \vec{\partial}_T^2\, \left[ B -
\frac{1}{b\, H_b}\, \Sigma + b\, \dot{E} \right] \nonumber\\
\hat{\chi} &=& -\frac{M_p}{a}\, \partial_1^2\, \left[ \chi -
\frac{1}{a\, H_b}\, \Sigma - a\, \left( \frac{b}{a}\, \left(
\tilde{B}+\frac{b}{a}\, E \right) \right)^{\bullet} \right]
\nonumber\\
\hat{\alpha}_1 &=& -\frac{1}{a}\, \left[ \alpha_1 + a\, M_p\,
\frac{\dot{B}_1 + H_a\, B_1}{H_b}\, \Sigma - b\, M_p\, B_1\,
\partial_1^2\, \left( \tilde{B} + \frac{b}{a}\, E \right) \right]
\nonumber\\
{\hat \alpha} &=& -\frac{1}{a}\, \partial_1\, \left[ \alpha - b\, M_p\,
B_1 \, \partial_1\, \left( \tilde{B} + \frac{b}{a}\, E \right) \right]
\nonumber\\
\hat{\alpha}_0 &=& \frac{1}{a}\, \partial_1\, \left[ \alpha_0 -
a\, M_p\, B_1\, \partial_1\, \left( \frac{b}{a}\, \left( \tilde{B}
+ \frac{b}{a}\, E \right) \right)^{\bullet} \right] \label{GI-2dS}
\end{eqnarray}
(where the dot and bullet denote time derivative, $H_a \equiv \dot{a} / a ,\,
H_b \equiv \dot{b} / b \,$, and $\vec{\partial}_T^2 \equiv \partial_2^2 + \partial_3^2$) 
in the $2$d scalar sector,
and
\begin{eqnarray}
\hat{B}_i &=& B_i + b\, \dot{E}_i \nonumber\\
\hat{\tilde{B}}_i &=& a\left( \tilde{B}_i + \frac{b}{a}\, E_i
\right) \nonumber\\
\hat{\alpha}_i &=& \frac{\alpha_i}{M_p} \label{GI-2dV}
\end{eqnarray}
in the $2$d vector sector. If there is also one scalar field, $\phi + \delta \phi$, there is the additional $2$d gauge invariant scalar mode
\begin{equation}
\hat {\delta \phi} = \delta \phi + \frac{\dot{\phi}}{H_b} \, \Sigma
\label{GI-infla}
\end{equation}
We note that the gauge invariant modes have the following mass dimensions:
$\left[ \hat{\tilde{B}}_i \right] = -1 ,\, \left[ \hat{B}_i \right] = \left[ \hat{\alpha}_i \right] = 0 ,\,
\left[ {\hat \Phi} \right] = \left[ {\hat \Psi} \right] =  \left[ \hat{\alpha}_1 \right] = \left[ \hat{\alpha} \right] = \left[ \hat {\delta \phi} \right] = 1 ,\, \left[ {\hat B} \right] = \left[ {\hat \chi} \right] = \left[ \hat{\alpha}_0 \right] = 2 \,$.

The above gauge invariant combinations  of the metric perturbations  do not reduce to the 
ones which are commonly given in the literature \cite{mfb} in the limit of isotropic background. However, we explicitly verified in \cite{hcp2} that  our choice and the more conventional one are equivalent. Our choice is motivated by the fact that it immediately identifies the nondynamical modes of the system: one can choose the gauge $B = E = \Sigma = E_i = 0$,~\footnote{This was the choice made in~\cite{gcp}, and it fixes completely the freedom associated to general coordinate transformations.} in which the gauge invariant combinations $\hat {\delta g}_{0\mu}$ and $\hat {\delta A}_0$ reduce to the corresponding non dynamical perturbations $\delta g_{0\mu}$ and $A_0$. These modes enter in the action without time derivatives, due to the specific form of the ``kinetic terms'' $F^2$ and $R$. Therefore, we expect that also the gauge invariant combinations $\hat {\delta g}_{0\mu}$ and $\hat {\delta A}_0$ are nondynamical, as the explicit computations reported in the next Subsections confirm.

As we mentioned, the two subsets of modes ($2$d scalars vs $2$d vectors) are decoupled from each other at the linearized level, and can be studied separately. We verified that, for the models of our interest, the $2$d vector modes do not develop any instability. For brevity, we do not report these computations here, and we focus on the more problematic $2$d scalar modes. In the next two Subsections we study the evolution of this system for the case of a single vector plus  a cosmological constant or a scalar inflaton.

\subsection{One vector plus a cosmological constant}
\label{1vec-lambda}

We study the simplest model in which a nonminimally coupled vector
field with a spatial vev provides an anisotropic expansion. Besides
the vector field, there is a vacuum energy $V_0$ which is responsible
for an overall accelerated expansion. The Subsection is divided into
three parts. We first present the model, and discuss the background
evolution. We then solve the linearized system of equations for the
perturbations, and find that it diverges at some point close to
horizon crossing. We finally compute the kinetic term $K$ of the
quadratic action of the perturbations, and show (i) that the
divergence of the linearized system takes place precisely when one
eigenvalue of $K$ vanishes, and (ii) that one of the perturbations is
a ghost for some moment of time.

\subsubsection{The model and the background solution}

The action of the model
\begin{equation}
S = \int d^4 x \left[ - \frac{1}{4} F_{\mu \nu} F^{\mu \nu} - V_0 - \frac{1}{2} \left( m^2 - \frac{R}{6} \right) A_\mu A^\mu \right]
\label{act-V0}
\end{equation}
gives the equations of motion
\begin{eqnarray}
&& G_{\mu\,\nu} = \frac{1}{M_p^2}\, \left[ -V_0\, g_{\mu\nu} +
T_{\mu\nu}^{(A)}
\right] \nonumber\\
&& T_{\mu\nu}^{(A)} \equiv F_{\mu}^{\,\,\, \sigma}\, F_{\nu\sigma}
- \frac{1}{4}\, F_{\alpha\beta}\, F^{\alpha\beta}\, g_{\mu\nu} +
\left( m^2 - \frac{R}{6} \right)\, A_{\mu}\, A_{\nu} -
\frac{1}{2}\, m^2\, A_{\alpha}\, A^{\alpha}\, g_{\mu\nu}
\nonumber\\
&& \qquad \quad -\frac{1}{6}\, \left( R_{\mu\nu} - \frac{1}{2}\,
R\, g_{\mu\nu} \right)\, A_{\alpha}\, A^{\alpha} - \frac{1}{6}\,
\left( g_{\mu\nu}\, \square - \nabla_{\mu}\, \nabla_{\nu}
\right)\, A_{\alpha}\, A^{\alpha}
\nonumber\\
&& \frac{1}{\sqrt{-g}}\, \partial_{\mu}\, \left[ \sqrt{-g}\,
F^{\mu\nu} \right] - \left( m^2 - \frac{R}{6} \right)\, A^{\nu} =
0 \label{field-eqns-cc}
\end{eqnarray}
If we assume that the background solutions  depend only on time, then the $\nu$ component
of the last of (\ref{field-eqns-cc}) gives $A^0 = 0 \,$. For a homogeneous background, the vev of the vector is everywhere in the same direction; we choose the coordinates such that this direction coincides with  the $x-$axis. We look for background solutions of the form (\ref{ansatz}), and we define
\begin{equation}
H_a \equiv \frac{\dot{a}}{a} \;\;\;,\;\;\; H_b \equiv \frac{\dot{b}}{b} \;\;\;,\;\;\;
H \equiv \frac{H_a + 2 \, H_b}{3} \;\;\;,\;\;\; h \equiv \frac{H_b - H_a}{3}
\end{equation}
Namely, $H$ is the average expansion rate, while $h$ parametrizes the departure from isotropy. The regime of small isotropy corresponds to $h \ll H \,$. In terms of these quantities, and for the background 
(\ref{ansatz}), equations (\ref{field-eqns-cc}) give \footnote{The first of (\ref{evolution-cc}) is the $00$ Einstein equation; the second is the linear combination of the $(11)$-$(22)$ Einstein equations; the third equation is the linear combination of the $(11) + 2 \times (22)$ Einstein equations . Finally, the fourth equation is the $x-$ component of the equations for the vector field (the $33$ Einstein equation coincides with the $22$ one, while the remaining equations are trivial).}
\begin{eqnarray}
&& H^2-h^2 = \frac{V_0}{3 \, M_p^2} + \frac{1}{6} \dot{B}_1^2-\frac{2}{3} h \, B_1 \, \dot{B}_1 + \frac{1}{6} \, B_1^2 \left( m^2 - 4 H h + 5 h^2 \right) \nonumber\\
&& \dot{h} + 3 H h = \frac{1}{3} B_1^2 \left( \dot{H} - \frac{\dot{h}}{2} \right) + \frac{1}{3} \dot{B}_1^2 + \frac{1}{3} \left( 2 H - 5 h \right) B_1 \dot{B}_1 + \frac{1}{3} \left( 3 H^2 - \frac{11}{2} H h + 5 h^2 - m^2 \right) B_1^2 \nonumber\\
&& 2 \dot{H} + 3 H^2 + 3 h^2 = \frac{V_0}{M_p^2} - \frac{1}{2} \dot{B}_1^2 - \frac{1}{3} B_1 \left[ \ddot{B}_1 + \left( 3 H - 2 h \right) \dot{B}_1 \right] + \frac{1}{6} \left( 4 H h - 5 h^2 + m^2 \right) B_1^2 \nonumber\\
&& \ddot{B}_1 + 3\, H\, \dot{B}_1 + \left( m^2 - 5 h^2 - 2\, h\, H
-2\, \dot{h}\right)\, B_1 = 0 \label{evolution-cc}
\end{eqnarray}
 One of the last three equations can be obtained from the other equations in (\ref{evolution-cc}) due to a nontrivial Bianchi identity. We see from the second of (\ref{evolution-cc}) that the anisotropy of the background is indeed supported by the vector vev $B_1$.

Below, when we study the perturbations of this model, we use the exact background equations 
(\ref{evolution-cc}). However, only for this discussion, we present the inflationary solution of 
(\ref{evolution-cc}) in the slow roll approximation (slow motion of $B_1$ towards zero, so that the anisotropy is not quickly damped away) and for small anisotropy. 
Specifically, we disregard the time derivatives of $H ,\, h$, and look for expansion series solutions of the type
\begin{eqnarray}
&& H = H_0 + c_H \, B_1^2 + {\rm O } \left( B_1^4 \right) \;\;,\;\;
h = c_h \, B_1^2 + {\rm O } \left( B_1^4 \right) \;\;,\;\;
\dot{B_1} \simeq c_B \, H_0 \, B_1 \;\;\;\;, \; {\rm with} \;\; B_1 ,\, c_B \ll 1
\label{slow-ansatz}
\end{eqnarray}
We insert these expression in the last equation of (\ref{evolution-cc}) and expand in $B_1$. This gives
$ 3 c_B H_0^2 + c_B^2 \, H_0^2 + m^2 = 0 \,$. Since, $c_B \ll 1$, this gives $c_B = - \frac{m^2}{3 \, H_0^2}$. Therefore, slow roll requires $m \ll H_0 \,$. We then insert (\ref{slow-ansatz}) in the first of (\ref{evolution-cc}), expand in $B_1$, and neglect $m$ vs. $H_0$. We find $H_0 = \sqrt{V_0} / \left( \sqrt{3} \, M_p \right) \;,\; c_H = m^2 / \left( 12 \, H_0 \right) \,$. Finally, we insert (\ref{slow-ansatz}) in the second and third of (\ref{evolution-cc}), expand in $B_1$, and again neglect $m$ vs. $H_0$. These equations then become identical, and give $c_h = H_0 / 3$. Therefore, 
\begin{equation}
H = H_0 + \frac{m^2}{12 \, H_0} \, B_1^2 + {\rm O } \left( B_1^4 \right) \;\;,\;\;
h = \frac{H_0}{3} \, B_1^2 +  {\rm O } \left( B_1^4 \right) \;\;,\;\;
\dot{B_1} = - \frac{m^2}{3 \, H_0} B_1 +  {\rm O } \left( B_1^3 \right) 
\label{slow}
\end{equation}
where $H_0$ is the expansion rate in absence of the vector field.

Before moving to the study of the perturbation, we note that the mass parameter $m$ of the vector field must satisfy
\begin{equation}
B_1 \, H_0^2 \la m^2 \ll H_0^2
\label{limm}
\end{equation}
The second condition is due to the slow roll requirement, as we already pointed out. The first condition comes from requiring that the ``mass term'' for $B_1$ in the last of (\ref{evolution-cc}) is positive, so that $B_1$ evolves towards the origin (so that the anisotropy decreases during inflation rather than increasing).

\subsubsection{Instability from the linearized equations} \label{subsect:eqnsV0}

We now expand  the field equations (\ref{field-eqns-cc}) at first order in the $2$d scalar perturbations of the metric and the vector field.~\footnote{As we mentioned, the $2$d vector perturbations are decoupled from the $2$d scalar ones; we have verified that the $2$d vectors do not develop any instability; for brevity, we do not present this computations here.} We denote the resulting linearized equations as
\begin{equation}
{\rm Eq}_{\mu\nu} : \delta \left( G_{\mu\nu} - \frac{1}{M_p^2}\,
\left( -V_0\, g_{\mu\nu} + T_{\mu\nu}^{(A)} \right) \right) = 0
\,\,\,\,\, , \,\,\,\,\, {\rm Eq}_{\mu} : \delta \left(
\frac{\delta S}{\delta A_{\mu}} \right) = 0 \label{perturbed}
\end{equation}
We then Fourier transform these equations.
Namely for each mode, we have
\begin{equation}
\delta\left( x \right) = \frac{1}{\left(2\pi\right)^3}\, \int d^3
k\, \delta\left(k\right)\, e^{-i k_L x - i k_{T2} y - i k_{T3} z}
\label{fourier}
\end{equation}
where $\delta$ denotes any of the perturbations (we denote 
the perturbations in coordinate and momentum space with the same symbol).
Modes with different comoving momenta are decoupled from each
other at the linearized level. We denote by $k_L$ the component of
the comoving momentum in the $x$-direction, and by
$\overrightarrow{k}_T$ the component of the momentum in the $y-z$
plane. The full comoving momentum is then given by
$k^2=k_L^2+\overrightarrow{k}_T^2$. The physical momentum is instead
\begin{equation}
p^2 = p_L^2 + \overrightarrow{p}_T^2 = \frac{k_L^2}{a^2} + \frac{\overrightarrow{k}_T^2}{b^2}
\label{defmom}
\end{equation}

The explicit expressions of the linearized equations for the scalar
sector are given in equations (\ref{app:linearized}) of
Appendix~\ref{appA}. We could express these equations solely in terms
of the gauge invariant modes defined in (\ref{GI-2dS}). This is a
nontrivial check on our algebra. Here, we disregard some of the
equations in (\ref{app:linearized}) which can be obtained from the
remaining ones (due to Bianchi identities). The (complete) set of
independent linearized equations which we choose to integrate is
\begin{eqnarray}
{\rm Eq}_{00}: && \left[ \left( 1 - \frac{B_1^2}{3} \right)\, H +
\left( 1 + \frac{B_1^2}{6} \right)\, h + \frac{1}{6}\, \dot{B}_1\,
B_1 \right] \dot{\hat\Psi} + \frac{1}{2}\, \left[ p_T^2 + \left(
m^2 - \frac{p_T^2 + 2 p_L^2}{6} + 5 h^2 - 4\, h\, H \right)\,
B_1^2 - 4\, h\, B_1\, \dot{B}_1 + \dot{B_1}^2 \right]
\hat\Psi \nonumber\\
&& +\left[ 3 H^2 - \left( 3 + \frac{5}{2}\, B_1^2 \right)\, h^2 -
\frac{1}{2}\, \dot{B}_1^2 + 2\, \left( B_1\, H + \dot{B}_1
\right)\, B_1\, h \right]\, \hat\Phi - \left[ \left( 1 +
\frac{B_1^2}{6} \right)\,\left( H + h \right) + \frac{1}{6}\,
B_1\, \dot{B}_1 \right]\, \hat\chi \nonumber\\
&& -\left[ \left(1 + \frac{B_1^2}{6} \right)\, \left( H -
\frac{1}{2}\, h \right) + \frac{1}{6}\, B_1\, \dot{B}_1 \right]\,
\hat{B} - \left[ \left( \frac{1}{2}\, H - h \right)\, B_1 +
\frac{1}{2}\, \dot{B}_1 \right]\, \hat{\alpha}_0  - \left(
\frac{1}{2}\, \dot{B}_1 - B_1\, h \right)\, \dot{\hat\alpha}_1
\nonumber\\
&& +\left[ \frac{1}{2}\, \left( \frac{p^2}{3} - m^2 - 5 h^2  +4\,
h\, H \right)\, B_1 + h\, \dot{B}_1 \right]\, \hat{\alpha}_1 = 0
\nonumber\\ \nonumber\\
{\rm Eq}_{01} : && \frac{1}{6}\, \left( \dot{\hat\alpha}_1 - B_1\,
\dot{\hat\Psi} \right)\, B_1 + \frac{B_1}{6}\, \left( B_1\, H -
2\, B_1\, h -2\, \dot{B}_1 \right)\, \hat\Psi + \frac{1}{6}\,
\left( - B_1\, H + 2\, h\, B_1 + \dot{B}_1 \right)\,
\hat{\alpha}_1 \nonumber\\
&& + \left( 1 + \frac{B_1^2}{6} \right)\, \left( H + h +
\frac{B_1\, \dot{B}_1}{6 + B_1^2} \right)\, \hat\Phi + \left(
\frac{p_T^2}{4 p_L^2} + \frac{p_T^2}{24 p_L^2}\, B_1^2 -
\frac{1}{3}\, {\cal D}_{\chi \chi} \right)\, \hat\chi - \frac{6 +
B_1^2}{24}\, \hat{B} - \frac{B_1}{3}\, {\cal D}_{\alpha_0
\alpha_0}\, \hat{\alpha}_0 = 0 \nonumber\\
{\rm Eq}_{0i} : && \frac{1}{2}\, \left( 1 - \frac{B_1^2}{6}
\right)\, \dot{\hat\Psi} + \frac{1}{6}\, B_1\, \dot{\hat\alpha}_1
+ \left[ \frac{B_1}{6}\, \left( H\, B_1 - 2\, \dot{B}_1 \right) -
\left( \frac{3}{2} + \frac{B_1^2}{12} \right)\, h \right]\,
\hat\Psi + \left[ \left( \frac{1}{3}\, H - \frac{7}{6}\, h
\right)\, B_1 + \frac{2}{3}\, \dot{B}_1 \right]\, \hat{\alpha}_1
\nonumber\\
&& + \left[ \left( h - \frac{H}{2}\right)\, B_1 -
\frac{\dot{B}_1}{2} \right]\, \hat\alpha + \left[ -
\left(\frac{1}{2} + \frac{B_1^2}{12} \right)\, h + \left( 1 +
\frac{B_1^2}{6} \right)\, H + \frac{1}{6}\, B_1\, \dot{B}_1
\right]\, \hat\Phi - \frac{1}{4}\, \left( 1 + \frac{B_1^2}{6}
\right)\, \hat\chi \nonumber\\
&& + \frac{p_L^2}{4 p_T^2}\, \left( 1 +
\frac{B_1^2}{6} \right)\, \hat{B} = 0 \nonumber\\ \nonumber\\
%
{\rm Eq}_{0} : && \dot{\hat\alpha}_1 + \left( 2\, h\, B_1 - H\,
B_1 - \dot{B}_1 \right)\, \hat\Psi + \left( H -2 h\right)\,
\hat{\alpha}_1 + \frac{p_T^2}{p_L^2}\, \dot{\hat\alpha} +
\frac{p_T^2}{p_L^2}\, \left( H -2 h \right)\, \hat\alpha + \left(
H\, B_1 -2 h\, B_1 + \dot{B}_1 \right)\, \hat\Phi - \frac{2}{3
B_1}\, {\cal D}_{\chi \chi}\, \hat\chi \nonumber\\
&& + \left( 1 + \frac{p_T^2}{p_L^2} - \frac{2}{3}\, {\cal
D}_{\alpha_0 \alpha_0} \right)\, \hat{\alpha}_0 = 0 \nonumber\\
\nonumber\\
{\rm Eq}_{11}: && \frac{1}{2}\, B_1\, \left( \ddot{\hat{\alpha}}_1
- 2\, B_1\, \ddot{\hat\Psi} \right) - \left( 3\, H\, B_1 + 2\,
\dot{B}_1 \right)\, B_1\, \dot{\hat\Psi} + \frac{1}{2}\, \left(
-H\, B_1 + 8\, h\, B_1 - \dot{B}_1 \right)\, \dot{\hat\alpha}_1 +
\left( {\cal M}_{\Psi \Psi} - p_T^2\, B_1^2 \right)\,
\hat\Psi \nonumber\\
&& + \left( {\cal M}_{\Psi \alpha_1} + \frac{p_T^2}{2}\, B_1
\right)\, \hat{\alpha}_1 + \left[ \left( 3 + \frac{1}{2}\, B_1^2
\right)\, h + \left(3-B_1^2\right)\, H + \frac{1}{2}\, B_1\,
\dot{B}_1 \right]\, \dot{\hat\Phi} \nonumber\\
&& + \left[ \left( -\frac{9}{4}\, m^2 + \frac{p_T^2 + 2 p_L^2}{4}
- \frac{15}{4}\, h^2 + 3\,h \, H\right)\, B_1^2 + 3\, h\, B_1\,
\dot{B}_1 - 3\, \left( \frac{p_T^2}{2}\, -\frac{V_0}{2 M_p^2} +
\frac{3}{2}\, h^2 - \frac{3}{2}\, H^2 + \frac{1}{4}\,
\dot{B}_1^2\right) \right]\, \hat\Phi \nonumber\\
&& - \left( \frac{3}{2} - \frac{1}{4}\, B_1^2 \right)\,
\dot{\hat{B}} + \frac{1}{2}\, B_1^2\, \dot{\hat\chi} + \left( 2 H
- h \right)\, B_1^2\, \hat\chi - \left[ \left( \frac{9}{2} +
\frac{1}{4}\, B_1^2 \right)\, h + \left( \frac{9}{2} -
\frac{5}{4}\, B_1^2 \right)\, H + \frac{1}{2}\, B_1\, \dot{B}_1
\right]\,\hat{B} \nonumber\\
&& + \frac{3}{2}\, \left( 2 h\, B_1 - H\, B_1 - \dot{B}_1
\right)\, \hat{\alpha}_0 = 0 \nonumber
\end{eqnarray}
\begin{eqnarray}
{\rm Eq}_{1}: && \ddot{\hat\alpha}_1 - \frac{1}{3}\, B_1\,
\ddot{\hat\Psi} + 3 H\, \dot{\hat\alpha}_1 + \left( \frac{8}{3}\,
B_1\, h - \frac{7}{3}\, B_1\, H - \dot{B}_1 \right)\,
\dot{\hat\Psi} + \left( {\cal M}_{\alpha_1 \alpha_1} + p_T^2
\right)\, \hat{\alpha}_1 -
\frac{B_1}{3}\, p_T^2\, \hat\Psi \nonumber\\
&& + \left( \dot{B}_1 - 2\, B_1\, h \right)\, \dot{\hat\Phi} +
\frac{B_1}{3}\, \left( \dot{\hat{B}} + \dot{\hat\chi} \right) +
\dot{\hat\alpha}_0 + \frac{B_1}{3}\, \left( p^2 - 6 m^2 \right)\,
\hat\Phi + \frac{1}{3}\, \left( 7\, B_1\, h + B_1\, H -3 \dot{B}_1
\right)\, \hat{B} + \frac{2}{3}\, \left( 2 H - h \right)\, B_1\,
\hat\chi \nonumber\\
&& - p_T^2\, \hat\alpha + 2\, \left( h + H \right)\,
\hat{\alpha}_0 = 0 \nonumber\\ \nonumber\\
{\rm Eq}_{i}: && \ddot{\hat\alpha} + 3\, \left( H -2 h \right)\,
\dot{\hat\alpha} + \left( {\cal M}_{\alpha \alpha} + p_L^2
\right)\, \hat\alpha - p_L^2\, \hat{\alpha}_1 +
\frac{p_L^2}{p_T^2}\, \left( B_1\, H - 2\, B_1\, h + \dot{B}_1
\right)\, \hat{B} + \dot{\hat\alpha}_0 + \left( 2 H -4 h \right)\,
\hat{\alpha}_0 = 0 \label{einstein-cc}
\end{eqnarray}
where ${\cal D}_{\chi\chi}, \, {\cal D}_{\alpha_0 \alpha_0}, \,
{\cal M}_{\Psi \Psi}, \, {\cal M}_{\Psi \alpha_1}, \, {\cal
M}_{\alpha_1 \alpha_1}, \, {\cal M}_{\alpha \alpha}$ depend on the
background quantities and are explicitly given in equations
(\ref{calD}) and (\ref{calM}) in Appendix~\ref{appA}. In writing
(\ref{einstein-cc}), we have also made use of the physical
momenta defined in eq. (\ref{defmom}).

We now solve the system of equations (\ref{einstein-cc})
numerically. We start by noting that the first four equations in
(\ref{einstein-cc}) contain at most a single time derivative for the
perturbations, and do not contain any time derivative of the modes $\{
\hat\Phi, \, \hat\chi, \, \hat{B}, \, \hat{\alpha}_0 \}$; these are
the nondynamical modes of the system, and these first four equations
can also be obtained by extremizing the quadratic action for the
perturbations (eq. (\ref{act-2dS-cc}) below) with respect to these
modes. These equations are precisely of the type (\ref{eq-nondyn})
given above where we discussed the linearized equations at a formal
level. Compare also with eq. (\ref{al0eq}) for the nondynamical mode
$\alpha_0$ in the simpler case studied in Section \ref{sec:novev}.

One can choose to solve these four constraint equations for $\{
\hat\Phi, \, \hat\chi, \, \hat{B}, \, \hat{\alpha}_0 \}$, and insert
the solutions in the remaining equations in (\ref{einstein-cc}). In
this way, we obtain a closed system of equations for the dynamical
modes.  A different but equivalent way to integrate the system
(\ref{einstein-cc}) is to differentiate the constraint equations, so
as to obtain differential equations for the nondynamical modes too.
Combined with the remaining equations in (\ref{einstein-cc}), one then
obtains a closed system that can be numerically integrated (the
constraint equations need to be imposed as initial condition).  In the
first method mentioned, one obtains a system with fewer but more
complicated equations. On the other hand, the second method results in
a larger set of less complicated equations.  Here, we adopt an
``intermediate'' method, which we have found to be convenient for the
numerical integration. We solve the second of (\ref{einstein-cc}) for
${\hat \chi}$:
\begin{eqnarray}
\hat\chi &=& \frac{p_L^2}{{\cal D}}\, \Bigg[ \left( 6 + B_1^2
\right)\, \hat{B} + 8 B_1\, {\cal D}_{\alpha_0 \alpha_0}\,
\hat{\alpha}_0 + 4\, \left( ( H -2 h)\, B_1 - \dot{B}_1 \right)\,
\hat{\alpha}_1 - 4\, B_1\, \left( ( H -2 h)\, B_1 - 2 \dot{B}_1
\right)\,
\hat{\Psi} \nonumber\\
&& \qquad\qquad\qquad\qquad\qquad\qquad - 4\, \left( 6 +
B_1^2\right)\, {\cal H}\, \hat\Phi - 4\, B_1\, \dot{\hat\alpha}_1
+ 4 B_1^2\, \dot{\hat\Psi} \Bigg] \label{solhatchi}
\end{eqnarray}
where the time dependent coefficients ${\cal D}_{\alpha_0 \alpha_0}, \, {\cal D}$  and ${\cal H}$
are given in equations (\ref{calD}) and (\ref{calH}) of Appendix~\ref{appA}. 

It is  important to verify that the second of (\ref{einstein-cc}) can indeed be solved in terms of ${\hat \chi}$; namely, that ${\cal D} \neq 0$. The easiest way to verify this  is to use the slow-roll solutions (\ref{slow}) in the expression (\ref{calH}) for ${\cal D}$, since we are working in a regime in which  these  slow-roll solutions are highly accurate. This gives
\begin{equation}
{\cal D} = 6 p_T^2 + \left( p_T^2 - 24 \, H^2 + 12 m^2   \right)\, B_1^2 + {\rm
O}\left( B_1^4 \right) 
\label{expserD}
\end{equation}
We can integrate out ${\hat \chi}$ provided that ${\cal D} \neq 0 \,$.  We see that ${\cal D}$ is positive at least as long as $p_T \ga 2 \, H \, B_1 \,$ (when the two terms become equal, the negative term dominates the ${\cal O} \left( B_1^2 \right)$ expression, an the expression cancels with the positive ${\cal O} \left( B_1^0 \right)$ term. At this point, the expansion series (\ref{expserD}) breaks down). As we show below, the computation we are performing shows that the linearized perturbations blow up when $p_L \approx \sqrt{12} \, H / B_1 \,$. Since integrating out ${\hat \chi}$ is a step of this computation, our result can be trusted as long as $p_T > 2 \, H \, B_1$ when $p_L \approx \sqrt{12} \, H / B_1$. Namely, we consider modes for which $p_T > \left( B_1^2 / \sqrt{3} \right) p_L \,$ at that moment. This is not a strong restriction, since $B_1 \ll 1\,$.

We insert the solution (\ref{solhatchi}) into the other
equations in (\ref{einstein-cc}), and  we differentiate the
three remaining constraint equations (we stress that we 
do not lose any information, provided that these equations
are imposed as initial conditions). In this way, 
we obtain a system of 6 differential equations, which can be
formally written as 
\begin{equation}
{\cal M}_{\kappa}\, \left( \begin{array}{c} \ddot{\hat\Psi} \\
\ddot{\hat\alpha}_1 \\ \ddot{\hat\alpha} \\ \dot{\hat\Phi} \\
\dot{\hat{B}} \\ \dot{\hat\alpha}_0 \end{array} \right) = \left(
\begin{array}{c} f_1 \\ f_2 \\ f_3 \\ f_4 \\ f_5 \\ f_6
\end{array} \right) \,\,\,\,\, , \,\,\,\,\, {\cal M}_{\kappa}
\equiv \left( \begin{array}{cccccc} \kappa_{11} & \kappa_{12} & 0
& \kappa_{14} & \kappa_{15} & \kappa_{16} \\
\kappa_{21} & \kappa_{22} & 0 & \kappa_{24} &
\kappa_{25} & \kappa_{26} \\
0 & 0 & 1 & 0 & 0 & 1 \\
\kappa_{41} & \kappa_{42} & 0 & \kappa_{44} & \kappa_{45} &
\kappa_{46} \\
\kappa_{51} & \kappa_{52} & 0 & \kappa_{54} & \kappa_{55} &
\kappa_{56} \\
\kappa_{61} & \kappa_{62} & \kappa_{63} & \kappa_{64} &
\kappa_{65} & \kappa_{66} \end{array} \right) \label{system}
\end{equation}
which correspond, respectively, to ${\rm
Eq}_{11}, \, {\rm Eq}_1, \, {\rm Eq}_i$, and to the
dime derivatives of ${\rm
Eq}_{00}, \, {\rm Eq}_{0i}, \,$ and $ {\rm Eq}_0$.

The elements of the matrix ${\cal M}_{\kappa}$ depend on the
background quantities. The functions on the right hand side of
(\ref{system}) $f_i$ are expressed as linear combinations of $\{
\dot{\hat\Psi}, \, \hat\Psi, \, \dot{\hat\alpha}_1, \,
\hat\alpha_1, \, \dot{\hat\alpha}, \, \hat\alpha, \, \hat\Phi, \,
\hat{B}, \, \hat{\alpha}_0 \} $ whose coefficients also depend on
the background quantities. The explicit expressions for the matrix
${\cal M}_{\kappa}$ and the coefficients $f_i$  are given in equations
(\ref{kappa6}) and (\ref{A6}) of
Appendix~\ref{appA}.

We invert ${\cal M}_{\kappa}$, and integrate the system numerically.
The determinant of ${\cal M}_{\kappa}$ vanishes
at some given time. As we approach that time, ${\cal M}_{\kappa}^{-1}$ diverges. 
Correspondingly, the numerical solutions of the system also diverge. 

By inserting the explicit expressions for the entries of ${\cal M}_\kappa$ given in Appendix~\ref{appA}, we find:
\begin{equation}
{\rm det}{\cal M}_{\kappa} = \frac{p_T^2\, \left( 6 + B_1^2
\right)^2}{864 {\cal D}}\, \left( h + H \right)^2\, \left[ 72\,
{\cal D}_{\alpha_0 \alpha_0} - 6\, B_1^2\, \left( 3 - 2 {\cal
D}_{\alpha_0 \alpha_0} \right) - B_1^4\, \left( 3 - 8 {\cal
D}_{\alpha_0 \alpha_0} \right) \right] \label{detM}
\end{equation}
Using the expressions for ${\cal D}_{\alpha_0 \alpha_0}$ and
${\cal D}$ given in (\ref{calD}) and (\ref{calH}), we then find that the determinant vanishes when
\begin{eqnarray}
p_L^2 = p_{L\, *}^2 &\equiv& \frac{1}{B_1^2\, \left( 6 + B_1^2
\right)}\, \Bigg\{ -18\, \left( 2 m^2 - 4 H^2 + 4 h^2 +
\dot{B}_1^2 \right) - \left( 6 m^2 - 12 H^2 - 72\, h\, H + 102\,
h^2 + 7 \dot{B}_1^2 \right)\, B_1^2 \nonumber\\
&& \qquad\qquad\qquad - \left( 4 H^2 - 28\, h\, H + 31\, h^2
\right)\, B_1^4 + 48\,h \, B_1\, \dot{B}_1 + 4 \left( 7 h - 2 H
\right)\, B_1^3\, \dot{B}_1 \Bigg\} \label{pL2st}
\end{eqnarray}
This expression for $p_{L*}$ is exact. Using the slow-roll solutions (\ref{slow}), we obtain the approximate expression:
\begin{equation}
p_{L*}^2 \simeq \frac{6 \left( 2 \, H_0^2 - m^2 \right)}{B_1^2} + {\rm O } \left( B_1^0 \right)
 \label{pL2st-small}
\end{equation}

To confirm that the linearized modes indeed blow up at $p_{L*}$, we
integrate the system (\ref{system}) for a specific choice of
parameters (the parameters we use have no particular relevance, and
are chosen only for illustrative purposes). We choose a small
anisotropy, $\left( h / H \right)_{\rm in} = {\rm O } \left( B_{1,{\rm
      in}}^2 \right) \simeq 10^{-2} \,$, and an initial value for the
momentum, $p_{\rm in} \simeq 10^3 \, H_{\rm in}$, so that the mode is
initially deeply inside the horizon. As we verify in
Appendix~\ref{appB}, the frequency for the modes is initially
adiabatically evolving, so that we can choose the initial conditions
for the dynamical modes ${\hat \Psi} ,\, {\hat \alpha}_1 ,\, {\hat
  \alpha}$ and their derivatives according to the adiabatic vacuum
prescription. We then use the constraint equations (the first, the
third, and the fourth of (\ref{einstein-cc})), to provide the initial
conditions for the three nondynamical modes ${\hat \Phi} ,\, {\hat B}
,\, {\hat \alpha}_0$ of the system.  The resulting evolution of the
``relativistic Newtonian potential'' ${\hat \Phi}$, and of two other
$2$d scalar modes are shown in Figure \ref{fig:plots}. We see that the
modes indeed diverge at some given time. We verified that $p_L =
p_{L*}$ at this moment. This is confirmed by the time evolution of one
of the eigenvalue of the kinetic matrix of the dynamical perturbations
(computed in \ref{subsect:ghostV0}) shown in the left panel of the
Figure. We see that the linearized solutions blow up precisely when
the system (\ref{system}), or equivalently, the kinetic matrix,
becomes singular.

\begin{figure}[h]
\centerline{
\includegraphics[width=0.4\textwidth,angle=-90]{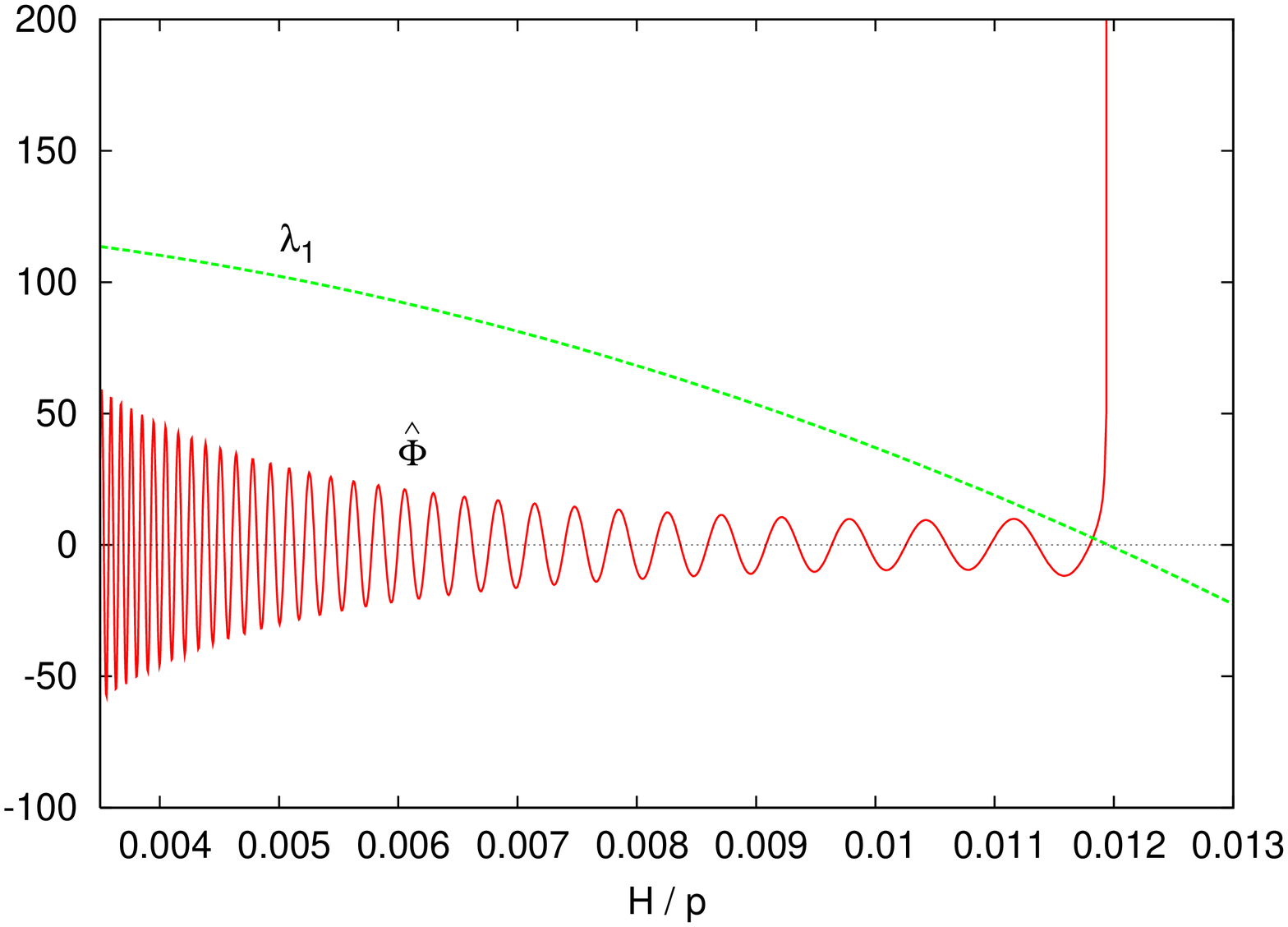}
\includegraphics[width=0.4\textwidth,angle=-90]{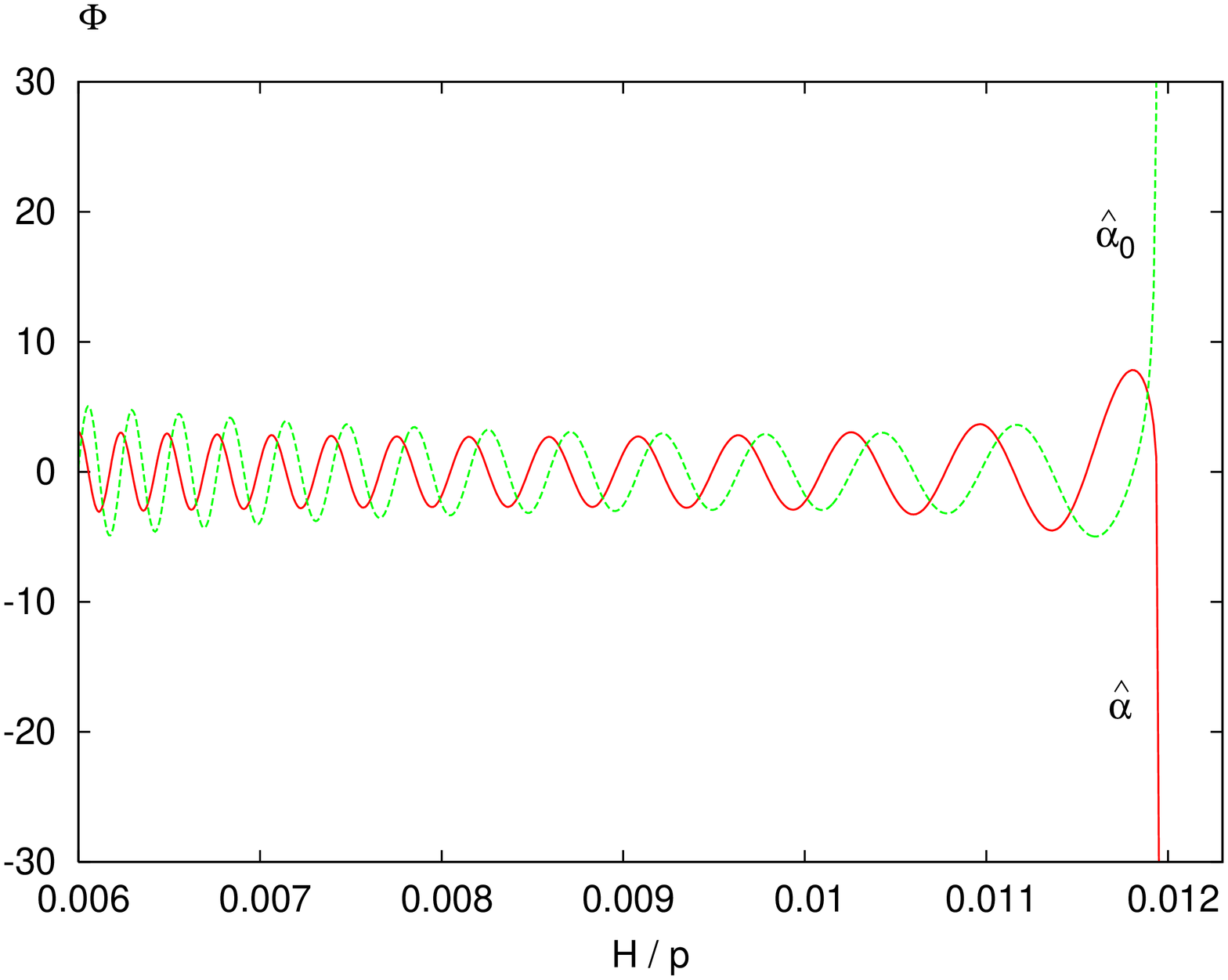}
} \caption{
Results from a numerical simulation with $m^2 = 0.1 \, H_0 \, B_{\rm in} = 0.1 ,\,
p_{L,{\rm in}} = 400 H_{\rm in} ,\, p_{T,{\rm in}} = 900 H_{\rm in} ,\, $ (corresponding to $H/p \simeq 10^{-3}$; the modes are initially in the adiabatic vacuum; only the final part of the evolution is shown in the two figures). Since $H/p$ grows during inflation, we use this quantity as ``time variable'' in the Figure.  Left panel: gauge invariant relativistic gravitational potential $\hat\Phi$. We show the real part in units of ${\hat \Psi}_{\rm in}$. We also show the eigenvalue $\lambda_1$ of the kinetic matrix (multiplied by $3 \times 10^5$, so that it is visible in the figure). We see that ${\hat \Phi}$
diverges when $\lambda_1 = 0 \,$. Right panel: real parts of the modes ${\hat \alpha}_0$ 
(in units of $100 \, H_0 \, {\hat \Psi}_{\rm in}$) and ${\hat \alpha}$ (in units of ${\hat \Psi}_{\rm in}$) . Also these modes (as all the modes of the system) diverge when $\lambda_1 = 0 \,$.} \label{fig:plots}
\end{figure}

\subsubsection{Ghost from the quadratic action} 
\label{subsect:ghostV0}

In order to understand the physical reasons behind the instability
we have just found, we compute the quadratic action for the 
perturbations. We expand the action (\ref{act-V0}) at quadratic order in the perturbations
(\ref{defn-perts-2d}). The resulting action can be expressed solely in terms of the gauge invariant modes defined in (\ref{GI-2dS})-(\ref{GI-2dV}). This provides a nontrivial check on our algebra. We also find that the action is the sum of two separate parts, one involving only the $2$d scalar modes, and one the $2$d vectors. We disregard this second piece in the following discussion. The action for the $2$d scalars reads
\begin{eqnarray}
S_{\rm 2dS} &=& \frac{1}{3}\, \int\, d^3k\, dt\, a\, b^2\, {\cal
L}_{\rm 2dS} \nonumber\\
{\cal L}_{2dS} &=& B_1^2\, \vert \dot{\hat\Psi} \vert^2 -
\frac{B_1}{2}\, \left( \dot{\hat\Psi}\, \dot{\hat\alpha}_1^* +
{\rm h.c.} \right) + \frac{3 p_T^2}{2 p_L^2}\, \vert
\dot{\hat\alpha} \vert^2 + \frac{3}{2}\, \vert \dot{\hat\alpha}_1
\vert^2 - \frac{3}{2}\, \left( 1 + \frac{1}{2}\, B_1^2 \right)\,
\left( H - 2\, h - \frac{2\, B_1\, \dot{B}_1}{2 + B_1^2} \right)\,
\left( \dot{\hat\Psi}^*\, \hat{\Psi} + {\rm h.c.} \right)
\nonumber\\
&-& \left( B_1\, h - \frac{1}{2}\, B_1\, H + \frac{1}{2}\,
\dot{B}_1 \right)\, \left( \dot{\hat\Psi}^*\, \hat{\alpha}_1 +{\rm
h.c.} \right) - \left[ \left( 3 + \frac{1}{2}\, B_1^2 \right)\, h
+ \left( 3 - B_1^2\right)\, H + \frac{1}{2}\, B_1\, \dot{B}_1
\right]\, \left( \dot{\hat\Psi}^*\, \hat{\Phi} + {\rm h.c.}
\right) \nonumber\\
&-& \frac{B_1^2}{2}\, \left( \dot{\hat\Psi}^*\, \hat{\chi} + {\rm
h.c.} \right) + \frac{1}{4}\, \left( 6 - B_1^2 \right)\, \left(
\dot{\hat\Psi}^*\, \hat{B} + {\rm h.c.} \right) + \frac{3 p_T^2}{2
p_L^2}\, \left( H - 2\,h \right)\, \left( \dot{\hat\alpha}^*\,
\hat{\alpha} + {\rm h.c.} \right) + \frac{3 p_T^2}{2 p_L^2}\,
\left( \dot{\hat\alpha}^*\, \hat{\alpha}_0 + {\rm h.c.} \right)
\nonumber\\
&-& \frac{3}{2}\, \left( B_1\, H - 2\, B_1\, h + \dot{B}_1
\right)\, \left( \dot{\hat\alpha}_1^*\, \hat{\Psi} + {\rm h.c.}
\right) + \frac{3}{2}\, \left( H - 2\, h \right)\, \left(
\dot{\hat\alpha}_1^*\, \hat{\alpha}_1 + {\rm h.c.} \right) +
\frac{3}{2}\, \left( \dot{B}_1 - 2\, B_1\, h \right)\, \left(
\dot{\hat\alpha}_1^*\, \hat{\Phi} + {\rm h.c.} \right) \nonumber\\
&+& \frac{B_1}{2}\, \left( \dot{\hat\alpha}_1^*\, \hat{\chi} +
{\rm h.c.} \right) + \frac{B_1}{2}\, \left( \dot{\hat\alpha}_1^*\,
\hat{B} + {\rm h.c.} \right) + \frac{3}{2}\, \left(
\dot{\hat\alpha}_1^*\, \hat{\alpha}_0 + {\rm h.c.} \right) -
\left( {\cal D}_{\Psi \Psi} + p_T^2\, B_1^2 \right)\, \vert
\hat\Psi \vert^2 \nonumber\\
&-& \left( {\cal D}_{\Psi \alpha_1} - \frac{1}{2}\, p_T^2\, B_1
\right)\, \left( \hat{\Psi}^*\, \hat{\alpha}_1 + {\rm h.c.}
\right) \nonumber\\
&+& \left[ \left( -\frac{3}{2}\, m^2 + \frac{2 p_L^2 + p_T^2}{4} -
\frac{15}{2}\, h^2 + 6\, h\, H \right)\, B_1^2 + 6\, h\, B_1\,
\dot{B}_1 - \frac{3}{2}\, \left( p_T^2 + \dot{B}_1^2 \right)
\right]\, \left( \hat{\Psi}^*\, \hat{\Phi} + {\rm h.c.} \right)
\nonumber\\
&-& \left( h\, B_1^2 - \frac{1}{2}\, B_1^2\, H + B_1\, \dot{B}_1
\right)\, \left( \hat{\Psi}^*\, \hat{\chi} + {\rm h.c.} \right) -
\left( \frac{9}{2}\, h + \frac{1}{4}\, (h-2 H)\, B_1^2 + B_1\,
\dot{B}_1 \right)\, \left(\hat{\Psi}^*\, \hat{B} + {\rm h.c.}
\right) \nonumber
\end{eqnarray}
\begin{eqnarray}
&-& \frac{3}{2}\, \left( B_1\, H - 2 B_1\, h + \dot{B}_1 \right)\,
\left( \hat{\Psi}^*\, \hat{\alpha}_0 + {\rm h.c.} \right) - \left(
{\cal D}_{\alpha \alpha} + \frac{3}{2}\, p_T^2 \right)\, \vert
\hat\alpha \vert^2 + \frac{3}{2}\, p_T^2\, \left( \hat{\alpha}^*\,
\hat{\alpha}_1 + {\rm h.c.} \right) \nonumber\\
&+& \frac{3}{2}\, \left( 2\, h\, B_1 - B_1\, H - \dot{B}_1
\right)\, \left( \hat{\alpha}^*\, \hat{B} + {\rm h.c.} \right) +
\frac{3 p_T^2}{2 p_L^2}\, \left( H -2\, h\right)\,
\left(\hat{\alpha}^*\, \hat{\alpha}_0 + {\rm h.c.} \right) -
\left( {\cal D}_{\alpha_1 \alpha_1} + \frac{3}{2}\, p_T^2
\right)\, \vert \hat{\alpha}_1 \vert^2 \nonumber \\
&+& \left[ \left( \frac{3}{2}\, m^2 - \frac{p^2}{2} +
\frac{15}{2}\, h^2 - 6\, h\, H \right)\, B_1 - 3 h\, \dot{B}_1
\right]\, \left( \hat{\alpha}_1^*\, \hat{\Phi} + {\rm h.c.}
\right) - \frac{1}{2}\, \left( H\, B_1 - 2 h\, B_1 - \dot{B}_1
\right)\, \left( \hat{\alpha}_1^*\, \hat{\chi} + {\rm h.c.}
\right) \nonumber\\
&+& \frac{1}{2}\, \left( 2\, H\, B_1 - 7 h\, B_1 + 4 \dot{B}_1
\right)\, \left( \hat{\alpha}_1^*\, \hat{B} + {\rm h.c.} \right) +
\frac{3}{2}\, \left(H - 2\, h \right)\, \left( \hat{\alpha}_1^*\,
\hat{\alpha}_0 + {\rm h.c.} \right) \nonumber\\
&+& \left[ \left( 9 + \frac{15}{2}\, B_1^2\right)\, h^2 - 9\, H^2
+ \frac{3}{2}\, \dot{B}_1^2 - 6\, h\, B_1\,  (H\, B_1 +
\dot{B}_1)\right]\, \vert \hat\Phi \vert^2 + \frac{1}{2}\, \left(
6 + B_1^2 \right)\, \left( H + h + \frac{B_1\, \dot{B}_1}{6 +
B_1^2} \right)\, \left( \hat{\Phi}^*\, \hat{\chi} + {\rm h.c.}
\right) \nonumber\\
&+& \frac{1}{2}\, \left( 6 + B_1^2\right)\, \left(H -\frac{1}{2}\,
h + \frac{B_1\, \dot{B}_1}{6 + B_1^2} \right)\, \left(
\hat{\Phi}^*\, \hat{B} + {\rm h.c.} \right) - \left( 3\,h\, B_1 -
\frac{3}{2}\, H\, B_1 - \frac{3}{2}\, \dot{B}_1 \right)\, \left(
\hat{\Phi}^*\, \hat{\alpha}_0 + {\rm h.c.} \right) \nonumber\\
&-& \left( {\cal D}_{\chi \chi} - \frac{6 + 8 B_1^2}{8 p_L^2}\,
p_T^2 \right)\, \vert \hat\chi \vert^2 - \frac{1}{8}\, \left( 6 +
B_1^2 \right)\, \left( \hat{\chi}^*\, \hat{B} + {\rm h.c.} \right)
- {\cal D}_{\chi \alpha_0}\, \left( \hat{\chi}^*\, \hat{\alpha}_0
+ {\rm h.c.} \right) + \frac{p_L^2}{p_T^2}\, \frac{6 + B_1^2}{8}\,
\vert \hat{B} \vert^2 \nonumber\\
&-& \left( {\cal D}_{\alpha_0 \alpha_0} - \frac{3 p^2}{2\, p_L^2}
\right)\, \vert \hat{\alpha}_0 \vert^2 \label{act-2dS-cc}
\end{eqnarray}
where the coefficients $\{ {\cal D}_{\Psi \Psi}, \, {\cal D}_{\Psi
\alpha_1}, \, {\cal D}_{\alpha \alpha}, \, {\cal D}_{\alpha_1
\alpha_1}, \, {\cal D}_{\chi \chi}, \, {\cal D}_{\alpha_0
\alpha_0}, \, {\cal D}_{\chi \alpha_0} \}$ depend on the
background quantities (and, hence, are functions of time), 
and their explicit forms are given in
equations (\ref{calD}) of Appendix~\ref{appA}. As a further check on our algebra, we explicitly verified that the extremization of this action with respect to the fields included in it reproduces the system of linearized equations (\ref{einstein-cc}).

In Section \ref{sec:ghost} we outlined at a formal level the
computation that we are now performing for the model
(\ref{act-V0}). We expressed the quadratic action in
eq. (\ref{formal-act}), where we distinguished between the dynamical
fields $Y_i$ and the nondynamical ones $N_i$. The action
(\ref{act-2dS-cc}) is the explicit form of the action
(\ref{formal-act}) for the model we are studying. The variables $\{
\hat\Phi, \, \hat\chi, \, \hat{B}, \, \hat\alpha_0 \}$ are the
nondynamical modes $\{ N_i \}$, since they enter into the action
(\ref{act-2dS-cc}) without time derivatives. Starting from the action
(\ref{act-2dS-cc}), we integrate out the nondynamical modes, and
compute the action for the dynamical modes, following the same steps
that lead from eq. (\ref{formal-act}) to
eq. (\ref{action-integrated}). In practice, we read the coefficients
$a_{ij} ,\, \dots ,\, f_{ij}$ from the action (\ref{act-2dS-cc}), by
comparing it with the formal expression (\ref{formal-act}); we then
compute the combinations $K = a - b^\dagger \, c^{-1} \, b \;,\;
\Lambda = d - b^\dagger \, c^{-1} \, f \;,\; \Omega^2 = - e +
f^\dagger \, c^{-1} \, f$ (cf. eqs. (\ref{klo})) that characterize the
action of the dynamical modes. This computation is a straightforward
algebraic manipulation; the resulting coefficients $K_{ij}, \,
\Lambda_{ij}, \, \Omega_{ij}^2$ are extremely lengthy, and we do not
report them here.

\begin{figure}
\centering
\includegraphics[width=0.5\textwidth,angle=-90]{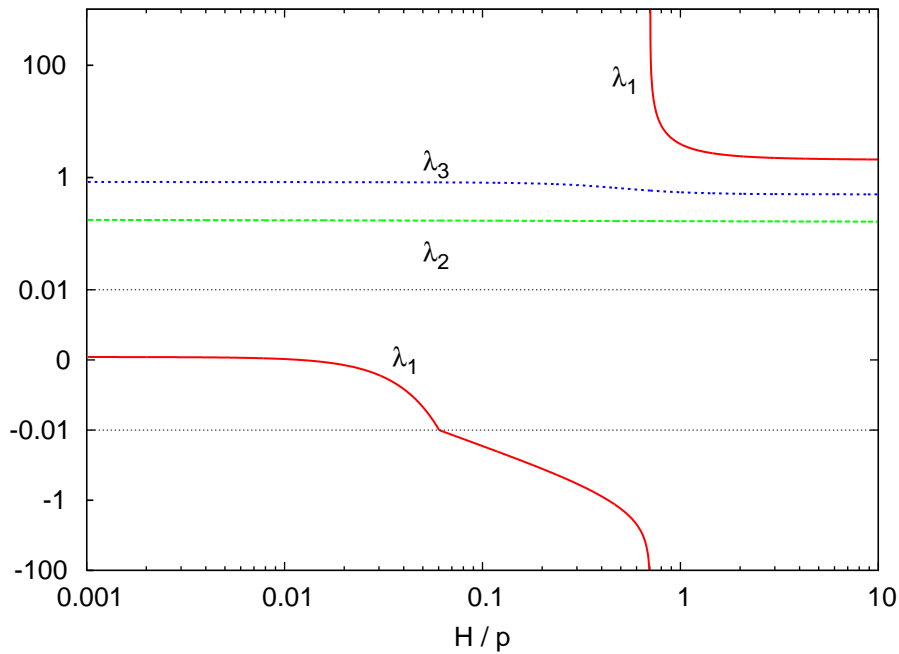}
\caption{Evolution of the eigenvalues of the kinetic matrix. The parameters and initial conditions are
as in Figure \ref{fig:plots}. Due to the wide range spanned by the eigenvalues, in the $y$ axis, we have used a linear scale inside the interval $\left[ - 0.01 ,\, 0.01 \right]$, and
a logarithmic scale outside.
}
\label{fig:cc-auto}
\end{figure}

The instability emerged from the linearized equations can be understood by studying the kinetic matrix $K$. In Figure \ref{fig:cc-auto} we show the three eigenvalues of this matrix, for the same numerical evolution (i.e., for the same parameters and initial conditions) as the one leading to Figure \ref{fig:plots}. We see that the two eigenvalues $\lambda_{2,3}$ are always positive, indicating that the two corresponding eigenmodes are well behaved positive-energy fields. On the contrary, the eigenvalue $\lambda_1$ vanishes close to horizon crossing. The system of linearized equations becomes singular at this point (cf. the formal equations (\ref{formal-eom}); they are singular if the matrix $K$ is noninvertible), and the linearized solutions diverge. We also see from the Figure that the eigenvalue $\lambda_1$ is negative for some time after this moment. The corresponding eigenmode is a ghost in this time interval.

Although the exact expression for $K$ is rather lengthy, it is possible to obtain a simple expansion series for its determinant, in the sub-horizon / early time limit:
\begin{equation}
{\rm det}\left({\frac{K}{a\, b^2}}\right) = \frac{p_T^6\,
B_1^2}{96\, p^6} - \frac{\left( 2 H_0^2 - m^2 \right) p_T^6}{16 \, p_L^2 \, p^6}
+ {\rm O } \left( \frac{H_0^4}{p^4} \right)
\label{detapprx}
\end{equation}
To obtain this expression, we first expanded the exact expression for the determinant in a power series of the momenta; we then simplified each term in the series by using the slow roll solutions (\ref{slow}), and finally kept for each term only the leading expression for $B_1^2 \ll 1 \,$ (which corresponds to small anisotropy, since $h / H = {\rm O } \left( B_1^2 \right) \,$). The terms of ${\rm O } \left( H_0^4 / p^4 \right)$ are parametrically suppressed with respect to the second term in (\ref{detapprx}) for $H_0 \ll p$. Therefore, the first two terms in (\ref{detapprx}) provide an accurate approximation of the determinant in the whole sub-horizon regime.

Eq.~(\ref{detapprx}) shows that the determinant is positive at sufficiently early times, and it then becomes negative in the later part of the sub-horizon regime. This confirms the behavior of $\lambda_1$ seen in Figure \ref{fig:cc-auto} (since $\lambda_{2,3} > 0$, the sign of the determinant coincides with that of $\lambda_1$). The determinant  vanishes when the two leading terms in (\ref{detapprx}) are (approximately) equal; this happens for $p_L^2 \simeq 6 \left( 2 H_0^2 - m^2 \right) / B_1^2 \,$. Not surprisingly, this is precisely the approximate value (\ref{pL2st-small}) of $p_L$ at which the linearized system (\ref{system}) becomes singular.

\begin{figure}
\centering
\includegraphics[width=0.5\textwidth]{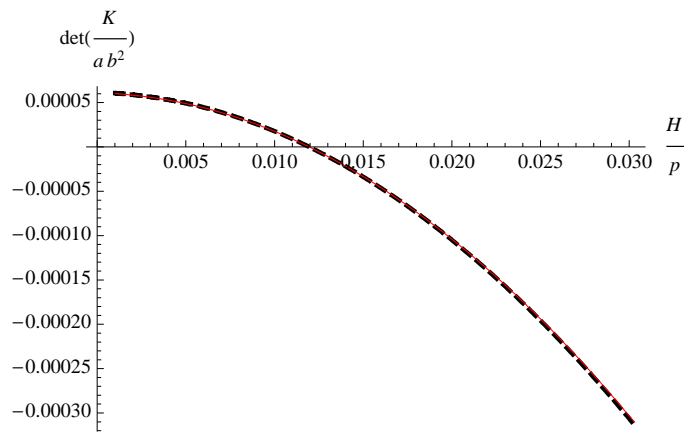}
\caption{Determinant of the kinetic matrix, for the same choice of parameters and initial conditions
as in the previous Figure. Compared with the previous figure, we show a close up of early times, around the point where the determinant vanishes. The black dashed curve shows the exact determinant, while the red curve shows the approximate expression given in eq. (\ref{detapprx}).}
\label{fig:cc-det}
\end{figure}

In Figure \ref{fig:cc-det}, we compare the approximate expression for the determinant - the first two terms in (\ref{detapprx}) - with the exact one (not reported here). We see that the approximated expression is extremely accurate in the range of our interest.

\subsection{One vector plus a scalar inflaton}
\label{1vec-infla}

The model proposed in ref. \cite{soda1} is very similar to the one we
have studied in the previous Subsection, with the only difference that
the vacuum energy that we have included is replaced there by a slowly
rolling inflaton field. We expect that the same instability that we
have found above is present also for this model. The study presented
in this Subsection confirms this. We first briefly review the model
and its slow roll solution. We then show that one of the $2$d scalar
modes is a ghost. More precisely, as in the case studied above, the
mode is well behaved in the deep UV regime, but it becomes a ghost
close to horizon crossing. When this happens, the system of linearized
equations becomes singular.

\subsubsection{The model and the background solution}

The model of ref. \cite{soda1} is characterized by the action:
\begin{equation}
S = \int d^4 x \left[ - \frac{1}{4} F_{\mu \nu} F^{\mu \nu} - \frac{1}{2} \partial_\mu \phi \, \partial^\mu \phi - V \left( \phi \right) - \frac{1}{2} \left( m^2 - \frac{R}{6} \right) A_\mu A^\mu \right]
\label{act-soda}
\end{equation}

The inflaton field $\phi$ replaces the vacuum energy $V_0$ that we considered in the action 
(\ref{act-V0}). In this way, one can have a graceful exit from inflation. The field equations following from this action  are
\begin{eqnarray}
&& G_{\mu\,\nu} = \frac{1}{M_p^2}\, \left[ T_{\mu\nu}^{(\phi)} +
T_{\mu\nu}^{(A)}
\right] \;\;\;,\;\;\;
T_{\mu\nu}^{(\phi)} \equiv \partial_{\mu} \phi \,
\partial_{\nu} \phi - g_{\mu\nu}\, \left( \frac{1}{2}\, \partial_{\alpha} \phi\, \partial^{\alpha} \phi + V(\phi)
\right)\nonumber\\
&& \frac{1}{\sqrt{-g}}\, \partial_{\mu}\, \left[ \sqrt{-g}\,
F^{\mu\nu} \right] - \left( m^2 - \frac{R}{6} \right)\, A^{\nu} =
0 \;\;\;,\;\;\;
\frac{1}{\sqrt{-g}}\, \partial_{\mu}\, \left[ \sqrt{-g}\,
g^{\mu\nu}\, \partial_{\nu}\, \phi \right] - V'\left( \phi \right)
= 0 \label{field-eqns-soda}
\end{eqnarray}
where $T_{\mu\nu}^{(A)}$ is defined as in (\ref{field-eqns-cc}). The background solution is again of the form (\ref{ansatz}), giving
\begin{eqnarray}
&& 3 H^2 - 3 h^2 = \frac{1}{M_p^2}\, \left( \frac{1}{2}\,
\dot\phi^2 + V\left(\phi\right) \right) + \frac{1}{2}\,
\dot{B}_1^2 - 2 h\, B_1\, \dot{B}_1 + \frac{1}{2}\, \left( m^2 + 5
h^2 - 4 h\, H \right)\, B_1^2
\nonumber\\
&& \dot{h} + 3 H h = \frac{1}{3} B_1^2 \left( \dot{H} - \frac{\dot{h}}{2} \right) + \frac{1}{3} \dot{B}^2 + \frac{1}{3} \left( 2 H - 5 h \right) B_1 \dot{B}_1 + \frac{1}{3} \left( 3 H^2 - \frac{11}{2} H h + 5 h^2 - m^2 \right) B_1^2 \nonumber\\
&& 2 \dot{H} + 3 H^2 + 3 h^2 = \frac{1}{M_p^2}\, \left(
-\frac{1}{2}\, \dot\phi^2 + V\left(\phi \right) \right) - \frac{1}{2} \dot{B}_1^2 - \frac{1}{3} B_1 \left[ \ddot{B}_1 + \left( 3 H - 2 h \right) \dot{B}_1 \right] + \frac{1}{6} \left( 4 H h - 5 h^2 + m^2 \right) B_1^2 \nonumber\\
&& \ddot{B}_1 + 3 H\, \dot{B}_1 + \left( m^2 - 5 h^2 - 2 h\, H - 2
\dot{h} \right)\, B_1 = 0 \nonumber\\
&& \ddot{\phi} + 3 H\, \dot{\phi} + V'\left( \phi \right) = 0
\label{evolution-soda-2}
\end{eqnarray}
where we have used the same combinations of Einstein equations we
have written in (\ref{evolution-cc}).

We assume that the inflaton field is in the slow roll regime,
$\dot\phi \approx - V'(\phi) / 3 H \,$, leading to inflation. For what
concerns the evolution of the vector field, and of the anisotropy, we
all the considerations done for the model (\ref{act-V0}) are valid
also in the present case, with the only difference that the vacuum
energy $V_0$ is now replaced by the slowly decreasing potential energy
of the inflaton. This leads to the slow roll and small anisotropy
($B_1 \ll 1$) background evolution
\begin{eqnarray}
&&H \approx H_0 + \frac{m^2}{12 \, H_0} \, B_1^2 + {\rm O } \left( B_1^4 \right) \;\;,\;\;
h \approx \frac{H_0}{3} \, B_1^2 +  {\rm O } \left( B_1^4 \right) \;\;,\;\;
\dot{B_1} \approx - \frac{m^2}{3 \, H_0} B_1 + {\rm O } \left( B_1^3 \right) \;\;,\;\;
\dot\phi \approx -\frac{V'(\phi)}{3 H_0} + {\rm O } \left( B_1^2 \right) \nonumber\\
&&H_0 \equiv \frac{\sqrt{V \left( \phi \right)}}{\sqrt{3} \, M_p}
\label{slow2}
\end{eqnarray}

\subsubsection{Ghost instability from the quadratic action}

We now study the perturbations of the background solution. We
again decompose the perturbations as in (\ref{defn-perts-2d}),
disregard the well behaved system of $2$d vector modes, and construct the gauge
invariant combinations (\ref{GI-2dS}) and (\ref{GI-infla}).

We proceed as in \ref{subsect:ghostV0} by expanding the action of the model at quadratic
order in the $2$d scalar modes, and by rewriting it solely in terms of the gauge invariant modes
(which provides a nontrivial check on our algebra). We find 
\begin{eqnarray}
S_{{\rm 2dS}} &=& \frac{1}{2}\, \int d^3 k\, dt\, a\, b^2\, {\cal
L}_{{\rm 2dS}} \nonumber\\
{\cal L}_{{\rm 2dS}} &\supset& \vert \dot{\hat{\delta\phi}}
\vert^2 + \frac{2}{3}\, B_1^2\, \vert \dot{\hat\Psi} \vert^2 +
\frac{p_T^2}{p_L^2}\, \vert \dot{\hat{\alpha}} \vert^2 + \vert
\dot{\hat{\alpha}}_1 \vert^2 - \frac{1}{3}\, B_1\, \left(
\dot{\hat\Psi}^*\, \dot{\hat{\alpha}}_1 + {\rm h.c.} \right) -
\frac{\dot\phi}{M_p}\, \left( \dot{\hat{\delta\phi}}^*\, \hat\Phi
+ {\rm h.c.} \right) \nonumber\\
&& + \frac{1}{3}\, \left( 2 \left( B_1^2 - 3 \right)\, H - \left(
6 +B_1^2 \right)\, h - B_1\, \dot{B}_1 \right)\, \left(
\dot{\hat\Psi}^*\, \hat\Phi + {\rm h.c.} \right) -
\frac{B_1^2}{3}\, \left( \dot{\hat\Psi}^*\, \hat\chi + {\rm h.c.}
\right) \nonumber\\
&& + \left( 1 - \frac{B_1^2}{6} \right)\, \left(
\dot{\hat\Psi}^*\, \hat{B} + {\rm h.c.} \right) +
\frac{p_T^2}{p_L^2}\, \left( \dot{\hat\alpha}^*\, \hat{\alpha}_0 +
{\rm h.c.} \right) + \left( \dot{B}_1 -2 B_1\, h \right)\, \left(
\dot{\hat\alpha}_1^*\, \hat\Phi + {\rm h.c.} \right) \nonumber\\
&& + \frac{B_1}{3}\, \left( \dot{\hat\alpha}_1^*\, \hat\chi + {\rm
h.c.} \right) + \frac{B_1}{3}\, \left( \dot{\hat\alpha}_1^*\,
\hat{B} + {\rm h.c.} \right) + \left( \dot{\hat{\alpha}}_1^*\,
\hat{\alpha}_0 + {\rm h.c.} \right) \nonumber\\
&& + \left( -6 H^2 + \left( 6 +5 B_1^2\right)\, h^2 + \dot{B}_1^2
-4 B_1\, \left( \dot{B}_1 + H\, B_1 \right)\, h +
\frac{\dot\phi^2}{M_p^2} \right)\, \vert \hat\Phi \vert^2
\nonumber\\
&& + \frac{1}{3}\, \left( \left( 6 + B_1^2 \right)\, \left( H + h
\right) + B_1\, \dot{B}_1 \right)\, \left( \hat{\Phi}^*\, \hat\chi
+ {\rm h.c.} \right) + \frac{1}{6}\, \left( \left(6+B_1^2\right)\,
(2 H-h)+2\, B_1\, \dot{B}_1 \right)\, \left( \hat{B}\,
\hat{\Phi}^* + {\rm h.c.} \right) \nonumber\\
&& + \left( \dot{B}_1 + \left( H - 2 h \right)\, B_1 \right)\,
\left( \hat{\Phi}^*\, \hat{\alpha}_0 + {\rm h.c.} \right) + \left(
\Delta + \frac{p_T^2}{12 p_L^2}\, \left( 6 + B_1^2 \right)
\right)\, \vert \hat\chi \vert^2 - \frac{1}{2}\, \left( 1 +
\frac{B_1^2}{6} \right)\, \left( \hat\chi^*\, \hat{B} + {\rm h.c.}
\right) \nonumber\\
&&+ \frac{\Delta}{B_1}\, \left( \hat\chi^*\, \hat{\alpha}_0 + {\rm
h.c.} \right) + \frac{p_L^2}{12 p_T^2}\, \left( 6 + B_1^2
\right)\, \vert \hat{B} \vert^2 + \left( \frac{\Delta}{B_1^2} +
\frac{p^2}{p_L^2} \right)\, \vert \hat{\alpha}_0 \vert^2 + \dots
\label{act-2dS-soda}
\end{eqnarray}
where we have defined
\begin{eqnarray}
&& \Delta = \frac{1}{p_L^2\, \left( 18 + 3 B_1^2 + 2 B_1^4
\right)}\, \Bigg\{ 3 \left( 6 m^2 + 12 h^2 - 12 H^2+ 3 \dot{B}_1^2
+ 3 \frac{\dot\phi^2}{M_p^2} \right)\, B_1^2 - 24 h\, B_1^3\,
\dot{B}_1 + 2\, \left( 2 H - 7 h \right)\, B_1^5\, \dot{B}_1
\qquad\qquad\qquad\qquad \nonumber\\
&& \qquad\qquad\qquad\qquad\qquad\qquad\quad + 3\, \left( m^2 + 17
h^2 - 12 h\, H - 2 H^2 + \frac{7}{6} \dot{B}_1^2 +
\frac{\dot\phi^2}{2 M_p^2} \right)\, B_1^4 + \left( \frac{31}{2}
h^2 - 14 h\, H + 2 H^2 \right)\, B_1^6 \Bigg\} \nonumber
\end{eqnarray}

In the action (\ref{act-2dS-soda}), we included only the terms that contribute to the kinetic matrix of the dynamical modes; the remaining terms, denoted by the dots, are omitted for brevity. Specifically, the terms included in (\ref{act-2dS-soda}) are those proportional to the coefficients $a_{ij} ,\, b_{ij} ,$ and $c_{ij}$ in eq. (\ref{formal-act}), where the quadratic action is written at a formal level. These are the only terms entering in the kinetic matrix $K$ of the dynamical modes, see eqs. (\ref{klo}) and (\ref{action-integrated}). It is immediate to compute $K$ from (\ref{act-2dS-soda}). However, the explicit entries of this matrix are very involved, and not illuminating. For this reason, we do not report them here. We can however compute the eigenvalues of this matrix numerically for any given choice of parameters (namely, for any background evolution, and momentum of the perturbation). In addition, it is possible to obtain a simple approximation for the determinant of $K$
\begin{equation}
{\rm det}\left({\frac{K}{a\, b^2}}\right) = \frac{p_T^6\,
B_1^2}{192\, p^6} - \frac{\left( 2 H_0^2 - m^2 \right) p_T^6}{32 \, p_L^2 \, p^6}
+ {\rm O } \left( \frac{H_0^4}{p^4} \right) \label{detapprx-2}
\end{equation}
where $H_0$ has been defined in eq. (\ref{slow2}). This expression is  accurate in the sub-horizon regime during inflation, in the limit of small anisotropy. More specifically, we obtained it using the same steps outlined after the analytic expression for the determinant of the model considered in the previous Subsection, eq. (\ref{detapprx}).  It is worth noting that (\ref{detapprx-2}) differs from (\ref{detapprx}) only by an extra factor of $1/2$; this is the original factor in the kinetic term of the inflaton (which is the only additional field in the model we are considering in this Subsection). This suggests that the perturbation 
${\hat \delta \phi}$ is decoupled from the other ones at leading order.

In figure \ref{fig:plots-infla} we present the results of a numerical evolution for a given set of parameters and initial conditions (starting from inflation, in the slow roll regime). The left panel shows the evolution of various background quantities. The right panel shows instead the determinant of the kinetic matrix 
for a mode with $p_L=100 H$, $p_T=200 H$ at the initial time (both the exact value, obtained from a numerical evaluation of the kinetic matrix, and the approximated value (\ref{detapprx-2}), are shown). We see that the determinant vanishes and becomes negative in the sub-horizon regime, when $H/p \sim B \,$.

\begin{figure}[h]
\centerline{
\includegraphics[width=0.4\textwidth,angle=-90]{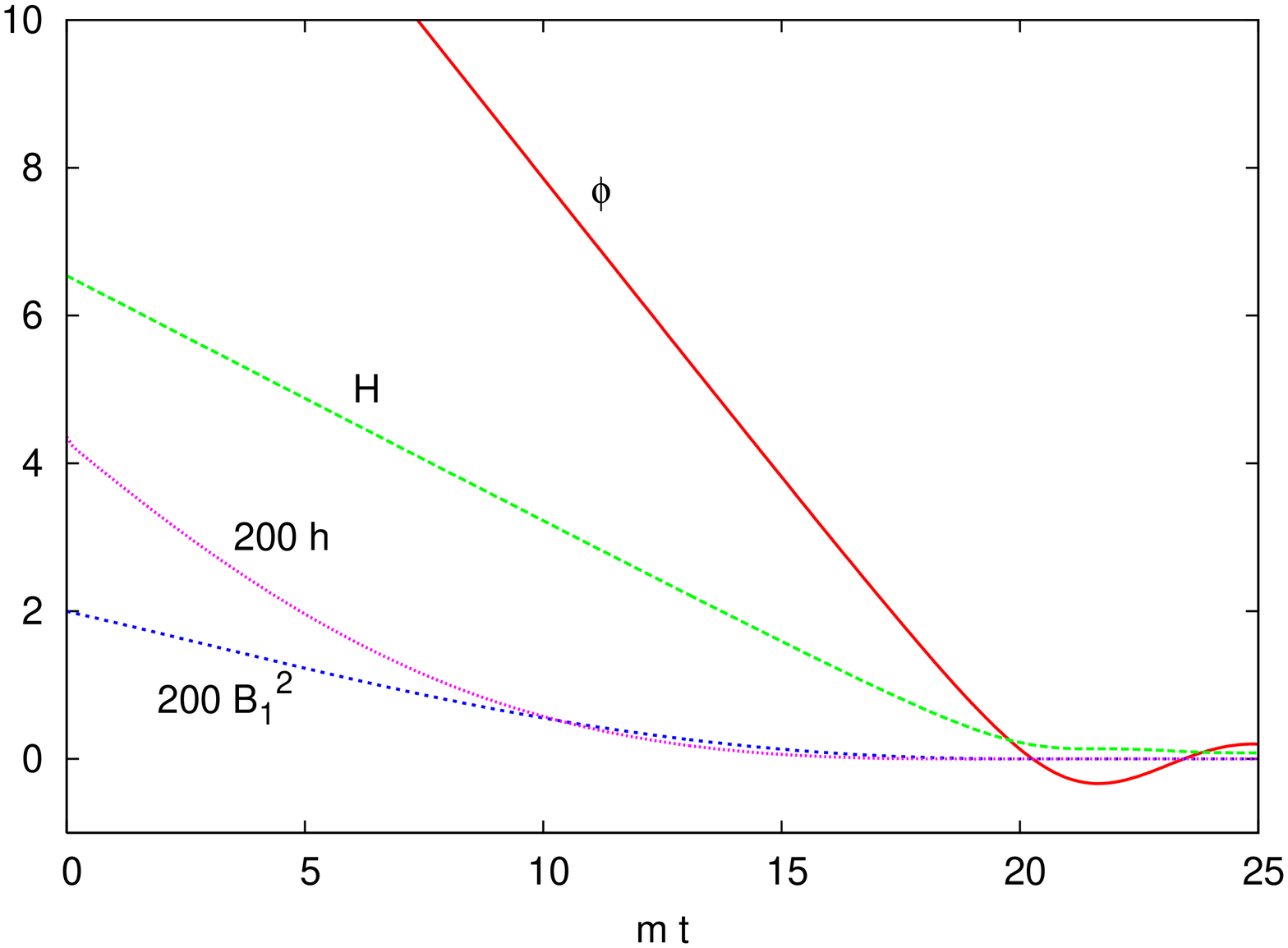}
\includegraphics[width=0.4\textwidth,angle=-90]{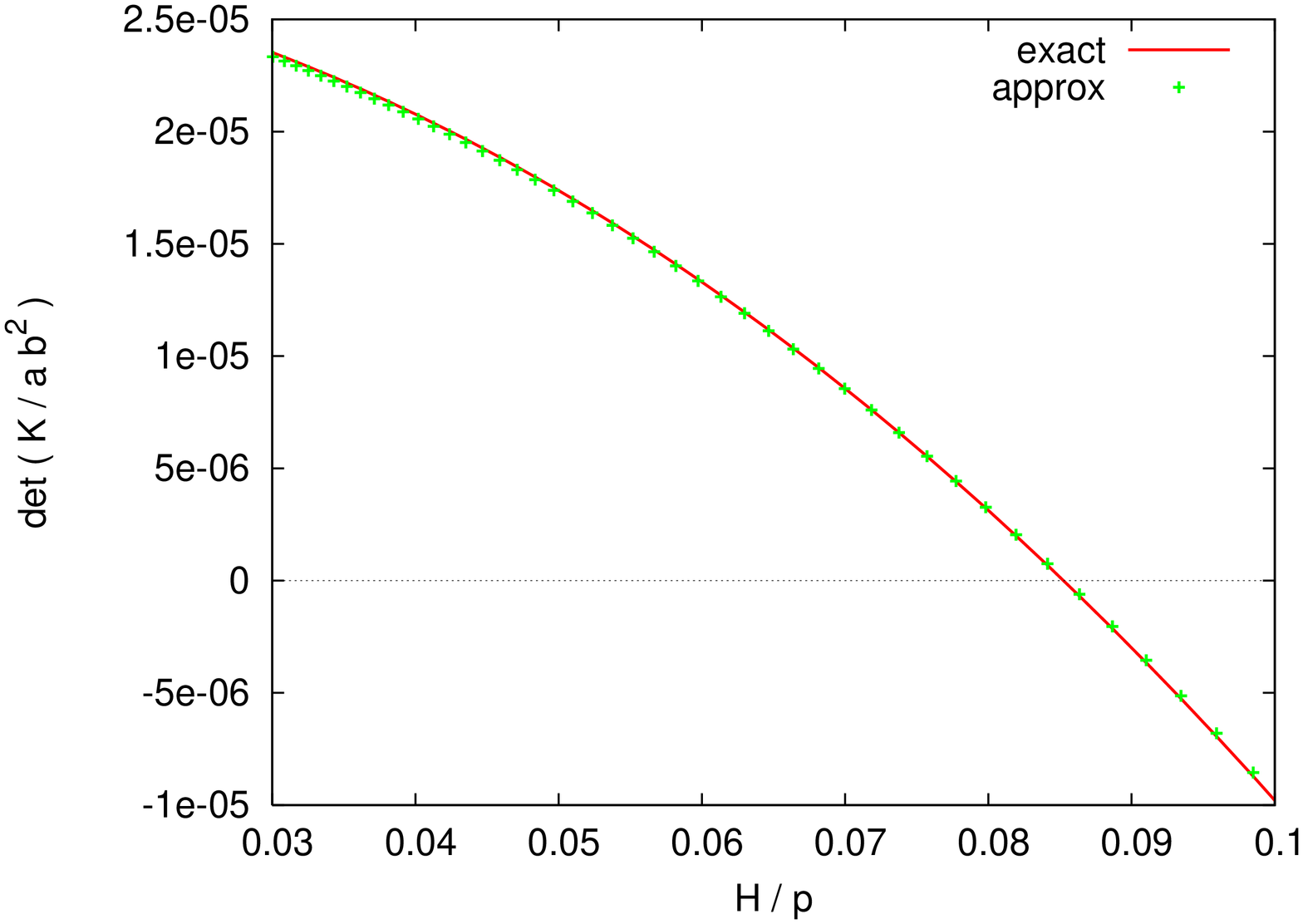}
} \caption{
Results from a numerical simulation starting at $t=0$ from $\phi =  16$ (providing about $60$ e-folds
of inflationary expansion), $B_1 = 0.1$ (providing a $\sim B_1^2 \simeq 1 \%$ anisotropy). More precisely, we have considered a massive inflaton potential, with the inflaton mass equal to $m \,$. Left panel: inflaton (in units of $M_p$), hubble parameters (in units of $m$), and dimensionless rescaled vector$B_1$. The anisotropic rate $h$ and the vector are rescaled so that they are visible in the figure. Right panel: determinant of the kinetic matrix of the perturbations, for modes with initial momenta $p_{L,{\rm in}} = 100 H_{\rm in} ,\, p_{T,{\rm in}} = 200 H_{\rm in} ,\, $. The red curve shows the exact determinant, while the green points show the approximate expression given in eq. (\ref{detapprx-2}). The determinant vanishes at the time $m \, t \simeq 0.16$.} \label{fig:plots-infla}
\end{figure}

In Figure \ref{fig:infla-auto} we show the evolution of the four eigenvalues of the kinetic matrix; notice that there is an additional dynamical mode, supported by the inflaton field, with respect to the model studied in the previous Subsection. However, as for that case, one mode is a ghost for some time. We know that the system of linearized equations for the perturbations become singular when the eigenvalue $\lambda_1$ crosses zero. We expect that also the solutions diverge at that point, analogously to the study of the previous Subsection.

\begin{figure}
\centering
\includegraphics[width=0.5\textwidth,angle=-90]{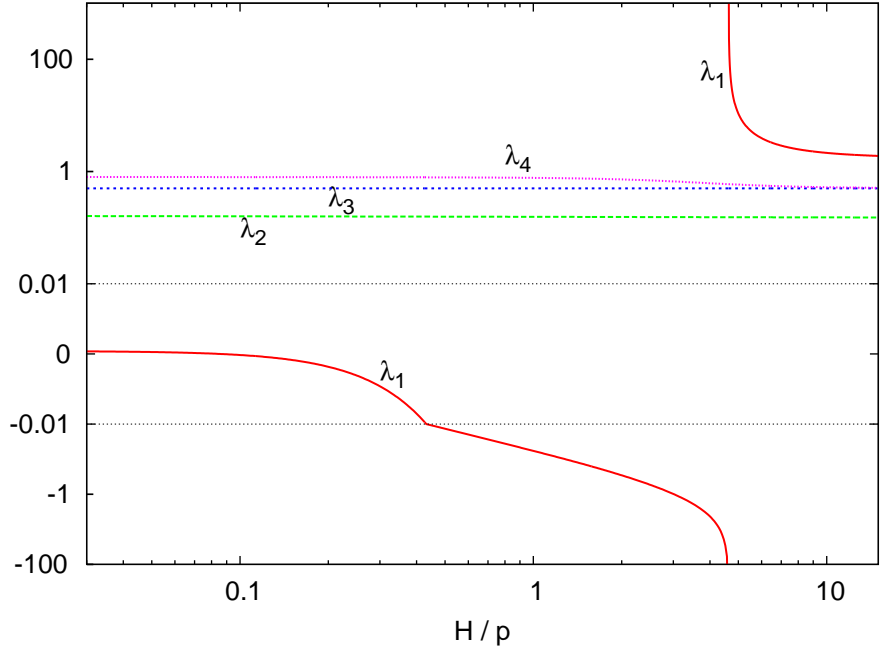}
\caption{Eigenvalues of the kinetic matrix, for the same choice of parameters and initial conditions
as in the right panel of Figure \ref{fig:plots-infla}. The eigenvalues $\lambda_{2,3,4}$ are always positive, so that the sign and the behavior of the determinant are determined by $\lambda_1$. This eigenvalue is initially positive, crosses zero at $H/p \simeq 0.082$ (see the previous figure), and diverges at $H/p \simeq 4.6$. In the $y$ axis, we have used a linear scale inside the interval $\left[ - 0.01 ,\, 0.01 \right]$, and a logarithmic scale outside.
}
\label{fig:infla-auto}
\end{figure}

\subsection{Vector inflation}
\label{vector-inflation}

We now study the simplest realization of the idea of vector inflation proposed in~\cite{mukhvect}. It is characterized by three mutually orthogonal vector fields nonminimally coupled to the curvature. The three fields have equal vev, providing a FRW background. Despite the background solution is isotropic, the model suffers of the same ghost instability as the models studied above. The discussion is divided in two parts. We first introduce the model, and discuss the background evolution. We then study the spectrum of perturbations around this background solution.

\subsubsection{The model and the background solution}

The action of the model is given in eq. (\ref{ac-gmv}), with $\xi = -1/6 \,$, and $N=3 \,$:
\begin{equation}
S = \int d^4 x \, \sqrt{-g} \,  \sum_{a=1}^{3}\, \left[ -\frac{1}{4}\,
F^{(a)}_{\mu\nu}\, F^{(a)\, \mu\nu} - \frac{1}{2}\, \left( - \frac{R}{6} + m^2 \right)\, A_{\mu}^{(a)}\, A^{(a)\mu} \right]
\label{ac-gmv2}
\end{equation}

The background vev of the three vectors is chosen as in eq. (\ref{3vev}), while the background geometry is $d s^2 = - d t^2 + a^2 \left( t \right) d 
\vec{x}^2 \,$. The system is governed by the equations
\begin{eqnarray}
&& G_{\mu\nu} = \frac{1}{M_p^2}\, \sum_{a=1}^{3}\,
T_{\mu\nu}^{(a)} \nonumber\\
&& T_{\mu\nu}^{(a)} \equiv F_{\mu}^{(a)\, \sigma}\,
F^{(a)}_{\nu\sigma} - \frac{1}{4}\, F^{(a)}_{\alpha\beta}\,
F^{(a)\, \alpha\beta}\, g_{\mu\nu} + \left( m^2 - \frac{R}{6}
\right)\, A_{\mu}^{(a)}\, A_{\nu}^{(a)} - \frac{1}{2}\, m^2\
A_{\alpha}^{(a)}\, A^{(a)\, \alpha}\, g_{\mu\nu} \nonumber\\
&& \qquad\qquad -\frac{1}{6}\, \left( R_{\mu\nu} - \frac{1}{2}\,
R\, g_{\mu\nu} \right)\, A_{\alpha}^{(a)}\, A^{(a)\, \alpha} -
\frac{1}{6}\, \left( g_{\mu\nu}\, \square - \nabla_{\mu}\,
\nabla_{\nu} \right)\, A_{\alpha}^{(a)}\, A^{(a)\,
\alpha} \nonumber\\
&& \frac{1}{\sqrt{-g}}\, \partial_{\mu}\, \left[ \sqrt{-g}\,
F^{(a)\, \mu\nu} \right] - \left( m^2 - \frac{R}{6} \right)\,
A^{(a)\, \nu} = 0 \label{field-eqns-mukh}
\end{eqnarray}

For the background under considerations, the $00$ Einstein equation and the $i-th$ spatial component of the equation for the $i-th$ vector give
\begin{eqnarray}
&& H^2 = \frac{1}{2}\, \left( \dot{B}^2 + m^2\, B^2 \right)
\nonumber\\
&& \ddot{B} + 3 H\, \dot{B} + m^2\, B = 0 \label{eqns-mukh}
\end{eqnarray}
As always for a FRW geometry, the diagonal $ii$ Einstein equations
does not provide additional information (due to a nontrivial Bianchi
identity). The remaining equations vanish identically. Upon the
identification $B = \phi_{\rm eff} / \left( \sqrt{3} M_p \right)$, we
recover the same equations as those of chaotic inflation driven by a
minimally coupled scalar field $\phi_{\rm eff}$ (where the suffix
stands for ``effective'') with potential $V = m^2 \phi_{\rm eff}^2 /
2$. Inflation is therefore characterized by the slow roll evolution
\begin{equation}
H^2 \approx \frac{m^2}{2}\, B^2 \,\,\,\, , \,\,\,\, \dot{B}
\approx -\frac{m^2}{3 H}\, B \label{slow-roll-mukh}
\end{equation}
which applies as long as the slow roll conditions ~\footnote{See \cite{vec-cond} for more detailed studies of initial and slow-roll conditions for vector inflation.}
\begin{eqnarray}
\epsilon &\equiv& \frac{M_p^2}{2} \left( \frac{\frac{d V}{d \phi_{\rm eff}}}{V} \right)^2 = \frac{2 \, M_p^2}{\phi_{\rm eff^2}} = \frac{2}{3 \, B^2} \ll 1 \nonumber\\
\eta &\equiv& M_p^2 \, \frac{\frac{d^2 V}{d \phi_{\rm eff}^2}}{V}  = \frac{2 \, M_p^2}{\phi_{\rm eff^2}} = \frac{2}{3 \, B^2} \ll 1 \nonumber\\
\end{eqnarray}
are valid. After that,  $B$ performs damped coherent oscillations around $B=0 \,$ (see for instance \cite{kls}):
\begin{equation}
B = \left( \frac{\sqrt{8}}{3 \, m \, t} + {\rm O } \left( \frac{1}{m^2 \, t^2} \right) \right) \, \sin \left( m \, t + \xi_0 \right)
\label{oscB}
\end{equation}
where $\xi_0$ is a phase (irrelevant for this discussion). The coherent oscillations provide a stage of effective matter domination ($w =0$ once averaged over a few oscillations) \cite{kls}. It is expected that the vectors then decay into the visible matter (either perturbatively, or nonperturbatively) giving rise to the radiation dominated stage of cosmology.

From the slow roll conditions, we see that $H^2 \gg m^2$ during most of the inflationary stage. At these times, the mass term for the vectors is negative, $- R / 6 + m^2 \simeq - 2 H^2 + m^2 < 0 \,$. However, $R$ decreases as $B$ rolls towards the origin, while $m$ remains constant. We find numerically that  $- R / 6 + m^2 = 0$ when $B \simeq 1.048 \,$. This happens towards the end of inflation.

\subsubsection{Ghost instability from the quadratic action}

We now study the perturbations of the model (\ref{ac-gmv2}) around the
background solution just discussed. Since the background geometry is
FRW, one may choose to adopt the standard decomposition of metric
perturbations, and decompose them into scalar, vector, and tensor
modes with respect to $3$d spatial rotations \cite{mfb}.  While this
is possible, one should however bear in mind that, contrary to the
standard case, these three groups of modes cannot be studied
separately even at the linearized level. Consider for instance the
tensor perturbations $h_{ij}^{TT}$, introduced as the traceless
($h_{ii}^{TT} = 0$), and transverse ($\partial_j \, h_{ij}^{TT} = 0$)
part of the spatial metric perturbations, $\delta g_{ij} = a^2 \,
h_{ij}^{TT} \,$ \cite{mfb}. While in the case of scalar field
inflation these modes are decoupled from the other perturbations, and
can be studied separately, in the case under consideration they are
coupled to the perturbations of the vector fields. Namely, we find the
following coupling in the quadratic action $\delta_2 S$ of the
perturbations:
\begin{equation}
\delta_2 S \supset - \frac{M_p}{2} \int d^4 x \, a^2 \left[ \left( \dot{B} + H \, B \right) \delta \dot{A}_j^{(i)} +
\left( 2 H^2 + \dot{H} - m^2 \right) B \, \delta A_j^{(i)} + \left( i \leftrightarrow j \right) \right] h_{ij}^{TT}
\label{coupledTT}
\end{equation}
As a consequence, the perturbations of the vector fields must be
included in the linearized equations of motion for the tensor metric
perturbations. In turns, the perturbations of the vector fields are
coupled to the scalar and vector perturbations of the metric. One then
finds that all the perturbations of the system need to be studied
together, even at the linearized
level.~\footnote{Ref.~\cite{mukhvect2} studied the tensor modes alone,
  arguing that the effect of the coupling to the $\delta A_\mu^{(a)}$
  perturbations can be ``averaged away'' in the limit of many
  vectors. However, each vector introduces a coupling of $h_{ij}^{TT}$
  with its own perturbations, and perturbations of different vectors
  cannot cancel against each other in the study of the spectrum (and,
  therefore, of the stability) of the theory.}

This makes the study extremely hard, and indeed no complete
computation exists up to date. The metric has $10$ perturbations,
while each vector introduces $4$ perturbations. Thus, in this simplest
realization of vector inflation, one starts with a system of $22$
coupled modes. Four perturbations can be removed by fixing the freedom
of general coordinate transformations, leading to a system of $18$
coupled modes. Not all these modes are dynamical. The nondynamical
modes originate from the initial perturbations $\delta g_{0\mu}$ and
$\delta A_0^{(a)}$, which enter in the quadratic action of the
perturbations without time derivatives. As we now show, it is possible
to choose gauge invariant perturbations which associate a gauge
invariant combination to each of the $\delta g_{0\mu}$ and $\delta
A_0^{(a)}$ modes. These seven gauge invariant combinations are also
nondynamical fields. We integrate them out as outlined in Section
\ref{sec:ghost}, and obtain the quadratic action for the dynamical
modes. From the study of the kinetic matrix of this system, we find
that three modes become ghosts for some time during inflation.

Since the usual decomposition in scalar/vector/tensor modes does not provide decoupled sets of perturbations, we do not employ it, and simply decompose the perturbations as
\begin{eqnarray}
&& g_{\mu\nu} = \left( \begin{array}{cccc} -1-2\Phi & a \chi_1 & a
\chi_2 & a \chi_3 \\
& a^2\, \left( 1 - 2 \Psi \right) & a^2\, \partial_1\,
\partial_2\, E_3 & a^2\, \partial_1\, \partial_3\, E_2 \\
& & a^2\, \left( 1 - 2 \Sigma \right) & a^2\, \partial_2\,
\partial_3\, E_1 \\
& & & a^2\, \left[ 1 - 2 \left( T - \Psi - \Sigma \right) \right]
\end{array} \right) \nonumber\\
&& A_{\mu}^{(a)} = A_{\mu}^{0\, (a)} + \alpha_{\mu}^{(a)} = a\,
M_p\, B\, \delta^{a}_{\mu} + \alpha_{\mu}^{(a)}
\label{metric-vector-mukh}
\end{eqnarray}
where $A_{\mu}^{0\, (a)}$ denotes background values of the vector
fields. The parametrization of the $\delta g_{33}$ component has chosen so that
$\delta g^i_i = - 2 \, T \,$. Moreover, for algebraic convenience, some $\delta g_{\mu \nu}$ entry
has been written with spatial derivatives (in practice, we assume that the momentum of the modes - after Fourier transforming - has nonvanishing components in all three directions, $k_x ,\, k_y ,\, k_z \neq 0 \,$).

We need to fix the gauge freedom associated with general coordinate transformations. 
Consider the infinitesimal transformation $x^\mu \rightarrow x^\mu + \xi^\mu$, under which the perturbations of the metric and the vector field transform as $\delta g_{\mu \nu} \rightarrow \delta g_{\mu \nu} - g_{\mu \nu \,,\alpha}^{(0)} \xi^\alpha - g_{\mu \alpha}^{(0)} \xi^\alpha_{,\nu} - g_{\alpha \nu}^{(0)} \xi^\alpha_{,\mu} \,$, and $\delta A_\mu^{(a)} \rightarrow \delta A_\mu^{(a)} - A_{\mu,\alpha}^{(a)} \xi^\alpha - A_\alpha^{(a)} \, \xi^\alpha_{,\mu} \,$. Due to the assumption of the modes we are studying, we need to consider infinitesimal transformations with nontrivial spatial dependence along all the three directions. We can therefore parametrize the transformation parameter as $\xi^\mu = \left( \xi^0 ,\, \partial_1 \xi^1 ,\, \partial_2 \xi^2 ,\, \partial_3 \xi^3 \right) \,$. The explicit transformations of the modes in 
(\ref{metric-vector-mukh}) are then 
\begin{eqnarray}
&&\Phi \rightarrow \Phi - \partial_0 \xi^0 \;\;\;,\;\;\; \chi_i \rightarrow \chi_i + \frac{1}{a} \, \partial_i \xi^0 - a \,\partial_0 \, \partial_i \, \xi^i \;\; ( {\rm no \; sum \; over \; } i ) \nonumber\\
&&E_1 \rightarrow E_1 - \xi_2 - \xi_3 \;\;,\;\;
E_2 \rightarrow E_2 - \xi_1 - \xi_3 \;\;,\;\;
E_3 \rightarrow E_3 - \xi_1 - \xi_2 \;\;,\;\; \nonumber\\
&&\Psi  \rightarrow \Psi + H \, \xi^0 + \partial_1^2 \xi^1 \;\;,\;\; 
\Sigma  \rightarrow \Sigma + H \, \xi^0 + \partial_2^2 \xi^2 \;\;,\;\;
T \rightarrow T + 3 \, H \, \xi^0 + \partial_i^2 \xi^i \nonumber\\
&&\alpha_\mu^{(i)} \rightarrow \alpha_\mu^{(i)} - \partial_0 \left( a \, M_p \, B \right) \xi^0 \, \delta_\mu^i
- a \, M_p \, B \, \partial_\mu \, \partial_i \xi^i \;\; ( {\rm no \; sum \; over \; } i ) 
\label{transf-mukh}
\end{eqnarray}
We consider the combinations
\begin{equation}
C^1 \equiv \frac{E_1 - E_2 - E_3}{2} \;\;,\;\;
C^2 \equiv \frac{E_2 - E_1 - E_3}{2} \;\;,\;\;
C^3 \equiv \frac{E_3 - E_1 - E_2}{2} \;\;,\;\;
C^0 \equiv \frac{1}{3 \, H} \left[ T - \partial_i^2 \, C^i \right]
\end{equation}
that transform as $C^\mu \rightarrow C^\mu + \xi^\mu \,$. Then, the modes
\begin{eqnarray}
{\hat \Phi} &\equiv& M_p \left( \Phi + \partial_0 \, C^0 \right) \nonumber\\
{\hat \chi}_i &\equiv& M_p \left( \chi_i - \frac{1}{a} \, \partial_i \, C^0 + a \, \partial_0 \, \partial_i \, C^i \right) \;\; ( {\rm no \; sum \; over \; } i ) \nonumber\\
{\hat \Psi} &\equiv& M_p \left( \Psi - H \, C^0 - \partial_1^2 \, C^1 \right) \nonumber\\
{\hat \Sigma} &\equiv& M_p \left( \Sigma - H \, C^0 - \partial_2^2 \, C^2 \right) \nonumber\\
{\hat \alpha}_\mu^{(i)} &\equiv& \alpha_\mu^{(i)} + \partial_0 \left( a \, M_p \, B \right) C^0 \, \delta_\mu^i
+ a \, M_p \, B \, \partial_\mu \, \partial_i C^i \;\; ( {\rm no \; sum \; over \; } i ) 
\label{gauge-inv-mukh}
\end{eqnarray}
are gauge invariant (all these modes have mass dimension $+1 \,$). 

We expanded the metric and vector fields according to (\ref{metric-vector-mukh}), and computed the quadratic action for the perturbations. We verified that the perturbations rearrange so that the quadratic action can be written solely in terms of the gauge invariant modes (\ref{gauge-inv-mukh}) (this is a nontrivial check on our algebra). In this way, we have eliminated the redundancy associated to general coordinate transformations.~\footnote{Notice that the procedure we adopted is equivalent to choose the gauge $E_1 = E_2 = E_3 = T = 0$. It is easy to see from (\ref{transf-mukh}) that this choice can always be made, and it completely fixes the gauge freedom.} The action in Fourier space reads
\begin{eqnarray}
\delta_2 S &=&  \int\, d^3 k\, dt\, a^3\, {\cal L}
\nonumber\\
{\cal L} &=& \frac{1}{2}\, \left(2 + B^2\right)\, \left(
\vert \dot{\hat \Psi} \vert^2 + \vert \dot{\hat \Sigma} \vert^2 \right) +
\frac{1}{2 a^2\, }\, \sum_{j,a=1}^3\, \vert
\dot{{\hat \alpha}}_j^{(a)} \vert^2 + \frac{1}{4}\, \left( 2 + B^2
\right)\, \left( \dot{{\hat \Psi}}^*\, \dot{{\hat \Sigma}} +
{\rm h.c.} \right) \nonumber\\
&& -\frac{1}{4}\, \left(2 + B^2 \right)\, \left[ i\, \dot{\hat \Psi}^*\,
\left( p_1\, {\hat \chi}_1 - p_3\, {\hat \chi}_3 \right) +
i \dot{\hat \Sigma}^*\, \left( p_2\, {\hat \chi}_2 - p_3\, {\hat \chi}_3 \right) + \rm{h.c.} \right] \nonumber\\
&& + \frac{1}{2 a\, }\, \sum_{j,a,=1}^3\, \left( i\, p_j\,
\dot{{\hat \alpha}}_j^{(a)*}\, {\hat \alpha}_0^{(a)} + {\rm h.c.} \right) -
\frac{1}{2\, a\, }\, \dot{B}\, \sum_{j=1}^{3}\, \left(
\dot{{\hat \alpha}}_j^{(j)*}\, {\hat \Phi} + {\rm h.c.} \right)+ \frac{B}{6 \,a
}\, \sum_{j,a,=1}^3\, \left( i p_j\, {\hat \chi}_j^*\, \dot{\hat \alpha}_a^{(a)}\,
 + {\rm h.c.} \right)
\nonumber\\
&& - \frac{3}{2}\, m^2\, B^2\, \vert {\hat \Phi} \vert^2 - \frac{1}{2}
\, \left( H\, B + \dot{B} \right)\, \sum_{j=1}^3\, \left( i
{\hat \Phi}^*\, p_j\, {\hat \alpha}_0^{(j)} + {\rm h.c.} \right) + \frac{1}{2}\,
\left[ B\, \dot{B} + \left(2+B^2\right)\, H \right]\,
\sum_{j=1}^{3}\, \left( i
{\hat \Phi}^*\, p_j\, {\hat \chi}_j + {\rm h.c.} \right) \nonumber\\
&& + \frac{1}{2}\, \left( p^2 + m^2 + H^2 - \frac{3}{2}\,
m^2\, B^2 \right)\, \sum_{a=1}^3\, \vert {\hat \alpha}_0^{(a)} \vert^2 -
\frac{B}{2}\, \left( H^2 + m^2 - \frac{3}{2}\, m^2\, B^2
\right)\, \sum_{j=1}^3\, \left(
{\hat \alpha}_0^{(j)*}\, {\hat \chi}_j + {\rm h.c.} \right) \nonumber\\
&& +\frac{1}{4}\, \sum_{j=1}^{3}\, \left[ p^2 + \frac{1}{2}\,
\left( p^2 + 4 m^2 \right)\, B^2 - 3\, m^2\, B^4 + 2\, H^2\, B^2
\right]\, \vert {\hat \chi}_j \vert^2 - \frac{1}{8}\,
\left(2+B^2\right)\, \sum_{i,j=1}^{3}\,  p_i\, p_j\,
{\hat \chi}_i\, {\hat \chi}_j^* + \dots  \label{action-full-mukh}
\end{eqnarray}
where we have used the physical momenta $p_i \equiv k_i/a$. Eq. (\ref{action-full-mukh}) actually is not the full quadratic action of the perturbations, but some terms (denoted by dots) are omitted. Let us clarify this. We find that no time derivatives of the combinations ${\hat \Phi} ,\, {\hat \chi}_i ,\, {\hat \alpha}_0^{(a)}$ enter in the quadratic action (neither in the terms shown here, nor in those omitted). These are the nondynamical gauge invariant modes  of the system.~\footnote{Notice that they corresponds to the nondynamical perturbations $\delta g_{0\mu}$ and $\delta A_0^{(a)}$ in the gauge $E_1 = E_2 = E_3 = T = 0$.} The remaining modes are dynamical. Eq.~(\ref{action-full-mukh}) is the action for all the gauge invariant modes of the system (both the dynamical, and the nondynamical ones). However, it contains only the terms that contribute to the kinetic matrix of the action for the dynamical modes, once the nondynamical modes have been integrated out.~\footnote{For clarity, compare with the formal discussion of Section \ref{sec:ghost}. The action 
(\ref{action-full-mukh}) given here corresponds to the formal action (\ref{formal-act}), with only the terms
proportional to the coefficients $a_{ij} ,\, b_{ij} ,$ and $c_{ij}$ included. Those are the only terms necessary to compute the kinetic matrix $K$ of the dynamical modes, cf. eqs. (\ref{klo}) and (\ref{action-integrated}).} These terms are given without any omission, nor approximation, so that eq. (\ref{action-full-mukh}) contains all the necessary information for the exact computation of the kinetic matrix of the dynamical modes.

Before integrating out the nondynamical modes, we can easily see that $5$ dynamical modes decouple from the remaining ones in the part of the action shown. There are $9$ dynamical modes in the spatial perturbations of the three vector fields, ${\hat \alpha}_j^{(a)} \,$. These perturbations enter in the action (\ref{action-full-mukh}) with a diagonal quadratic term (the second term, $\propto \vert {\hat \alpha}_j^{(a)} \vert^2$). Then, they are coupled with the remaining modes only in the third line of (\ref{action-full-mukh})). We see that, out of the nine modes ${\hat \alpha}_j^{(a)} \,$, only the four linear combinations $p_j \dot{\hat \alpha}_j^{(a)}$ ($a = 1,2,3$) and $\dot{\hat \alpha}_j^{(j)}$ are involved in these couplings. The remaining $5$ linear combinations are decoupled. Therefore, we can rotate the fields ${\hat \alpha}_j^{(a)} \,$ into the coupled and decoupled linear combinations:
\begin{eqnarray}
{\hat \alpha}_1^{(1)} &=&  \frac{p_{23}^2}{\sqrt{2} \, p^2} \, {\hat v}_1 + \frac{p_1}{p} \, {\hat v}_2 + \frac{p_{12} \, p_{23}}{p \, \sqrt{p_{12}^2 + p_{23}^2}} \, {\hat v}_8 - \frac{p_{13} \, p_{23}^2}{\sqrt{2} \, p^2 \, \sqrt{p_{12}^2 + p_{23}^2}} \, {\hat v}_9 \nonumber\\
{\hat \alpha}_2^{(1)} &=& - \frac{p_1 \, p_2}{\sqrt{2} \, p^2} \, {\hat v}_1 + \frac{p_2}{p} \, {\hat v}_2 + \frac{p_3}{p_{23}} \, {\hat v}_5 - \frac{p_1 \, p_2 \, p_{12}}{p_{23} \, p \, \sqrt{p_{12}^2 + p_{23}^2}} \, {\hat v}_8 + \frac{p_1 \, p_2 \, p_{13}}{\sqrt{2} \, p^2 \, \sqrt{p_{12}^2 + p_{23}^2}} \, {\hat v}_9 \nonumber\\
{\hat \alpha}_3^{(1)} &=& - \frac{p_1 \, p_3}{\sqrt{2} \, p^2} \, {\hat v}_1 + \frac{p_3}{p} \, {\hat v}_2 - \frac{p_2}{p_{23}} \, {\hat v}_5 - \frac{p_1 \, p_{12} \, p_3}{p_{23} \, p \, \sqrt{p_{12}^2 + p_{23}^2} }  \, {\hat v}_8 + \frac{p_1 \, p_3 \, p_{13}}{\sqrt{2} \, p^2 \, \sqrt{p_{12}^2 + p_{23}^2}} \, {\hat v}_9 \nonumber\\
{\hat \alpha}_1^{(2)} &=& - \frac{p_1 \, p_2}{\sqrt{2} \, p^2} \, {\hat v}_1 + \frac{p_1}{p} \, {\hat v}_3 - \frac{p_3}{p_{13}} \, {\hat v}_6 - \frac{p_1 \, p_2 \, \sqrt{p_{12}^2 + p_{23}^2}}{\sqrt{2} \, p_{13} \, p^2} \, {\hat v}_9 \nonumber\\
{\hat \alpha}_2^{(2)} &=& \frac{p_{13}^2}{\sqrt{2} \, p^2} \, {\hat v}_1 + \frac{p_2}{p} \, {\hat v}_3 + \frac{p_{13} \, \sqrt{p_{12}^2 + p_{23}^2}}{\sqrt{2} \, p^2} \, {\hat v}_9 \nonumber\\
{\hat \alpha}_3^{(2)} &=& - \frac{p_2 \, p_3}{\sqrt{2} \, p^2} \, {\hat v}_1 + \frac{p_3}{p} \, {\hat v}_3 + \frac{p_1}{p_{13}} \, {\hat v}_6 - \frac{p_2 \, p_3 \, \sqrt{p_{12}^2 + p_{23}^2}}{\sqrt{2} \, p_{13} \, p^2} \, {\hat v}_9 \nonumber\\
{\hat \alpha}_1^{(3)} &=& - \frac{p_1 \, p_3}{\sqrt{2} \, p^2} \, {\hat v}_1 + \frac{p_1}{p} \, {\hat v}_4 + \frac{p_2}{p_{12}} \, {\hat v}_7 + \frac{p_1 \, p_3 \, p_{23}}{p_{12} \, p \, \sqrt{p_{12}^2+p_{23}^2}} \, {\hat v}_8 + \frac{p_1 \, p_3 \, p_{13}}{\sqrt{2} \, p^2 \, \sqrt{p_{12}^2 + p_{23}^2}} \, {\hat v}_9 \nonumber\\
{\hat \alpha}_2^{(3)} &=& - \frac{p_2 \, p_3}{\sqrt{2} \, p^2} \, {\hat v}_1 + \frac{p_2}{p} \, {\hat v}_4 - \frac{p_1}{p_{12}} \, {\hat v}_7 + \frac{p_2 \, p_3 \, p_{23}}{p_{12} \, p \, \sqrt{p_{12}^2+p_{23}^2}} \, {\hat v}_8 + \frac{p_2 \, p_3 \, p_{13}}{\sqrt{2} \, p^2 \, \sqrt{p_{12}^2 + p_{23}^2}} \, {\hat v}_9 \nonumber\\
{\hat \alpha}_3^{(3)} &=& \frac{p_{12}^2}{\sqrt{2} \, p^2} \, {\hat v}_1 + \frac{p_3}{p} \, {\hat v}_4 - \frac{p_{12} \, p_{23}}{p \, \sqrt{p_{12}^2 + p_{23}^2}} \, {\hat v}_8 - \frac{p_{12}^2 \, p_{13}}{\sqrt{2} \, p^2 \, \sqrt{p_{12}^2 + p_{23}^2}} \, {\hat v}_9
\label{rotation}
\end{eqnarray}
where for brevity we have defined $p_{ij} \equiv \sqrt{p_i^2 + p_j^2} \,$.

The combinations ${\hat v}_1 ,\, \dots ,\, {\hat v}_4$ are the coupled modes, while 
${\hat v}_5 ,\, \dots ,\, {\hat v}_9$ are decoupled from the remaining perturbations. One can check that the matrix relating the vector $\left\{{\hat \alpha}_1^{(1)} ,\, {\hat \alpha}_2^{(1)} ,\,  {\hat \alpha}_3^{(1)},\, {\hat \alpha}_1^{(2)} ,\, \dots, \, {\hat \alpha}_3^{(3)} \right\}$ to the vector $\left\{{\hat v}_1 ,\, \dots ,\, {\hat v}_9 \right\}$ is  orthogonal. Therefore, the quadratic term $\propto \sum_{j,a} \vert {\hat \alpha}_j^{(a)} \vert^2$ becomes $\propto \sum_i \vert {\hat v}_i \vert^2$. As a consequence, the modes ${\hat v}_5 ,\, \dots ,\, {\hat v}_9$ enter  only through the term
\begin{equation}
{\cal L} \supset \frac{1}{2 \, a^2}\, \sum_{i=5}^9\, \vert
\dot{\hat v}_i \vert^2
\label{dec-kin}
\end{equation}
and are therefore decoupled from the other modes in the part of the action shown in (\ref{action-full-mukh}) (we stress that this part is all we need to compute the kinetic term for the dynamical modes). 
Since the coefficient in front of each term in the sum (\ref{dec-kin}) is positive, we see that these modes are well behaved (positive energy) excitations. Let us now focus on the coupled system of the remaining modes (which includes the remaining $6$ dynamical modes, and the $7$ nondynamical modes). The next step is to
integrate out the nondynamical modes as outlined formally in
equations form (\ref{formal-act}) to (\ref{action-integrated}). Doing
so, we end up with a rather long kinetic matrix for the six dynamical modes $\{ {\hat \Psi},
\, {\hat \Sigma}, \, {\hat v}_i \}$  ($i=1,2,3,4$). We were unable to perform analytical simplifications of this matrix, as those leading to the expressions (\ref{detapprx}) and (\ref{detapprx-2}) for the simpler models studied above. However, we can evaluate this matrix numerically for any given choice of parameters.

More specifically, we show the results of a numerical evolution starting from $B_{in}=10$, 
$\dot{B}_{\rm in} = - \sqrt{2} \, m /3$ according to the slow roll conditions (\ref{slow-roll-mukh}), and with momentum of the modes initially satisfying $p_1=100\, H , \, p_2=80\, H , \, p_3=60\, H$  (these values have be chosen only for illustrative purposes and have no particular relevance, other that we require the mode to be well inside the horizon initially; we verified that other choices of the momenta lead to the same conclusions as those shown here). We show in Figures \ref{fig:mukh} and \ref{fig:mukh2} the evolution of the eigenvalues of the kinetic matrix (the second Figure is a close-up of the first one where the eigenvalue $\lambda_3$ crosses zero) close to horizon crossing.~\footnote{The overall factor $a^3$ appearing in (\ref{action-full-mukh}) has not been included in this computation. This is irrelevant for our discussion, since this factor simply rescales all eigenvalues by a common positive number.} We see that two eigenvalues $\lambda_{1,2}$ are initially negative; the corresponding eigenvalues are ghosts at these times. These eigenvalues become positive without crossing zero (they diverge). A third eigenvalue $\lambda_3$ is initially positive, but it becomes negative crossing zero. It then diverges and becomes again positive. As the formal eqs. (\ref{formal-eom}) show, the linearized system of equations becomes singular when this eigenvalue crosses zero. We expect that also the linearized solutions diverge at this point (cf. the explicit solutions given in the model of Subsection \ref{1vec-lambda}).

\begin{figure}
\centering
\includegraphics[width=0.52\textwidth,angle=-90]{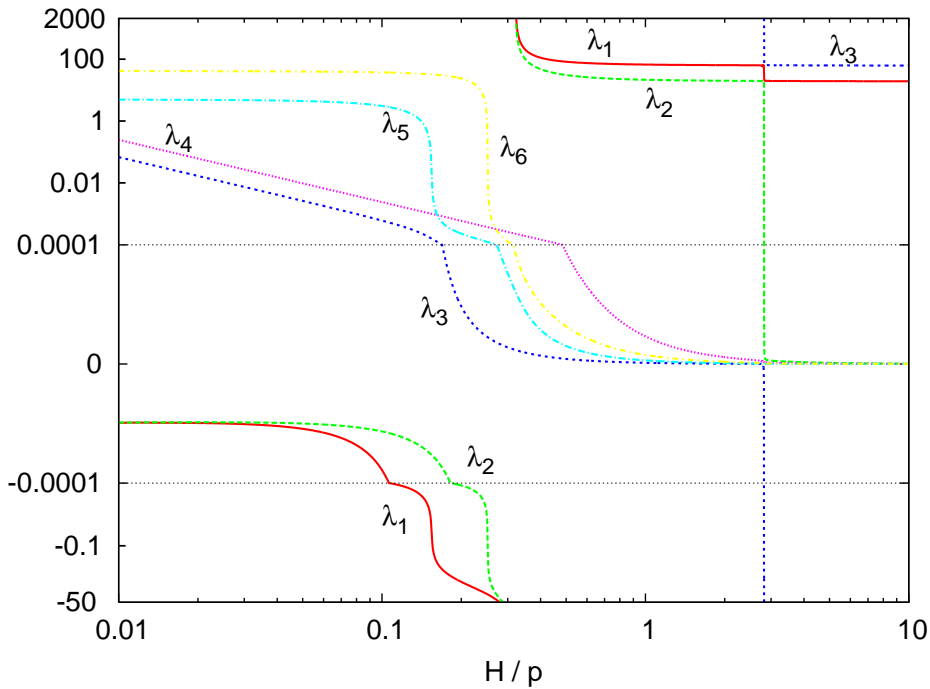}
\caption{Eigenvalues of the kinetic matrix for the dynamical
and gauge invariant perturbations ${\hat \Psi} ,\, {\hat \Sigma} ,\,
{\hat v}_{1,\dots,4}$ for the model of vector inflation (\ref{ac-gmv2}), and for one
specific choice of initial conditions (see the main text). Since $H/p$ increases with time during the stage shown, we use it as a ``time variable'' in this and the next Figure. The eigenmodes
corresponding to $\lambda_{1,2}$ are ghosts at the earliest times shown (low
$H/p$). The mode corresponding to $\lambda_3$ becomes a ghost close
to horizon crossing. This eigenvalues (and the determinant of the kinetic matrix)
crosses zero at this point, signaling an instability of the system also at
the linearized level. The kinetic matrix, and its eigenvalues, are dimensionless.
Notice that we use linear units in the interval $\left[ - 0.0001, 0.0001 \right]$, 
and logarithmic units outside.}\label{fig:mukh}
\end{figure}

\begin{figure}[h]
\centerline{
\includegraphics[width=0.52\textwidth,angle=-90]{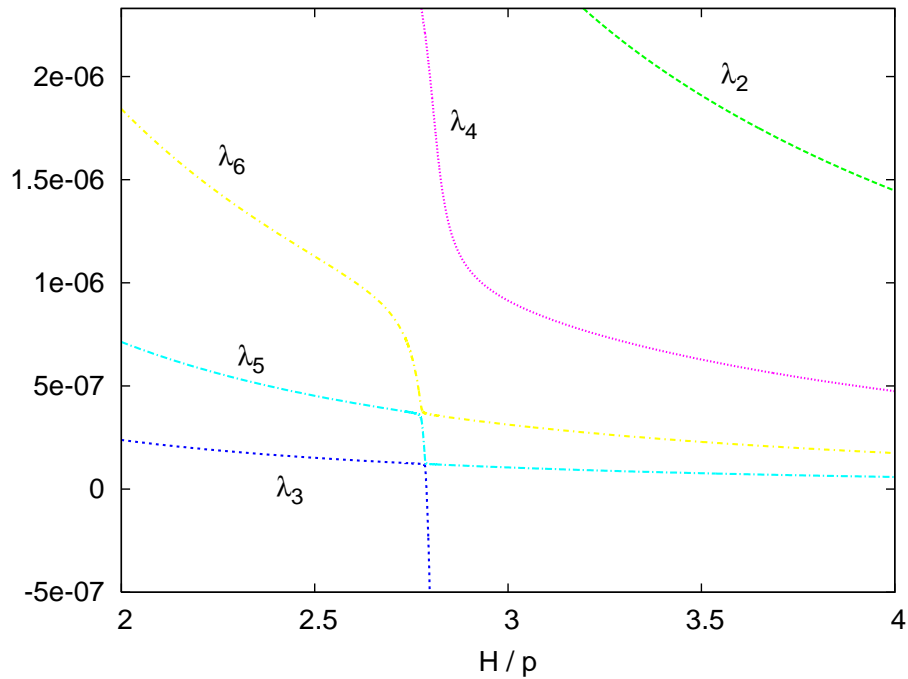}
} \caption{Close up  of the previous Figure where $\lambda_3$ vanishes. A further close up shows 
that $\lambda_5$ and $\lambda_6$, and, later, $\lambda_3$ and $\lambda_5$ do not cross each other, although they appear to do so in the Figure.} \label{fig:mukh2}
\end{figure}

The study so far concentrated on the nature of the modes at horizon crossing, in an inflationary regime for which the total mass term of the vectors was negative, $- R / 6 + m^2 \simeq - 2 H^2 + m^2 < 0 \,$. 
However, as we discussed after eq. (\ref{oscB}), $- R / 6 + m^2$ vanishes towards the end of inflation.
We studied the behavior of the eigenvalues of the kinetic matrix also around this point. We considered the same background evolution as for the previous plot, but smaller values of the momenta, so that $H/p$ is not exponentially small at the times shown (we want to avoid that our results are affected by numerical inaccuracies). As shown in Figure \ref{fig:muk-m0}, we find that two eigenvalues cross zero precisely when the total mass vanishes. Also at this point, the system of equations for the eigenvalues becomes singular We expect that the linearized solutions diverge also at this point.

\begin{figure}[h]
\centerline{
\includegraphics[width=0.5\textwidth,angle=-90]{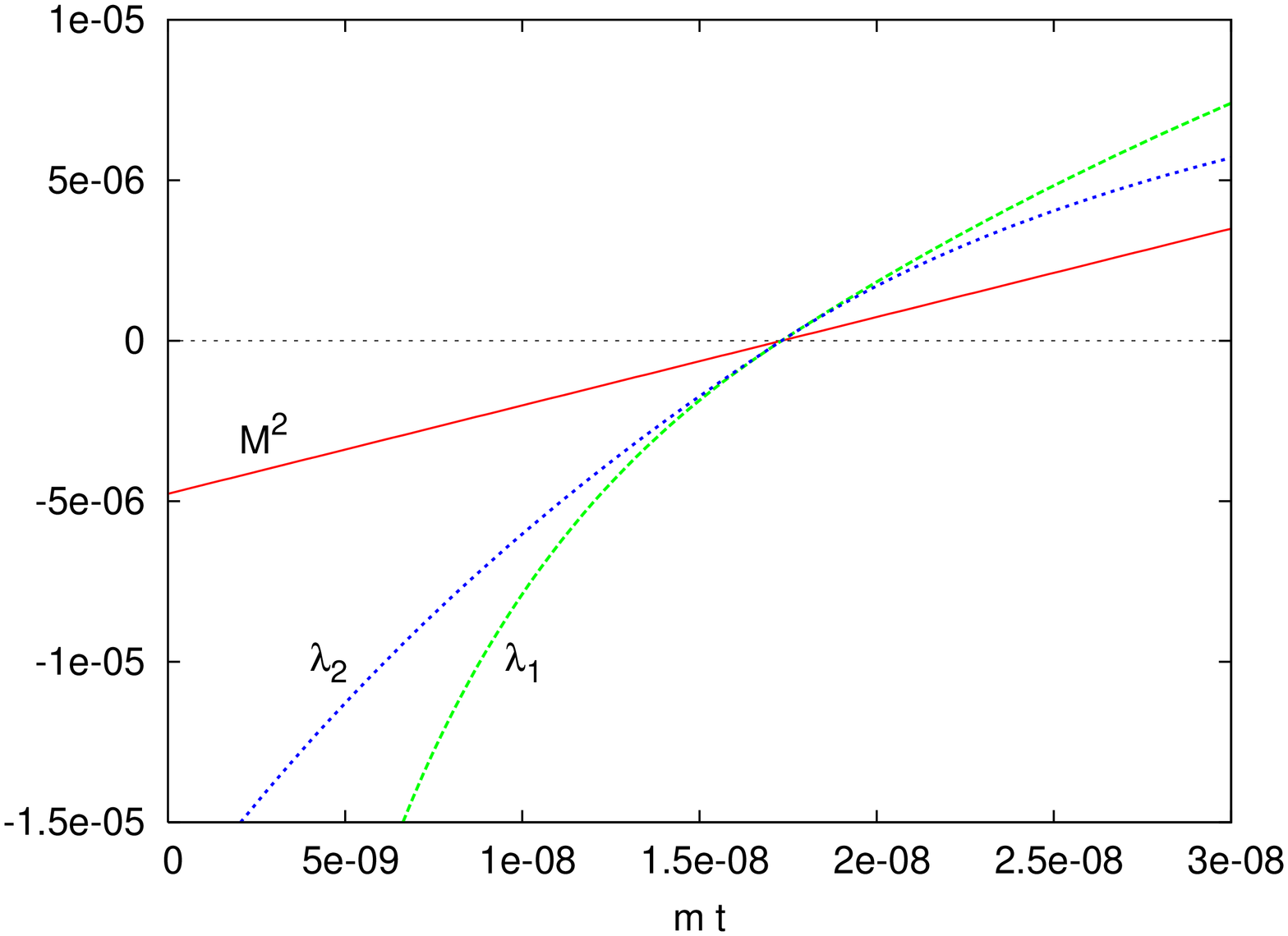}
} \caption{
Two eigenvalues vanish when the mass term $M^2 = m^2 - R /6$ of the vector
fields vanishes. The mass $M^2$ is shown in units of $m^2/300 \,$. We have $B \simeq 1.048$ when the total mass vanishes (this occurs still during inflation). The mode has been chosen to be outside the horizon when the total mass vanishes ($H/p \simeq 1872$ at the time shown, and $k_1:k_2:k_3 = 10:8:4$).
} \label{fig:muk-m0}
\end{figure}

It is interesting to compare the behavior of the eigenvalues shown in these Figures with that obtained for  the previous models. We find that the two eigenvalues $\lambda_{1,2}$ behave precisely as in the case of zero vev studied in the previous Section: they are negative in the deep sub-horizon regime, they diverge close to horizon crossing, and they cross zero when the total mass vanishes. On the contrary, the eigenvalue $\lambda_3$ behaves precisely as in the cases of a single vector with nonvanishing vev studied above (cf. Figures \ref{fig:cc-auto} and \ref{fig:infla-auto}): it is positive in the deep sub-horizon regime, it crosses zero close to horizon crossing, and it remains negative for some time afterwards. It appears from these behaviors that the mixing with gravity affects only one linear combination of the three ghosts.

We conclude the present discussion with some remarks on the previous
study \cite{mukhvect3} of perturbations of vector inflation. We
already discussed in the Introduction while the computations of
\cite{mukhvect3} - being limited to the linearized equations for the
perturbations in either the sub-horizon or the super-horizon regime -
cannot show the ghost instabilities found here. Here we want to reply
to some specific comments on our previous works \cite{hcp1,hcp2}
contained in \cite{mukhvect3}. Most of these remarks are answered by
the fact that, in our previous works, we only provided arguments for
the presence of ghosts in the model of vector inflation, deferring the
explicit computation to the present work. It is mentioned in
\cite{mukhvect3} that the ghost may be an artifact of having expanded
the vector field in transverse and longitudinal part according to
$A_\mu^T + \partial_\mu \phi$. This would introduce additional time
derivatives, which could affect our findings. This Stuckelberg
decomposition was introduced in \cite{hcp1,hcp2} only as the simplest
way to elucidate the problem. The computations presented here do not
introduce additional time derivatives in the parametrization of the
perturbations, and lead to the same conclusions as the much simpler
Stuckelberg analysis. It was also pointed out in \cite{mukhvect3} that
the complete computation, with gravity perturbations also included,
was in order. This is precisely what is performed here. Another
comment of \cite{mukhvect3} was that the model of vector inflation
should only be regarded as an effective field theory, valid only at
small energies. The (unknown) UV completion of this theory may be
without ghosts. We agree with this claim.~\footnote{Apart from the
  fact that we are no aware of any well behaved UV completion of a
  theory with ghosts.} This was precisely the point raised in our
previous works. As we have seen here, the ghost instabilities appear
during most of the sub-horizon regime, and close to horizon
crossing. Any UV completion needs to be relevant at these
stages. Therefore, the effective theory of vector inflation cannot be
used for the study of cosmological perturbations in the sub-horizon
regime. Since this regime is crucial to obtain phenomenological
results, this invalidates any prediction of the effective model
\cite{mukhvect}. Finally, it was argued in \cite{mukhvect3} that the
instability may simply be due to the growth of the scale factors, and
to a wrong rescaling of fields. This is not the case, since the
linearized system of equations - and, most likely, its solutions -
diverges at some finite moments of time, while the scale factor
remains finite.

\section{Discussion}

\label{sec:discussion}

Although the paradigm of inflation is well supported by the
observational data, we still do not know the actual particle physics
mechanism behind the inflationary expansion, and the generation of
cosmological perturbations. It is customary to parametrize our
ignorance in terms of scalar fields: they may be fundamental fields,
or simply order parameters which provide an effective description of
some degrees of freedom in the theory (for instance, the
brane-antibrane separation in some string motivated models of
inflation). However, it may well be possible that these two key
elements of cosmology are due to higher spin fields. Vector fields are
probably the simplest possibility after scalars. They can in principle
leave a distinct signature from the scalar case, since a nonvanishing
spatial vev of a vector breaks the isotropy of space. This can provide
anisotropic expansion, and / or generate a spectrum of primordial
perturbations that breaks statistical isotropy. The main obstacle
faced by explicit realizations of this idea is that, in the standard
case, vector fields decrease too quickly due to the expansion of the
universe. Therefore, suitable modifications of the standard action
need to be made.

Recently, a class of models was considered in which the vector is
nonminimally coupled to the curvature, ${\cal L} \supset R / 12 \,
A^\mu \, A_\mu \,$. Indeed, for this specific coupling, the vev of the
vector evolves as that of a minimally coupled scalar field; this
offers the possibility of realizing an inflationary background, with a
controllable anisotropy \cite{mukhvect,soda1}. In addition, the
transverse perturbations of the vector behave as the perturbations of
a minimally coupled scalar; this is the basis of the mechanism of
vector curvaton \cite{Dimopoulos} for the generation of a nearly scale
invariant spectrum of primordial perturbations. To our knowledge, the
analogy between the $ R / 12 \, A^\mu \, A_\mu \,$ coupling and the
minimally couple scalar field first appeared in the 1987 work by
Turner and Widrow \cite{Turner:1987bw}, where it was exploited for the
generation of a primordial magnetic field during inflation. The
renewed interest in this mechanism is mostly motivated by some
features in the WMAP data that hint for a small break of statistical
isotropy.

All of the mentioned works suggested new interesting mechanisms, and a
complete check of the stability of these proposals was beyond their
scope. It is tempting to assume that these models should be well
behaved due to the strong analogy with the minimally coupled scalar
field case. There is however a crucial difference between these two
cases, and between the case of a minimally vs a nonminimally couple
vector; it is due to the longitudinal vector polarization. In the
above works, the vector has a U(1) invariant kinetic term ${\cal L}
\supset - 1/4 F_{\mu \nu} F^{\mu \nu} \,$. If only this term was
present, the vector would only have the two transverse
polarizations. However, the nonminimal coupling to the curvature
breaks the U(1) symmetry, and gives rise to an additional,
longitudinal, polarization. The nature of this mode is controlled by
the sign of the mass term, which, for these mechanisms to work, needs
to be negative. For the scalar case, a negative mass squared means
that the field is a tachyon; for a vector field, a negative mass
squared means that the longitudinal polarization is a ghost.

Motivated by this initial consideration, we studied the stability of
this class of theories. We did so by computing the free action for the
dynamical (physically propagating) modes of such theories, around the
background solutions considered in the various proposals. The sign of
the eigenvalues of the kinetic matrix of these action indicates
whether the corresponding eigenmode is a positive or negative energy
excitation. Our computations confirmed that there is a ghost for each
nonminimally coupled vector field in the model. As we already
mentioned in the Introduction, theories with ghosts are consistent
only as effective theories, valid below some energy scale
$\Lambda$. Inflationary predictions heavily rely on the initial
conditions; for instance, the vector curvaton mechanism of
\cite{Dimopoulos} results in a scale invariant spectrum because of the
specific choice of initial adiabatic vacuum. This choice is made in
the quantized theory for the perturbations, which is performed in the
deep UV regime (energy $\gg H^{-1}$) \cite{mfb}. In presence of a
cut-off, the initial adiabatic vacuum cannot be imposed at arbitrarily
early times, and, depending on the precise numerical value of the
cut-off, it may not be possible to impose it at all. This casts doubts
on the phenomenological predictions obtained for such models.

In fact, all theories with an explicit mass $M$ for the vector require a cut-off which makes them invalid
at high energies, irrespectively of the sign of the mass
term. We can see this
based on the behavior of massive vector fields at high energies. The explicit mass breaks the gauge invariance in a hard way. It is
well known that, in such case, the interactions of the longitudinal
bosons violate unitarity at a scale which is parametrically set by
$M$, leading to a quantum theory out of control. For the present
models, $M$ is the Hubble rate or below, so that the entire
sub-horizon regime may be ill-defined. At high energies, the longitudinal mode will also interact 
strongly with the other fields in the theory (this will renormalize the
coupling constants). Then, depending on exactly when this happens, the quantum
theory of the perturbations may be out of control throughout the
entire short wavelength regime. If this is the case, all initial conditions would become
unjustified, and the theory would lose its predictive power. Although this problem
is present for both positive and negative mass terms, a theory which has a hard
vector mass and a ghost is more problematic than a theory with
only a hard vector mass. The most immediate UV completion of a theory
with a hard mass is through a higgs mechanism. The mass would be then
due to the vev of a scalar field that becomes dynamical above the
scale $M$. In this way the theory remains under control also in the
short wavelength regime, and one can apply all the standard
computations valid for scalar fields during inflation. However, if
$M^2$ needs to have the wrong sign, the scalar field in this UV
completed theory needs to be a ghost. In fact, we are not aware of any well 
behaved UV completion of a theory with a ghost.

We stress that this instability is not related to the classical
behavior of the solutions of the linearized equations for the
perturbations, and it would be present even if the latter remained
finite. However, we argued that, for the models considered here, also
the linearized solutions diverge. This is a second instability that
adds up to the one we have just discussed. This instability appears
because the eigenvalues corresponding to the ghosts do not remain
negative over the whole evolution, but change sign, and cross zero at
some finite moment of time $t_*$ (there are two such moments in the
model of vector inflation considered here). We showed that the
linearized equations are singular at $t_*$, and we expect that the
linearized solutions diverge for $t \rightarrow t_* \,$. While it is
possible that the divergency does not occur at the full nonlinear
level, this instability also invalidates all the phenomenological
signatures of these models which are based on the linearized
computation (as for instance the primordial spectrum of
perturbations). We solved the linearized equations only in the
simplest cases of a vector field with no vev, and of a vector field
with vev plus a cosmological constant. We did not solve them for the
models \cite{mukhvect,soda1}. We have shown however that the
linearized equations become singular also in these cases. It is
important to stress that even if, due to some unexpected cancellation,
the solutions to these equations would remain finite, this would not
eliminate the ghost instability that we have discussed in the two
previous paragraphs, and that we have proven to be present for all the
models studied here.

We conclude the Discussion with some remarks on models different from
those considered here, and in our previous works \cite{hcp1,hcp2}. The
instability we pointed out motivates the study of such models, as for
instance vector fields with nonstandard kinetic
terms~\cite{acw,nonst-kin,soda2} (although, some of these proposals
are also unstable), nonabelian vectors with nonminimal coupling to the
curvature~\cite{nicola}~\footnote{The stability analysis performed
  here applies also to the nonabelian model of \cite{nicola} if the
  vectors have no vev (since, in this case, the nonabelian structure
  does not affect the quadratic action for the perturbations);
  however, the case with a nonvanishing vev requires a separate
  investigation.}, spinors~\cite{spin}, or higher $p-$forms
\cite{nemanja,pforms,pforms2} ~\footnote{The vector case is $p=1$;
  Ref. \cite{pforms2} generalized the arguments of \cite{hcp1,hcp2},
  and pointed out that also the $p=2$ case contains ghosts.}. Of
particular interest, in our opinion, is a class of models
characterized by a function of a scalar field multiplying the kinetic
terms of the vectors, $I \left( \phi \right) F^{\mu \nu} F_{\mu \nu}
\,$, but no potential term for the vector \cite{soda2,nomass}. U(1)
invariance is preserved in these models, and the problematic
longitudinal mode is absent. A complete study of the cosmological
perturbations (conducted along the lines described here) is the next
necessary step for obtaining the phenomenological predictions of these
models.

\begin{acknowledgments}

We thank  N. Bartolo, D.H. Lyth, C. Pitrou, J. Soda, and, particularly, N. Kaloper for very useful discussions. The work of B.H. and M.P. was partially supported by the DOE grant DE-FG02-94ER-40823.

\end{acknowledgments}

\appendix

\section{Details of the linearized computation of Subsection \ref{1vec-lambda}}
\label{appA}

This appendix contains the details of the stability analysis of the model of Subsection
\ref{1vec-lambda}. The explicit forms of the linearized equations (\ref{perturbed})
are 
\begin{eqnarray}
{\rm Eq}_{00} :&& -\frac{2}{M_p} \Bigg\{ \left[ \left(1 -
\frac{1}{3} B_1^2 \right)\, H + \left( 1 + \frac{1}{6} B_1^2
\right)\, h + \frac{1}{6} B_1\, \dot{B}_1 \right]\, \dot{\hat\Psi}
\qquad\qquad\qquad\qquad\qquad\qquad\qquad\qquad\qquad\qquad\qquad\qquad
\nonumber\\
&& \qquad\quad + \frac{1}{2} \left[ p_T^2 + \left( m^2 -
\frac{p_T^2 + 2 p_L^2}{6} + 5 h^2 - 4 h\, H \right)\, B_1^2 - 4
h\, H\, B_1\, \dot{B}_1 + \dot{B}_1^2 \right]\, \hat{\Psi}
\nonumber\\
&& \qquad\quad + \left[ 3 H^2 - \left( 3 + \frac{5}{2} B_1^2
\right)\, h^2 - \frac{1}{2} \dot{B}_1^2 + 2 \left( B_1\, H +
\dot{B}_1 \right)\, B_1\, h \right]\, \hat\Phi - \left[ \left(1 +
\frac{1}{6} B_1^2 \right)\, \left( H + h \right) + \frac{1}{6}
B_1\, \dot{B}_1 \right]\, \hat\chi \nonumber\\
&& \qquad\quad - \left[ \left( 1 + \frac{1}{6} B_1^2 \right)\,
\left( H - \frac{1}{2} h \right) + \frac{1}{6} B_1\, \dot{B}_1
\right]\, \hat{B} - \left[ \left( \frac{1}{2} H - h \right)\, B_1
+ \frac{1}{2} \dot{B}_1 \right]\, \hat\alpha_0 - \left(
\frac{1}{2} \dot{B}_1 - B_1\, h \right)\, \dot{\alpha}_1
\nonumber\\
&& \qquad\quad + \left[ \frac{1}{2} \left( \frac{1}{3} p^2 - m^2 -
5 h^2 + 4 h\, H \right)\, B_1 + h\, \dot{B}_1 \right]\,
\hat\alpha_1 \Bigg\} = 0 \nonumber\\
{\rm Eq}_{01} : && -2\, \frac{i p_L\, a}{M_p}\, \Bigg\{
\frac{1}{6} \left( \dot{\alpha}_1 - B_1\, \dot{\hat\Psi} \right)\,
B_1 + \frac{B_1}{6}\, \left( B_1\, H - 2 B_1\, h - 2 \dot{B}_1
\right)\, \hat\Psi + \frac{1}{6} \left( -B_1\, H + 2 h\, B_1 +
\dot{B}_1
\right)\, \hat\alpha_1 \nonumber\\
&& \qquad\qquad + \left(1 + \frac{1}{6} B_1^2 \right)\, \left( H +
h + \frac{B_1\, \dot{B}_1}{6 + B_1^2} \right)\, \hat\Phi + \left(
\frac{p_T^2}{4 p_L^2} + \frac{p_T^2}{24 p_L^2} B_1^2 - \frac{1}{3}
\cal{D}_{\chi\chi} \right)\, \hat\chi - \frac{6+B_1^2}{24} \hat{B}
- \frac{B_1}{3} {\cal D}_{\alpha_0 \alpha_0}\, \hat\alpha_0
\Bigg\} = 0 \nonumber\\
{\rm Eq}_{0i} : && -2\, \frac{i\, p_{Ti}\, b}{M_p} \Bigg\{
\frac{1}{2} \left( 1 - \frac{B_1^2}{6} \right) \dot{\hat\Psi} +
\frac{1}{6} B_1\, \dot{\hat\alpha}_1 + \left[ \frac{B_1}{6} \left(
H B_1 - 2 \dot{B}_1 \right) - \left( \frac{3}{2} +
\frac{B_1^2}{12} \right) h \right]\, \hat\Psi + \left[ \left(
\frac{1}{3} H - \frac{7}{6} h \right)\, B_1 + \frac{2}{3}
\dot{B}_1 \right]\, \hat\alpha_1 \nonumber\\
&& \qquad\qquad + \left[ \left( h - \frac{1}{2} H \right) B_1 -
\frac{1}{2} \dot{B}_1 \right]\, \hat\alpha + \left[ - \left(
\frac{1}{2} + \frac{B_1^2}{12} \right) h + \left( 1 + \frac{1}{6}
B_1^2 \right) H + \frac{1}{6} B_1\, \dot{B}_1 \right] \hat\Phi -
\frac{1}{4} \left( 1 + \frac{B_1^2}{6} \right)\, \hat\chi
\nonumber\\
&& \qquad\qquad + \frac{p_L^2}{4 p_T^2}\, \left( 1 + \frac{1}{6}
B_1^2 \right)\, \hat{B} \Bigg\} = 0 \nonumber\\
{\rm Eq}_{11} : && \frac{2 a^2}{3 M_p} \Bigg\{ \frac{1}{2} B_1\,
\left( \ddot{\hat\alpha}_1 - 2 B_1\, \ddot{\hat\Psi} \right) -
\left( 3 H\, B_1 + 2 \dot{B}_1 \right) B_1\, \dot{\hat\Psi} +
\frac{1}{2} \left( -H B_1 + 8 h\, B_1 - \dot{B}_1 \right)\,
\dot{\hat\alpha}_1 + \left( {\cal M}_{\Psi \Psi} - p_T^2\, B_1^2
\right)\, \hat\Psi \nonumber\\
&& \qquad\quad + \left( {\cal M}_{\Psi \alpha_1} + \frac{p_T^2}{2}
B_1 \right)\, \hat\alpha_1 + \left[ \left( 3 + \frac{1}{2} B_1^2
\right) h + \left( 3 - B_1^2 \right) H + \frac{1}{2} B_1 \dot{B}_1
\right]\, \dot{\hat\Phi} \nonumber\\
&& \qquad\quad + \left[ \left( -\frac{9}{4}\, m^2 + \frac{p_T^2 +
2 p_L^2}{4} - \frac{15}{4}\, h^2 + 3\,h \, H\right)\, B_1^2 + 3\,
h\, B_1\, \dot{B}_1 - 3\, \left( \frac{p_T^2}{2}\, -\frac{V_0}{2
M_p^2} + \frac{3}{2}\, h^2 - \frac{3}{2}\, H^2 + \frac{1}{4}\,
\dot{B}_1^2\right) \right]\, \hat\Phi \nonumber\\
&& \qquad\quad - \left( \frac{3}{2} - \frac{1}{4}\, B_1^2
\right)\, \dot{\hat{B}} + \frac{1}{2}\, B_1^2\, \dot{\hat\chi} +
\left( 2 H - h \right)\, B_1^2\, \hat\chi - \left[ \left(
\frac{9}{2} + \frac{1}{4}\, B_1^2 \right)\, h + \left( \frac{9}{2}
- \frac{5}{4}\, B_1^2 \right)\, H + \frac{1}{2}\, B_1\, \dot{B}_1
\right]\,\hat{B} \nonumber\\
&& \qquad\quad + \frac{3}{2}\, \left( 2 h\, B_1 - H\, B_1 -
\dot{B}_1 \right)\, \hat{\alpha}_0 \Bigg\} = 0 \nonumber\\
{\rm Eq}_{1i} : && \frac{a\,b\, p_L\, p_{Ti}}{3 M_p} \Bigg\{
\frac{3}{p_L^2} \left[ \left( 2 h - H \right) B_1 - \dot{B}_1
\right]\, \dot{\hat\alpha}_1 + {\cal M}_{\Psi \alpha}\, \hat\alpha
+ \frac{6 + B_1^2}{4 p_T^2}\, \dot{\hat{B}} + \frac{6 + B_1^2}{4
p_T^2}\, \left( 3 H + 6 h + 2 \frac{B_1\, \dot{B}_1}{6 + B_1^2}
\right)\, \hat{B} + \frac{6 + B_1^2}{4 p_L^2}\, \dot{\hat\chi}
\nonumber\\
&& \qquad\qquad + \frac{6 + B_1^2}{4 p_L^2}\, \left( 3 H - 6 h + 2
\frac{B_1\, \dot{B}_1}{6 + B_1^2} \right)\, \hat\chi + B_1^2\,
\hat\Psi + \frac{1}{2} \left( 6 + B_1^2 \right)\, \hat\Phi - B_1\,
\hat\alpha_1 - \frac{3}{p_L^2} \left[ \left( H - 2 h\right)\, B_1
+ \dot{B}_1 \right]\, \hat\alpha_0 \Bigg\} = 0 \nonumber
\end{eqnarray}
\begin{eqnarray}
{\rm Eq}_{ij} : && \frac{b^2}{6 M_p} \Bigg\{ \, \Bigg[ \left( 6 -
B_1^2 \right)\, \ddot{\hat\Psi} + \left[ \left( 18 - B_1^2 \right)
H - \left( 18 + B_1^2 \right) h - 6 B_1\, \dot{B}_1 \right]\,
\dot{\hat\Psi} + \left[ {\cal M}_{\Sigma \Psi} + \left( 6 - B_1^2
\right)\, p_T^2 - 2 B_1^2\, p_L^2 \right]\, \hat\Psi + 2 B_1\,
\ddot{\hat\alpha} \nonumber\\
&& \qquad + 2 \left[ \left( 5 H - 7 h \right) B_1 + 5
\dot{B}_1 \right]\, \dot{\hat\alpha}_1 - \left( {\cal M}_{\Sigma
\alpha_1} - 2 p^2\, B_1 \right)\, \hat\alpha_1 + \left( 6 + B_1^2
\right)\, \left( 2 H - h + 2 \frac{B_1\, \dot{B}_1}{6 + B_1^2}
\right)\, \dot{\hat\Phi} \nonumber\\
&& \qquad - \left[ \left( p^2 + 30 h^2 - 24 h\, H \right)\,
B_1^2 - 24 h\, B_1\, \dot{B}_1 + 6 \left( p^2 + 6 (h^2 - H^2) +
\dot{B}_1^2 \right)\, \right]\, \hat\Phi - \left( 6 + B_1^2
\right)\, \dot{\hat\chi} \nonumber\\
&& \qquad - \left( 6 + B_1^2 \right)\, \left( 3 H - 3 h + 2
\frac{B_1\, \dot{B}_1}{6 + B_1^2} \right)\, \hat\chi + 6 \left[
\left( H -2 h \right)\, B_1 + \dot{B}_1 \right]\, \hat\alpha_0
- \left( 6 + B_1^2 \right) \left( \dot{\hat B}^2 + \left( 3 H + 2 \frac{B_1 \dot{B}_1}{6 + B_1^2}
\right) {\hat B} \right) \Bigg]\, \delta_{ij} \nonumber\\
&& \qquad + p_{Ti}\, p_{Tj}\, \left[ \frac{6 +
B_1^2}{p_T^2}\, \left( \dot{\hat{B}} + \left( 3 H + 2 \frac{B_1\,
\dot{B}_1}{6 + B_1^2} \right)\, \hat{B} \right) - \left( 6 - B_1^2
\right)\, \hat\Psi + \left( 6 + B_1^2 \right)\, \hat\Phi - 2 B_1\,
\hat{\alpha}_1 \right]\, \Bigg\} = 0 \nonumber\\
{\rm Eq}_0 : && i\, p_L\, \Bigg\{ \dot{\hat\alpha}_1 + \left( 2\,
h\, B_1 - H\, B_1 - \dot{B}_1 \right)\, \hat\Psi + \left( H -2
h\right)\, \hat{\alpha}_1 + \frac{p_T^2}{p_L^2}\, \dot{\hat\alpha}
+ \frac{p_T^2}{p_L^2}\, \left( H -2 h \right)\, \hat\alpha +
\left( H\, B_1 -2 h\, B_1 + \dot{B}_1 \right)\, \hat\Phi \nonumber\\
&& \qquad - \frac{2}{3 B_1}\, {\cal D}_{\chi \chi}\, \hat\chi +
\left( 1 + \frac{p_T^2}{p_L^2} - \frac{2}{3}\, {\cal
D}_{\alpha_0 \alpha_0} \right)\, \hat{\alpha}_0 \Bigg\} = 0 \nonumber\\
{\rm Eq}_1 : && \frac{1}{a}\, \Bigg\{ \ddot{\hat\alpha}_1 -
\frac{1}{3}\, B_1\, \ddot{\hat\Psi} + 3 H\, \dot{\hat\alpha}_1 +
\left( \frac{8}{3}\, B_1\, h - \frac{7}{3}\, B_1\, H - \dot{B}_1
\right)\, \dot{\hat\Psi} + \left( {\cal M}_{\alpha_1 \alpha_1} +
p_T^2 \right)\, \hat{\alpha}_1 - \frac{B_1}{3}\, p_T^2 \, \hat\Psi \nonumber\\
&& \qquad + \left( \dot{B}_1 - 2\, B_1\, h \right)\,
\dot{\hat\Phi} + \frac{B_1}{3}\, \left( \dot{\hat{B}} +
\dot{\hat\chi} \right) + \dot{\hat\alpha}_0 + \frac{B_1}{3}\,
\left( p^2 - 6 m^2 \right)\, \hat\Phi + \frac{1}{3}\, \left( 7\,
B_1\, h + B_1\, H -3 \dot{B}_1 \right)\, \hat{B} \nonumber\\
&& \qquad + \frac{2}{3}\, \left( 2 H - h \right)\, B_1\, \hat\chi
- p_T^2\, \hat\alpha + 2\, \left( h + H \right)\, \hat{\alpha}_0
\, \Bigg\} = 0 \nonumber\\
{\rm Eq}_{i} : && \frac{p_{Ti}}{p_L\, b}\, \Bigg\{
\ddot{\hat\alpha} + 3\, \left( H -2 h \right)\, \dot{\hat\alpha} +
\left( {\cal M}_{\alpha \alpha} + p_L^2 \right)\, \hat\alpha -
p_L^2\, \hat{\alpha}_1 + \frac{p_L^2}{p_T^2}\, \left( B_1\, H -
2\, B_1\, h + \dot{B}_1 \right)\, \hat{B} + \dot{\hat\alpha}_0 +
\left( 2 H -4 h \right)\, \hat{\alpha}_0 \Bigg\} = 0 \nonumber\\
\label{app:linearized}
\end{eqnarray}
where the index $i$ on the equations spans over the $i=2,3$ isotropic spatial directions, and
where the momenta are defined in eqs. (\ref{fourier}) and (\ref{defmom}).

Only the $2$d scalar perturbations are included in this computation. More in general, if we include both the $2$d scalar and $2$d vector modes, the perturbed equations carrying $i$ or $j$ indices split in two separate parts, one for the $2$d scalar, and one for the $2$d vector modes. Although we have not written the $2$d vector parts of the above equations, we have explicitly checked
that they are decoupled from the $2$d scalar parts. The $2$d
vector parts are not related to the instability we have
demonstrated in the main text; therefore we do not discuss them
here. Another property to be noted in the system
(\ref{app:linearized}) is that not all the equations are
independent. Using the perturbed Bianchi identities, it is
possible to obtain eqs.   ${\rm Eq}_{ij}, \, {\rm Eq}_{1i}$ from the
remaining ones  in (\ref{app:linearized}). This remaining equations are the set of
linearized equations (\ref{einstein-cc})  that we have chosen to solve in the main
text.~\footnote{One could have chosen a different but equivalent subset of  independent equations.
Our choice is related to the fact that the linearized equations of motion
obtained by extremizing the quadratic
action (\ref{act-2dS-cc}) precisely gives the  set of equations
(\ref{einstein-cc}).}

For brevity, we have grouped some of the extended terms that depend on the
background quantities, which appear both in the action and in the
linearized equations. We have  denoted them with calligraphic letters
${\cal D}, \, {\cal M}$. These  terms are
\begin{eqnarray}
{\cal D}_{\Psi \Psi} &=& \frac{1}{18 + 3 B_1^2 + 2 B_1^4}\, \Bigg[
\left( \frac{3}{2}\, m^2 + \frac{195}{8}\, h^2 - \frac{39}{2}\,
h\, H \right)\, B_1^6 + 18\, \left( 3 m^2 - 6 H^2 - \frac{51}{4}\,
h^2 + 15 h\, H \right)\, B_1^2 - \frac{39}{2}\, h\, \dot{B}_1\,
B_1^5 \nonumber\\
&& \qquad\qquad\qquad\qquad + \frac{9}{2}\, \left( 6 h - 7 H
\right)\, \dot{B}_1\, B_1^3 + 27\, \left( 8 h - 3 H \right)\,
B_1\, \dot{B}_1 + 27\, \left( 3 H^2 -3 h^2 - \frac{5}{2}\,
\dot{B}_1^2 \right) \nonumber\\
&& \qquad\qquad\qquad\qquad + 3\, \left( 6 m^2 + \frac{51}{4}\,
h^2 - \frac{51}{4}\, H^2 + \frac{13}{8}\, \dot{B}_1^2 \right)\,
B_1^4 \Bigg] \nonumber\\
{\cal D}_{\Psi \alpha_1} &=& \frac{-1}{18 + 3 B_1^2 + 2 B_1^4}\,
\Bigg[ \left( \frac{45}{4}\, h^2 - 9 h\, H \right)\, B_1^5 + 27\,
\left( 2 h - H \right)\, \dot{B}_1 + 3\, \left( H - 5 h \right)\,
\dot{B}_1\, B_1^4 - \frac{9}{2}\, \left( 6 h + H \right)\,
\dot{B}_1\, B_1^2 \nonumber\\
&& \qquad\qquad\qquad\quad +9\, \left( 3 m^2 - 6 h^2 + 12\,h \, H
-9 H^2 + \frac{3}{2}\, \dot{B}_1^2 \right)\, B_1 + 3\, \left(
\frac{3}{2}\, m^2 + \frac{39}{2}\, h^2 - 12 h\, H - \frac{9}{2}\,
H^2 + \frac{7}{4}\, \dot{B}_1^2 \right)\, B_1^3 \Bigg] \nonumber\\
{\cal D}_{\alpha \alpha} &=& \frac{3 p_T^2}{p_L^2\, \left( 18 +3
B_1^2 + 2 B_1^4 \right)}\, \Bigg[ \left( \frac{15}{4}\, h^2 - 3
h\, H \right)\, B_1^4 - 12\, h\, B_1\, \dot{B}_1 + \left( 2 H - 7
h\right)\, \dot{B}_1\, B_1^3 \nonumber\\
&& \qquad\qquad\qquad\qquad\qquad + 9\, \left( m^2 -2 h^2 + 4 h\,
H -3 H^2 + \frac{1}{2}\, \dot{B}_1^2 \right) + \left(
\frac{3}{2}\, m^2 + \frac{39}{2}\, h^2 - 12\, h\, H -
\frac{9}{2}\, H^2 + \frac{7}{4}\, \dot{B}_1^2 \right)\, B_1^2
\Bigg] \nonumber\\
{\cal D}_{\alpha_1 \alpha_1} &=& \frac{p_L^2}{p_T^2}\, {\cal
D}_{\alpha \alpha} \nonumber\\
{\cal D}_{\chi \chi} &=& \frac{-1}{p_L^2\, \left( 18 + 3 B_1^2 + 2
B_1^4 \right)}\, \Bigg[ \left( \frac{93}{4}\, h^2 - 21\, h\, H + 3
H^2 \right)\, B_1^6 - 36\, h\, \dot{B}_1\, B_1^3 + 3\, \left( 2 H
- 7 h\right)\, \dot{B}_1\, B_1^5 \nonumber\\
&& \qquad\qquad\qquad\qquad\qquad + 9\, \left( 3 m^2 + 6 h^2 - 6
H^2 + \frac{3}{2}\, \dot{B}_1^2 \right)\, B_1^2 + 3\, \left(
\frac{3}{2}\, m^2 + \frac{51}{2}\, h^2 - 18\, h\, H - 3 H^2 +
\frac{7}{4}\, \dot{B}_1^2 \right)\, B_1^4 \Bigg] \nonumber\\
{\cal D}_{\alpha_0 \alpha_0} &=& \frac{1}{B_1^2}\, {\cal D}_{\chi \chi} \nonumber\\
{\cal D}_{\chi \alpha_0} &=& \frac{1}{B_1}\, {\cal D}_{\chi \chi}
\label{calD}
\end{eqnarray}
and
\begin{eqnarray}
{\cal M}_{\Psi \Psi} &=& \frac{3}{18 + 3 B_1^2 + 2 B_1^4}\, \Bigg[
\left( m^2 + 5 h^2 - 4 h\, H \right)\, B_1^6 - 4 h\, B_1^5\,
\dot{B}_1 - 2 \left( 4 H - 19 h \right)\, B_1^3\, \dot{B}_1 + 24
\left( H - 2 h \right)\, B_1\, \dot{B}_1 + 3 \dot{B}_1^2
\nonumber\\
&& \qquad\qquad\qquad\qquad+ \left( \frac{11}{2} m^2 -
\frac{91}{2} h^2 + 46 h\, H - 12 H^2 + \dot{B}_1^2 \right)\, B_1^4
- \left( 3 m^2 + 21 h^2 + 12 h\, H - 36 H^2 + \frac{23}{2}
\dot{B}_1^2 \right)\,  B_1^2 \Bigg] \nonumber\\
{\cal M}_{\Psi \alpha_1} &=& \frac{1}{18 + 3 B_1^2 + 2 B_1^4}\,
\Bigg[ - \left( 2 m^2 + 5 h^2 - 4 h\, H \right)\, B_1^5 - 36\,
\left( H - 2 h \right)\, \dot{B}_1 + 4 \left( H - h \right)\,
B_1^4\, \dot{B}_1 + 6 \left( 3 H - 16 h \right)\, B_1^2\,
\dot{B}_2 \nonumber\\
&& \qquad\qquad\qquad\qquad - 9 \left( m^2 - 13 h^2 + 12 h\, H - 2
H^2 - \frac{1}{9} \dot{B}_1^2 \right)\, B_1^3 + 18 \left( m^2 +
h^2 + 4 h\, H - 6 H^2 + \frac{5}{3} \dot{B}_1^2 \right)\, B_1
\Bigg] \nonumber\\
{\cal M}_{\Psi \alpha} &=& \frac{3}{2 p_L^2\, \left( 18 + 3 B_1^2
+ 2 B_1^4 \right)}\, \Bigg[ -3 \left( 4 H - 5 h \right)\, h\,
B_1^5 + 4 \left( H - 5 h \right)\, B_1^4\, \dot{B}_1 - 36 \left( H
-2 h \right)\, \dot{B}_1 - 6 \left( 6 h + H \right)\, B_1^2\,
\dot{B}_1 \nonumber\\
&& \qquad\qquad\qquad\qquad\quad + 36 \left( m^2 - 2 h^2 + 4 h\, H
- 3 H^2 + \frac{1}{2} \dot{B}_1^2 \right)\, B_1 + 6 \left( m^2 +
13 h^2 - 8 h\, H - 3 H^2 + \frac{7}{6} \dot{B}_1^2 \right)\, B_1^3
\Bigg] \nonumber
\end{eqnarray}
\begin{eqnarray}
{\cal M}_{\alpha_1 \alpha_1} &=& \frac{2}{18 + 3 B_1^2 + 2
B_1^4}\,\Bigg[ \left( m^2 + 10 h^2 - 8 h\, H \right)\, B_1^4 - 8
h\, B_1^3\, \dot{B}_1 - 6 \left( 2 H - 5 h \right)\, B_1\,
\dot{B}_1 + 9 \left( m^2 - 5 h^2 + 4 h\, H - \frac{2}{3}
\dot{B}_1^2 \right) \nonumber\\
&& \qquad\qquad\qquad\qquad + 3 \left( \frac{5}{2} m^2 -
\frac{13}{2} h^2 + 10 h\, H - 6 H^2 + \dot{B}_1^2 \right)\, B_1^2
\Bigg] \nonumber\\
{\cal M}_{\alpha \alpha} &=& \frac{2}{18 + 3 B_1^2 + 2 B_1^4}\,
\Bigg[ \left( m^2 + 22 h^2 - 14 h\, H \right)\, B_1^4 - 8 h\,
B_1^3\, \dot{B}_1 - 6 \left( 2 H - 5 h \right)\, B_1\, \dot{B}_1 +
9 \left( m^2 + 7 h^2 - 2 h\, H - \frac{2}{3} \dot{B}_1^2 \right)
\nonumber\\
&& \qquad\qquad\qquad\qquad + 3 \left( \frac{5}{2} m^2 -
\frac{1}{2} h^2 + 7 h\, H - 6 H^2 + \dot{B}_1^2 \right)\, B_1^2
\Bigg] \nonumber\\
{\cal M}_{\Sigma \Psi} &=& \frac{12}{18 + 3 B_1^2 + 2 B_1^4}\,
\Bigg[ \left( m^2 + 5 h^2 - 4 h\, H \right)\, B_1^6 - 4 h\,
B_1^5\, \dot{B}_1 + 2 \left( 7 h - 4 H \right)\, B_1^3\, \dot{B}_1
+ 6 \left( 7 h - 2 H \right)\, B_1\, \dot{B}_1 - 15 \dot{B}_1^2
\nonumber\\
&& \qquad\qquad\qquad\qquad + \left( 15 m^2 - 57 h^2 + 60 h\, H -
18 H^2 - \frac{5}{2}\, \dot{B}_1^2 \right)\, B_1^2 + \left(
\frac{11}{2} m^2 - \frac{1}{2} h^2 + 10\, h\, H - 12 H^2 +
\dot{B}_1^2 \right)\, B_1^4 \Bigg] \nonumber\\
{\cal M}_{\Sigma \alpha_1} &=& \frac{4}{18 + 3 B_1^2 + 2 B_1^4}\,
\Bigg[ 2 \left( m^2 + \frac{5}{2} h^2 - 2 h\, H \right)\, B_1^5 +
\frac{3}{2} \left( h - 6 H\right)\, B_1^2\, \dot{B}_1 + 9 \left( 7
h - 2 H \right)\, \dot{B}_1 + \left( 2 H - 11 h \right)\, B_1^4\,
\dot{B}_1 \nonumber\\
&& \qquad\qquad\qquad\qquad + 36 \left( m^2 - \frac{7}{2} h^2 + 4
h\, H - \frac{3}{2} H^2 + \frac{1}{6} \dot{B}_1^2 \right)\, B_1 +
9 \left( m^2 + 2 h^2 - 2 H^2 + \frac{5}{9}\, \dot{B}_1^2 \right)\,
B_1^3 \Bigg] \label{calM}
\end{eqnarray}

As we have discussed in the main text, we proceed by solving the
second of (\ref{einstein-cc}) for the mode $\hat\chi$, and
inserting the solution back into the rest of the equations. Next,
we differentiate equations ${\rm Eq}_{00}, \, {\rm Eq}_{0i}, \,
{\rm Eq}_{0}$ (with the solution for $\hat\chi$ given in
(\ref{solhatchi}) inserted in them) in order to obtain first order
differential equations for the modes $\hat\Phi, \, \hat{B}, \,
\hat\alpha_0$. Combined with ${\rm Eq}_{11}, \, {\rm Eq}_{1}, \,
{\rm Eq}_{i}$ (again with the solution  (\ref{solhatchi}) inserted in them), these equations
form the set of equations to be solved numerically, summarized in
matrix form in (\ref{system}). Here we give the detailed
expressions of the terms appearing in this system of equations.

We first define the following useful combinations of background
dependent terms which frequently appear in the linearized system:
\begin{eqnarray}
{\cal D} &=& \left(6 + B_1^2\right)\, p_T^2 - 8 p_L^2\, {\cal
D}_{\chi \chi} \nonumber\\
{\cal H} &=& H + h + \frac{B_1\, \dot{B}_1}{6 + B_1^2}
\label{calH}
\end{eqnarray}
where ${\cal D}_{\chi\chi}$ is defined in (\ref{calD}). The time
derivative of ${\cal D}$ and ${\cal H}$ are also useful, which is
explicitly given by
\begin{eqnarray}
\dot{\cal D} &=& -2\, p_T^2\, \left( 6 + B_1^2 \right)\, \left(
H+h\right) + 16\, p_L^2\, \left( H - 2 h\right)\, {\cal D}_{\chi
\chi} + 2\, p_T^2\, B_1\, \dot{B}_1 - 8\, p_L^2\, \dot{\cal
D}_{\chi \chi} \nonumber\\
\dot{\cal H} &=& - 2\, \frac{B_1\, \dot{B}_1}{6 + B_1^2}\, {\cal
H} + \dot{H} + \dot{h} + 2\, \left( H + h \right)\, \frac{B_1\,
\dot{B}_1}{6 + B_1^2} + \frac{\dot{B}_1^2 + B_1\, \ddot{B}_1}{6 +
B_1^2}
\end{eqnarray}
where $\ddot{B}_1, \, \dot{H}, \, \dot{h}$ are obtained from
(\ref{evolution-cc}):
\begin{eqnarray}
\dot{H} &=& \frac{2}{18 + 3 B_1^2 + 2 B_1^4}\, \Bigg[ \left( m^2 -
\frac{35}{4} h^2 + 7 h\, H - 3 H^2 \right)\, B_1^4 + 12 h\, B_1\,
\dot{B}_1 + \left( 7 h - 2 H \right)\, B_1^3\, \dot{B}_1 -
\frac{9}{2}\, \left( 6 h^2 + \dot{B}_1^2 \right) \nonumber\\
&& \qquad\qquad\qquad\qquad - \frac{3}{2}\, \left( 18 h^2 - 12 h\,
H + \frac{7}{6} \dot{B}_1^2 \right) B_1^2 \Bigg] \nonumber\\
\dot{h} &=& \frac{1}{18 + 3 B_1^2 + 2 B_1^4}\, \Bigg[ 3 \left( 2 H
- 5 h \right)\, h\, B_1^4 + 8 h\, B_1^3\, \dot{B}_1 + 6 \left( 2 H
- 5 h \right)\, B_1\, \dot{B}_1 + 6 \left( \dot{B}_1^2 - 9 h\, H
\right) \nonumber\\
&& \qquad\qquad\qquad\qquad - 6 \left( m^2 - 2 h^2 + \frac{11}{2}
h\, H - 3 H^2 + \frac{1}{2} \dot{B}_1^2 \right)\, B_1^2 \Bigg]
\nonumber\\
\ddot{B}_1 &=& \frac{1}{18 + 3 B_1^2 + 2 B_1^4}\, \Bigg[ -2 \left(
m^2 + 10 h^2 - 8 h\, H \right)\, B_1^5 + 2 \left( 8 h - 3 H
\right)\, B_1^4\, \dot{B}_1 - 54 H\, \dot{B}_1 + 15 \left( H - 4 h
\right)\, B_1^2\, \dot{B}_1 \nonumber\\
&& \qquad\qquad\qquad\qquad - 18 \left( m^2 - 5 h^2 + 4 h\, H -
\frac{2}{3} \dot{B}_1^2 \right)\, B_1 - 15 \left( m^2 -
\frac{13}{5} h^2 + 4 h\, H - \frac{12}{5} H^2 + \frac{2}{5}
\dot{B}_1^2 \right)\, B_1^3 \Bigg]
\end{eqnarray}
We also need the explicit expressions for $\dot{{\cal
D}}_{\chi \chi}$ and $\dot{\cal D}_{\alpha_0 \alpha_0}$:
\begin{eqnarray}
\dot{{\cal D}}_{\chi \chi} &=& 2 \left[ H - 2 h - \frac{3 + 4
B_1^2}{18 + 3 B_1^2 + 2 B_1^4}\, B_1\, \dot{B}_1 \right]\, {\cal
D}_{\chi \chi} \nonumber\\
&& - \frac{9}{p_L^2\, \left( 18 + 3 B_1^2 + 2 B_1^4 \right)}\,
\Bigg[ 2 \left( \frac{31}{4} h^2 - 7 h\, H + H^2 \right)\, B_1^5\,
\dot{B}_1 - 12 h\, B_1^2\, \dot{B}_1^2 + \frac{5}{3} \left( 2 H -
7 h \right)\, B_1^4\, \dot{B}_1^2 \nonumber\\
&& \qquad\qquad\qquad\qquad\qquad\quad + 6 \left( m^2 + 2 h^2 - 2
H^2 + \frac{1}{2} \dot{B}_1^2 \right)\, B_1\, \dot{B}_1 + 4 \left(
\frac{m^2}{2} + \frac{17}{2} h^2 - 6 h\, H - H^2 + \frac{7}{12}
\dot{B}_1^2 \right)\, B_1^3\, \dot{B}_1 \nonumber\\
&& \qquad\qquad\qquad\qquad\qquad\quad - 4 \dot{h}\, B_1^3\,
\dot{B}_1 + \left( \frac{2}{3} \dot{H} - \frac{7}{3} \dot{h}
\right)\, B_1^5\, \dot{B}_1 + \left( \frac{31}{6} h\, \dot{h} -
\frac{7}{3}\, \left( H\, \dot{h} + h\, \dot{H} \right) +
\frac{2}{3} H\, \dot{H} \right)\, B_1^6 - 4 h\, B_1^3\,
\ddot{B}_1 \nonumber\\
&& \qquad\qquad\qquad\qquad\qquad\quad + \left( \frac{2}{3} H -
\frac{7}{3} h \right)\, B_1^5\, \ddot{B}_1 + \left( 12 \left( h\,
\dot{h} - H\, \dot{H} \right) + 3 \dot{B}_1\, \ddot{B}_1 \right)\,
B_1^2 \nonumber\\
&& \qquad\qquad\qquad\qquad\qquad\quad + \left( 17 h\, \dot{h} - 6
\left( H\, \dot{h} + h\, \dot{H} \right) - 2 H\, \dot{H} +
\frac{7}{6}\, \dot{B}_1\, \ddot{B}_1 \right)\, B_1^4 \Bigg]
\nonumber\\
\dot{{\cal D}}_{\alpha_0 \alpha_0} &=& \frac{1}{B_1^2}\, \left(
\dot{\cal D}_{\chi \chi} - 2 \frac{\dot{B}_1}{B_1}\, {\cal
D}_{\chi \chi} \right)
\end{eqnarray}
Now we give the explicit form of the matrix ${\cal M}_{\kappa}$
used in (\ref{system}):
\begin{eqnarray}
&& \kappa_{11} = -B_1^2 + \frac{2 p_L^2\, B_1^4}{{\cal D}} \nonumber\\
&& \kappa_{12} = \frac{B_1}{2} - \frac{2 p_L^2\, B_1^3}{{\cal D}} \nonumber\\
&& \kappa_{14} = \left( 3 + \frac{B_1^2}{2} \right)\, h + \left( 3
- B_1^2\right)\, H + \frac{1}{2}\, B_1\, \dot{B}_1 - \frac{2
p_L^2\, B_1^2\, \left(6 + B_1^2\right)}{{\cal D}}\, {\cal H} \nonumber\\
&& \kappa_{15} = -\frac{3}{2} + \frac{B_1^2}{4} + \frac{p_L^2\,
B_1^2\, \left(6 + B_1^2\right)}{2 {\cal D}} \nonumber\\
&& \kappa_{16} = \frac{4 p_L^2\, B_1^3\, {\cal D}_{\alpha_0
\alpha_0}}{{\cal D}} \nonumber 
\end{eqnarray}
\begin{eqnarray}
&& \kappa_{21} = - \frac{B_1}{3} + \frac{4 p_L^2\,
B_1^3}{3 {\cal D}} \nonumber\\
&& \kappa_{22} = 1 - \frac{4 p_L^2\, B_1^2}{3 {\cal D}} \nonumber\\
&& \kappa_{24} = -2\, B_1\, h + \dot{B}_1 - \frac{4 p_L^2\, B_1\,
\left( 6 + B_1^2\right)}{3 {\cal D}}\, {\cal H} \nonumber\\
&& \kappa_{25} = \frac{1}{3}\, B_1\, \left( 1 + \frac{p_L^2\,
\left(6+B_1^2\right)}{{\cal D}} \right)  \nonumber\\
&& \kappa_{26} = 1 + \frac{8 p_L^2\, B_1^2\, {\cal D}_{\alpha_0
\alpha_0}}{3 {\cal D}} \nonumber 
\end{eqnarray}
\begin{eqnarray}
&& \kappa_{41} = \left(1 + \frac{B_1^2}{6}\right)\, h + \left(1 -
\frac{B_1^2}{3} \right)\, H + \frac{1}{6}\, B_1\, \dot{B}_1 -
\frac{2 p_L^2\, B_1^2\, \left(6 +B_1^2\right)}{3 {\cal D}}\, {\cal H} \nonumber\\
&& \kappa_{42} = B_1\, h - \frac{\dot{B}_1}{2} + \frac{2 p_L^2\,
B_1\, \left(6 + B_1^2\right)}{3 {\cal D}}\, {\cal H} \nonumber\\
&& \kappa_{44} = 3 H^2 - \left( 3 + \frac{5}{2}\, B_1^2 \right)\,
h^2 - \frac{1}{2}\, \dot{B}_1^2 + 2\, B_1\, \left( B_1\, H +
\dot{B}_1 \right)\, h + \frac{2 p_L^2\, \left(6 +
B_1^2\right)^2}{3 {\cal D}}\, {\cal H}^2 \nonumber\\
&& \kappa_{45} = -\frac{1}{2}\, \left[ \left( 1 + \frac{B_1^2}{6}
\right)\, \left( 2 H - h\right) + \frac{1}{3}\, B_1\, \dot{B}_1 +
\frac{p_L^2\, \left(6 + B_1^2\right)^2}{3 {\cal D}}\, {\cal H}\right] \nonumber\\
&& \kappa_{46} = B_1\, \left( h - \frac{H}{2} \right) -
\frac{\dot{B}_1}{2} - \frac{4 p_L^2\, B_1\,
\left(6+B_1^2\right)}{3 {\cal D}}\, {\cal H}\, {\cal D}_{\alpha_0
\alpha_0} \nonumber 
\end{eqnarray}
\begin{eqnarray}
&& \kappa_{51} = \frac{1}{2}\, \left( 1 - \frac{1}{6}\, B_1^2
\right) - \frac{p_L^2\, B_1^2\, \left(6 + B_1^2\right)}{6 {\cal
D}} \nonumber\\
&& \kappa_{52} = \frac{1}{6}\, \left( 1 + \frac{p_L^2\, \left(6 +
B_1^2\right)}{{\cal D}} \right)\, B_1 \nonumber\\
&& \kappa_{54} = \frac{1}{12}\, \left( 6 + B_1^2 \right)\,
\left(2H - h\right) + \frac{1}{6}\, B_1\, \dot{B}_1 +
\frac{p_L^2\, \left(6
+ B_1^2\right)^2}{6 {\cal D}}\, {\cal H} \nonumber\\
&& \kappa_{55} = -\frac{p_L^4\, \left(6 + B_1^2\right)}{3 p_T^2\,
{\cal D}}\, {\cal D}_{\chi \chi} \nonumber\\
&& \kappa_{56} = -\frac{p_L^2\, B_1\, \left(6 + B_1^2\right)}{3\,
{\cal D}}\, {\cal D}_{\alpha_0 \alpha_0} \nonumber 
\end{eqnarray}
\begin{eqnarray}
&& \kappa_{61} = - \frac{8 p_L^2\, B_1}{3
{\cal D}}\, {\cal D}_{\chi \chi} \nonumber\\
&& \kappa_{62} = 1 + \frac{8 p_L^2}{3 {\cal D}}\, {\cal D}_{\chi \chi} \nonumber\\
&& \kappa_{63} = \frac{p_T^2}{p_L^2} \nonumber\\
&& \kappa_{64} = B_1\, \left( H -2 h\right) + \dot{B}_1 + \frac{8
p_L^2\, \left(6 + B_1^2\right)}{3 B_1\, {\cal D}}\, {\cal H}\,
{\cal D}_{\chi \chi} \nonumber\\
&& \kappa_{65} = - \frac{2 p_L^2\, \left(6 + B_1^2\right)}{3 B_1\,
{\cal D}}\, {\cal D}_{\chi \chi} \nonumber\\
&& \kappa_{66} = 1 + \frac{p_T^2}{p_L^2} - \frac{2 p_T^2\, \left(6
+ B_1^2\right)}{3 {\cal D}}\,  {\cal D}_{\alpha_0 \alpha_0}
\label{kappa6}
\end{eqnarray}
Finally, we explicitly write down the right hand side of
(\ref{system}) involving the functions $f_i$ where $i=1 \dots 6$.
Each $f_i$ is a linear combination of the variables
\begin{equation}
F_i \equiv \{ \dot{\hat\Psi}, \, \dot{\hat\alpha}_1, \,
\dot{\hat\alpha}, \, \hat\Psi, \, \hat\alpha_1, \, \hat\alpha, \,
\hat\Phi, \, \hat{B}, \, \hat\alpha_0 \}
\end{equation}
The explicit forms of the functions $f_1, \dots f_6$ can then be
expressed as
\begin{equation}
f_{i} = \sum_j {\cal A}_{ij}\, F_j
\end{equation}
The coefficients ${\cal A}_{ij}$ depend entirely on the background
quantities. They are given by
\begin{eqnarray}
&& {\cal A}_{11} = \left( 2\, \dot{B}_1 + 3 H\, B_1 \right)\, B_1
- \frac{2 p_L^2\, B_1^4}{{\cal D}}\, \left( H + 4 h + 4
\frac{\dot{B}_1}{B_1} - \frac{\dot{\cal D}}{\cal D} \right)
\nonumber\\
&& {\cal A}_{12} = \left( \frac{H}{2} - 4 h \right)\, B_1 +
\frac{\dot{B}_1}{2} + \frac{2 p_L^2\, B_1^3}{{\cal D}}\, \left( H
+ 4 h + 2 \frac{\dot{B}_1}{B_1} - \frac{\dot{\cal D}}{{\cal D}}
\right) \nonumber\\
&& {\cal A}_{13} = 0 \nonumber\\
&& {\cal A}_{14} = p_T^2 B_1^2 - {\cal M}_{\Psi \Psi} + \frac{4
p_L^2\, B_1^4}{{\cal D}}\, \left[ \frac{1}{2}\, \left( \dot{H} -2
\dot{h} \right) + \left( H + h - \frac{\dot{\cal D}}{2 {\cal D}}
\right)\, \left( H - 2 h - 2\frac{\dot{B}_1}{B_1} \right) -
\frac{\dot{B}_1^2}{B_1^2} - \frac{\ddot{B}_1}{B_1} + \left(H-2 h
\right)\, \frac{\dot{B}_1}{B_1} \right] \nonumber\\
&& {\cal A}_{15} = -\frac{p_T^2}{2}\, B_1 - {\cal M}_{\Psi
\alpha_1} - \frac{2 p_L^2\, B_1^3}{{\cal D}}\, \left[ \dot{H} - 2
\dot{h} + 2\, \left( H + h - \frac{\dot{\cal D}}{2 {\cal D}}
\right)\, \left( H -2 h - \frac{\dot{B}_1}{B_1} \right) + \left( H
-2 h \right)\, \frac{\dot{B}_1}{B_1} - \frac{\ddot{B}_1}{B_1}
\right] \nonumber\\
&& {\cal A}_{16} = 0 \nonumber\\
&& {\cal A}_{17} = \frac{3}{2}\, \left( p_T^2 - \frac{V_0}{M_p^2}
+ 3 h^2 - 3 H^2 + \frac{\dot{B}_1^2}{2} \right) + \frac{1}{4}\,
\left( 9 m^2 - 2 p_L^2 - p_T^2 + 15 h^2 - 12 h\, H \right) B_1^2 -
3 h\, B_1\, \dot{B}_1 \nonumber\\
&& \qquad + \frac{2 p_L^2\, B_1^2\, \left( 6 + B_1^2\right)}{{\cal
D}}\, \left[ \dot{\cal H} + 2 \frac{B_1\, \dot{B}_1}{6 + B_1^2}\,
{\cal H} + 2 \left( H + h - \frac{\dot{\cal D}}{2 {\cal D}}
\right)\, {\cal H} \right] \nonumber\\
&& {\cal A}_{18} = \frac{9}{2}\, \left( 1 +
\frac{B_1^2}{18}\right)\, h + \frac{9}{2}\, \left( 1 - \frac{5
B_1^2}{18} \right)\, H + \frac{1}{2}\, B_1\, \dot{B}_1 - p_L^2\,
\left( 6 + B_1^2\right)\, \left( {\cal H} - \frac{\dot{\cal D}}{2
{\cal D}} \right)\, \frac{B_1^2}{{\cal D}} \nonumber\\
&& {\cal A}_{19} = \frac{3 B_1}{2}\, \left( H - 2 h +
\frac{\dot{B}_1}{B_1} \right) - \frac{4 p_L^2\, B_1^3}{{\cal D}}\,
{\cal D}_{\alpha_0 \alpha_0}\, \left[ 2 H + 2 h +
\frac{\dot{B}_1}{B_1} + \frac{\dot{\cal D}_{\alpha_0
\alpha_0}}{{\cal D}_{\alpha_0 \alpha_0}} - \frac{\dot{\cal
D}}{{\cal D}} \right] \nonumber 
\end{eqnarray}
\begin{eqnarray}
&& {\cal A}_{21} = \frac{1}{3}\, \left( 7 H - 8 h \right)\, B_1 +
\dot{B}_1 - \frac{4 p_L^2}{3}\, \left( H + 4 h + 4
\frac{\dot{B}_1}{B_1} - \frac{\dot{\cal D}}{{\cal D}} \right)\,
\frac{B_1^3}{{\cal D}} \nonumber\\
&& {\cal A}_{22} = -3 H + \frac{4 p_L^2}{3}\, \left( H + 4 h + 2
\frac{\dot{B}_1}{B_1} - \frac{\dot{\cal D}}{{\cal D}} \right)\,
\frac{B_1^2}{{\cal D}} \nonumber\\
&& {\cal A}_{23} = 0 \nonumber\\
&& {\cal A}_{24} = \frac{p_T^2 B_1}{3} - \frac{8 p_L^2\, B_1^3}{3
{\cal D}}\, \left[ \left( 2 H + 2 h - \frac{\dot{\cal D}}{{\cal
D}} \right)\, \left( h- \frac{H}{2} + \frac{\dot{B}_1}{B_1}
\right) + \dot{h} - \frac{\dot{H}}{2} - \left( H - 2 h\right)\,
\frac{\dot{B}_1}{B_1} + \frac{\dot{B}_1^2}{B_1^2} +
\frac{\ddot{B}_1}{B_1} \right]
\nonumber\\
&& {\cal A}_{25} = - p_T^2 - {\cal M}_{\alpha_1 \alpha_1} -
\frac{4 p_L^2\, B_1^2}{3 {\cal D}}\, \left[ \left( 2 H + 2 h -
\frac{\dot{\cal D}}{{\cal D}} \right)\, \left( H - 2 h -
\frac{\dot{B}_1}{B_1} \right) + \dot{H} - 2 \dot{h} + \left( H - 2
h\right)\, \frac{\dot{B}_1}{B_1} - \frac{\ddot{B}_1}{B_1} \right]
\nonumber\\
&& {\cal A}_{26} = p_T^2 \nonumber\\
&& {\cal A}_{27} = \left( 2 m^2 - \frac{p^2}{3} \right)\, B_1 +
\frac{4 p_L^2\, B_1\, \left( 6 + B_1^2 \right)}{3\, {\cal D}}\,
\Bigg[ 2 \left( H + h - \frac{\dot{\cal D}}{2 {\cal D}} +
\frac{B_1\, \dot{B}_1}{6 + B_1^2} \right)\,
{\cal H} + \dot{\cal H}  \Bigg] \nonumber\\
&& {\cal A}_{28} = - \left( \frac{H}{3} + \frac{7 h}{3} \right)\,
B_1 + \dot{B}_1 - \frac{2\, p_L^2\, B_1\, \left( 6 +
B_1^2\right)}{3\, {\cal D}}\, \left( {\cal H} - \frac{\dot{\cal
D}}{2 {\cal D}} \right)
\nonumber\\
&& {\cal A}_{29} = -2 (H + h) - \frac{16\, p_L^2\, B_1^2\, {\cal
D}_{\alpha_0 \alpha_0}}{3\, {\cal D}}\, \left[ H + h -
\frac{\dot{\cal D}}{2 {\cal D}} + \frac{\dot{B}_1}{2 B_1} +
\frac{\dot{\cal D}_{\alpha_0 \alpha_0}}{2 {\cal D}_{\alpha_ 0
\alpha_0}} \right] \nonumber 
\end{eqnarray}
\begin{eqnarray}
&& {\cal A}_{31} = {\cal A}_{32} = {\cal A}_{34} = {\cal A}_{37} = 0 \nonumber\\
&& {\cal A}_{33} = -3\, \left( H - 2 h \right) \nonumber\\
&& {\cal A}_{35} = p_L^2 \nonumber\\
&& {\cal A}_{36} = -p_L^2 - {\cal M}_{\alpha \alpha} \nonumber\\
&& {\cal A}_{38} = -\frac{p_L^2}{p_T^2}\, \left( (H-2 h)\, B_1 + \dot{B}_1 \right) \nonumber\\
&& {\cal A}_{39} = -2\, \left( H - 2 h \right) \nonumber 
\end{eqnarray}
\begin{eqnarray}
&& {\cal A}_{41} = \left( \frac{2}{3} H - \frac{1}{3} h \right)\,
B_1\, \dot{B}_1 - \frac{2}{3} \dot{B}_1^2 - \frac{1}{6} B_1\,
\ddot{B}_1 - \frac{p_T^2}{2} - \left( m^2 - \frac{2 p_L^2 +
p_T^2}{6} + 5 h^2 - 4 h\, H \right)\, \frac{B_1^2}{2} + 2 h\,
B_1\, \dot{B}_1 \nonumber\\
&& \qquad\qquad - \left( 1 + \frac{B_1^2}{6} \right)\, \dot{h} -
\left( 1 - \frac{B_1^2}{3} \right)\, \dot{H} + \frac{2 p_L^2\,
B_1^2\, \left( 6 +B_1^2\right)}{3 {\cal D}}\, \Bigg[ \left(-3 H +
6 h + 4 \frac{\dot{B}_1}{B_1} - \frac{\dot{\cal D}}{{\cal D}} +
2\,
\frac{B_1\, \dot{B}_1}{6 + B_1^2} \right)\, {\cal H} + \dot{\cal H} \Bigg] \nonumber\\
&& {\cal A}_{42} = \frac{1}{2}\, \left( m^2 - \frac{p^2}{3} + 5
h^2 - 4 h\, H - 2 \dot{h} \right)\, B_1 - 2 h\, \dot{B}_1 +
\frac{1}{2}\, \ddot{B}_1 \nonumber\\
&& \qquad\qquad - \frac{2 p_L^2\, B_1\, \left( 6 +B_1^2\right)}{3
{\cal D}}\, \left[ \left(-3 H + 6 h + 2 \frac{\dot{B}_1}{B_1} -
\frac{\dot{\cal D}}{{\cal D}} + 2\, \frac{B_1\, \dot{B}_1}{6 +
B_1^2} \right)\, {\cal H} + \dot{\cal H} \right]
\nonumber\\
&& {\cal A}_{43} = 0 \nonumber\\
&& {\cal A}_{44} = \left( H + h \right) p_T^2 - \left[ \frac{2
p_L^2 + p_T^2}{6} H - \frac{4 p_L^2 - p_T^2}{6} h - 2 \left(
\dot{H} h + H \dot{h} \right) + 5 h \dot{h} \right] B_1^2 - \left[
m^2 - \frac{2 p_L^2 + p_T^2}{6} - 4 H h - 2 \dot{h} + 5
h^2 \right]\, B_1  \dot{B}_1 \nonumber\\
&& \qquad\qquad - \dot{B}_1\, \ddot{B}_1 + 2\, \left( \dot{B}_1^2
+ B_1\, \ddot{B}_1 \right)\,h + \frac{2 p_L^2\, B_1^2 \left( 6 +
B_1^2 \right)}{3 {\cal D}}\, \left( 2 h - H + 2
\frac{\dot{B}_1}{B_1} \right)\, \Bigg[ \left( 4 h - 2 H -
\frac{\dot{\cal D}}{{\cal D}} + 2\, \frac{B_1\, \dot{B}_1}{6 +
B_1^2} \right)\, {\cal H} + \dot{\cal H} \Bigg]
\nonumber\\
&& \qquad\qquad + \frac{2 p_L^2\, B_1^2 \left( 6 + B_1^2\right)}{3
{\cal D}}\, {\cal H}\, \left[ 2 \dot{h} - \dot{H} + 2
\frac{\dot{B}_1^2}{B_1^2} + 2 \left( 2 h - H \right)\,
\frac{\dot{B}_1}{B_1} + 2 \frac{\ddot{B}_1}{B_1}
\right] \nonumber\\
&& {\cal A}_{45} = -h\,\ddot{B}_1 + \left[ \frac{1}{2}\, \left(
m^2 - \frac{p^2}{3} \right) + \frac{5}{2}\, h^2 - 2 h\, H -
\dot{h} \right]\, \dot{B}_1 + \left[ \left( \frac{p^2}{3} - 2\,
\dot{h} \right)\, H + \left( \frac{-2 p_L^2 + p_T^2}{3} + 5
\dot{h} -2
\dot{H} \right)\, h \right]\, B_1 \nonumber\\
&& \qquad\qquad + \frac{2 p_L^2\, B_1 \left( 6 + B_1^2 \right)}{3
{\cal D}}\, \left( 2 h - H + \frac{\dot{B}_1}{B_1} \right)\,
\Bigg[ \left( 2 H - 4 h + \frac{\dot{\cal D}}{{\cal D}} - 2\,
\frac{B_1\, \dot{B}_1}{6 + B_1^2} \right)\, {\cal H} - \dot{\cal
H} \Bigg]
\nonumber\\
&& \qquad\qquad - \frac{2 p_L^2\, B_1 \left( 6 + B_1^2\right)}{3
{\cal D}}\, {\cal H}\, \left[ 2 \dot{h} - \dot{H} + (2 h - H )\,
\frac{\dot{B}_1}{B_1} +
\frac{\ddot{B}_1}{B_1} \right] \nonumber\\
&& {\cal A}_{46} = 0 \nonumber\\
&& {\cal A}_{47} = -6 H\, \dot{H} + \dot{B}_1\, \ddot{B}_1 - 2 h\,
\left( 2 H\, B_1\, \dot{B}_1 + B_1^2\, \dot{H} + \dot{B}_1^2 +
B_1\, \ddot{B}_1 \right) - 2\, B_1\, \left( B_1\, H + \dot{B}_1
\right)\, \dot{h} + \left( 6 + 5 B_1^2\right)\,h \, \dot{h} + 5\,
B_1\, \dot{B}_1\, h^2 \nonumber\\
&& \qquad\qquad + \frac{4 p_L^2\, \left( 6 + B_1^2\right)^2}{3
{\cal D}}\, {\cal H}\, \left[ \left( H - 2 h + \frac{\dot{\cal
D}}{2 {\cal D}} - 2\, \frac{B_1\, \dot{B}_1}{6 + B_1^2} \right)\,
{\cal H} -
\dot{\cal H} \right] \nonumber\\
&& {\cal A}_{48} = \frac{1}{6}\, \left[ \left( 2 H - h \right)\,
B_1\, \dot{B}_1 + \dot{B}_1^2 + B_1\, \ddot{B}_1 \right] -
\frac{1}{12}\, \left( 6 + B_1^2 \right)\, \left( \dot{h} - 2\,
\dot{H} \right) - \frac{p_L^2\, \left( 6 + B_1^2 \right)^2}{3
{\cal D}}\, \left( H - 2 h\right)\, {\cal H} \nonumber\\
&& \qquad\qquad + \frac{p_L^2\, \left( 6 + B_1^2 \right)^2}{6
{\cal D}}\, \left[ \left( 4\, \frac{B_1\, \dot{B}_1}{6 + B_1^2} -
\frac{\dot{\cal D}}{{\cal D}} \right)\, {\cal H} + \dot{\cal H}
\right] \nonumber\\
&& {\cal A}_{49} = \frac{1}{2}\, \ddot{B}_1 + \frac{1}{2}\, \left(
H - 2 h \right)\, \dot{B}_1 + \frac{1}{2}\, \left( \dot{H} - 2
\dot{h} \right)\, B_1 \nonumber\\
&& \qquad\qquad - \frac{8 p_L^2\, B_1\, \left( 6 + B_1^2\right)}{3
{\cal D}}\, {\cal D}_{\alpha_0 \alpha_0}\, \left[ {\cal H}\,
\left( H - 2 h + \frac{\dot{\cal D}}{2 {\cal D}} -
\frac{\dot{B}_1}{2 B_1} - \frac{\dot{\cal D}_{\alpha_0
\alpha_0}}{2 {\cal D}_{\alpha_0 \alpha_0}} \right) - \frac{1}{2}\,
\dot{\cal H} - \frac{B_1\, \dot{B}_1}{6 + B_1^2}\, {\cal H}
\right] \nonumber 
\end{eqnarray}
\begin{eqnarray}
{\cal A}_{51} &=& -\frac{1}{6}\, B_1\, \left( B_1\, H - 3\,
\dot{B}_1 \right) + \frac{1}{2}\, \left( 3 + \frac{B_1^2}{6}
\right)\, h + \frac{p_L^2\, B_1^2\, \left( 6 + B_1^2 \right)}{3
{\cal D}}\, \left[ 3 h - \frac{3}{2}\, H + \frac{B_1\,
\dot{B}_1}{6 + B_1^2} + 2\, \frac{\dot{B}_1}{B_1} -
\frac{\dot{\cal D}}{2 {\cal D}} \right] \nonumber\\
{\cal A}_{52} &=& - \frac{B_1}{6}\, \left( 2 H - 7 h \right) -
\frac{5}{6}\, \dot{B}_1 + \frac{p_L^2\, B_1\, \left( 6 +
B_1^2\right)}{3 {\cal D}}\, \left[ \frac{3}{2}\, H - 3 h -
\frac{B_1\, \dot{B}_1}{6 + B_1^2} - \frac{\dot{B}_1}{B_1} +
\frac{\dot{\cal D}}{2 {\cal D}} \right] \nonumber\\
{\cal A}_{53} &=& B_1\, \left( \frac{H}{2} - h \right) +
\frac{\dot{B}_1}{2} \nonumber\\
{\cal A}_{54} &=& \frac{1}{6}\, h\, B_1\, \dot{B}_1 -
\frac{1}{6}\, \left( B_1\, H - 2\, \dot{B}_1 \right)\, \dot{B}_1 -
\frac{1}{6}\, B_1\, \left( H\, \dot{B}_1 + B_1\, \dot{H} - 2\,
\ddot{B}_1 \right) + \frac{1}{2}\, \left( 3 + \frac{1}{6}\, B_1^2
\right)\, \dot{h} \nonumber\\
&& + \frac{p_L^2\, B_1^2\, \left( 6 + B_1^2 \right)}{3 {\cal D}}\,
\left( 2 h - H + 2 \frac{\dot{B}_1}{B_1} \right)\, \left( 2 h - H
+ \frac{B_1\, \dot{B}_1}{6 + B_1^2} - \frac{\dot{\cal D}}{2 {\cal
D}} \right) \nonumber\\
&& + \frac{p_L^2\, B_1^2\, \left( 6 + B_1^2\right)}{6 {\cal D}}\,
\left[ 2 \dot{h} - \dot{H} + 2\, \left( 2 h - H \right)\,
\frac{\dot{B}_1}{B_1} + 2\, \frac{\dot{B}_1^2}{B_1^2} + 2\,
\frac{\ddot{B}_1}{B_1} \right] \nonumber\\
{\cal A}_{55} &=& \frac{1}{6}\, \left( 7 h - 2 H \right)\,
\dot{B}_1 + \frac{1}{6}\, \left( 7 \dot{h} - 2 \dot{H} \right)\,
B_1 - \frac{2}{3}\, \ddot{B}_1 - \frac{p_L^2\, B_1\, \left( 6 +
B_1^2 \right)}{3 {\cal D}}\, \left( 2 h - H + \frac{B_1\,
\dot{B}_1}{6 + B_1^2} - \frac{\dot{\cal D}}{2 {\cal D}} \right)\,
\left( 2 h - H + \frac{\dot{B}_1}{B_1} \right) \nonumber\\
&& - \frac{p_L^2\, B_1\, \left( 6 + B_1^2\right)}{6 {\cal D}}\,
\left[ 2 \dot{h} - \dot{H} + \left( 2 h - H \right)\,
\frac{\dot{B}_1}{B_1} + \frac{\ddot{B}_1}{B_1} \right] \nonumber\\
{\cal A}_{56} &=& \left( \frac{\dot{H}}{2} - \dot{h} \right)\, B_1
+ \left( \frac{H}{2} - h \right)\, \dot{B}_1 +
\frac{\ddot{B}_1}{2} \nonumber\\
{\cal A}_{57} &=& \frac{1}{6} \left[ B_1 \left( h - 2H \right) -
\dot{B}_1 \right] \dot{B}_1 - \frac{1}{6} B_1 \ddot{B}_1 +
\frac{1}{12} \left( 6 + B_1^2 \right) \left( \dot{h} - 2 \dot{H}
\right) + \frac{p_L^2 \left(6 + B_1^2\right)^2}{3 {\cal D}}\,
\left[ H - 2 h - 2 \frac{B_1\, \dot{B}_1}{6 + B_1^2} +
\frac{\dot{\cal D}}{2 {\cal D}} - \frac{\dot{\cal H}}{2 {\cal H}}
\right]  {\cal H} \nonumber\\
{\cal A}_{58} &=& \frac{p_L^4\, \left( 6 + B_1^2\right)\, {\cal
D}_{\chi \chi}}{3 p_T^2\, {\cal D}^2}\, \left[ 12 h\, \left( {\cal
D} + 4 p_L^2\, {\cal D}_{\chi \chi} \right) - 16 p_L^2\,
\frac{B_1\, \dot{B}_1}{6 + B_1^2}\, {\cal D}_{\chi \chi} + \left(
6 + B_1^2\right)\, p_T^2\, \frac{\dot{\cal D}_{\chi \chi}}{{\cal
D}_{\chi \chi}} \right] \nonumber\\
{\cal A}_{59} &=& \frac{p_L^2\, B_1\, {\cal D}_{\alpha_0
\alpha_0}}{3 {\cal D}^2}\, \Bigg[ p_T^2\, B_1^3\, \dot{B}_1 + 12
\left( 3 p_T^2 -4 p_L^2\, {\cal D}_{\chi \chi} \right)\,
\frac{\dot{B}_1}{B_1} + 12 \left( p_T^2 - 2 p_L^2\, {\cal D}_{\chi
\chi} \right)\, B_1\, \dot{B}_1 + p_T^2\, \left( 6 h +
\frac{\dot{\cal D}_{\alpha_0 \alpha_0}}{{\cal D}_{\alpha_0
\alpha_0}} \right)\, B_1^4 \nonumber\\
&& + 12 \left( 18 p_T^2\, h + 4 p_L^2\, \dot{\cal D}_{\chi \chi} +
\left( 3 p_T^2 - 4 p_L^2\, {\cal D}_{\chi \chi} \right)\,
\frac{\dot{\cal D}_{\alpha_0 \alpha_0}}{{\cal D}_{\alpha_0
\alpha_0}} \right) + 4\, \left( 18 p_T^2\, h + 2 p_L^2\, \dot{\cal
D}_{\chi \chi} + \left( 3 p_T^2 - 2 p_L^2\, {\cal D}_{\chi \chi}
\right)\, \frac{\dot{\cal D}_{\alpha_0 \alpha_0}}{{\cal
D}_{\alpha_0 \alpha_0}} \right) B_1^2 \Bigg] \nonumber
\end{eqnarray}
\begin{eqnarray}
{\cal A}_{61} &=& -\frac{4 p_T^4}{3 {\cal D}^2}\, \Bigg\{
\frac{3}{4}\, \left( 2 h - H \right)\, B_1^5 - \frac{3}{4}\,
\dot{B}_1\, B_1^4 -9\, \left( 3 - 4 \frac{p_L^2}{p_T^2}\, {\cal
D}_{\chi \chi} \right)\, \dot{B}_1 - \left( 9 -10\,
\frac{p_L^2}{p_T^2}\, {\cal D}_{\chi \chi} \right)\, B_1^2\,
\dot{B}_1 + 2 h\, \left( 9 -20\, \frac{p_L^2}{p_T^2}\, {\cal
D}_{\chi \chi} \right)\, B_1^3 \nonumber\\
&& + \left[ - 9 \left( 3 + B_1^2\right) + 14\, \left( 6 + B_1^2
\right)\, \frac{p_L^2}{p_T^2}\, {\cal D}_{\chi \chi} - 64\,
\frac{p_L^4}{p_T^4}\, {\cal D}_{\chi \chi}^2 \right]\, H\, B_1 +
2\, \left[ 27 - 120\,\frac{p_L^2}{p_T^2}\, {\cal D}_{\chi \chi} +
64\, \frac{p_L^4}{p_T^4}\, {\cal D}_{\chi \chi}^2 \right]\, h\,
B_1 \nonumber\\
&& - 2\, \left( 6 + B_1^2\right)\, B_1\, \frac{p_L^2}{p_T^2}\,
\dot{\cal D}_{\chi \chi} \Bigg\} \nonumber\\
{\cal A}_{62} &=& -\frac{4 p_T^4}{3 B_1\, {\cal D}^2}\, \Bigg\{
\frac{3}{4}\, \left( H -2 h\right)\, B_1^5 + 4\,
\frac{p_L^2}{p_T^2}\, \left( 3 - 4 \frac{p_L^2}{p_T^2}\, {\cal
D}_{\chi \chi} \right)\, {\cal D}_{\chi \chi}\, \dot{B}_1 - 2
\frac{p_L^2}{p_T^2}\, {\cal D}_{\chi \chi}\, B_1^2\, \dot{B}_1
\nonumber\\
&& - 2 \left[ 9 \left( 3 + B_1^2 \right) - 20 \left( 6 + B_1^2
\right) \frac{p_L^2}{p_T^2}\, {\cal D}_{\chi \chi} + 64
\frac{p_L^4}{p_T^4}\, {\cal D}_{\chi \chi}^2 \right] h\, B_1 +
\left[ 9 \left( 3 + B_1^2 \right) -14 \left( 6 + B_1^2 \right)
\frac{p_L^2}{p_T^2}\, {\cal D}_{\chi \chi} + 64
\frac{p_L^4}{p_T^4}\, {\cal D}_{\chi \chi}^2 \right] H\, B_1
\nonumber\\
&& + 2 \left( 6 + B_1^2 \right)\, \frac{p_L^2}{p_T^2}\, B_1\,
\dot{\cal D}_{\chi \chi} \Bigg\} \nonumber\\
{\cal A}_{63} &=& - \frac{p_T^2}{p_L^2}\, \left( H - 8 h \right)
\nonumber
\end{eqnarray}
\begin{eqnarray}
{\cal A}_{64} &=& \ddot{B}_1 + \left( H -2 h \right)\, \dot{B}_1 +
\left( \dot{H} -2 \dot{h} \right)\, B_1 - \frac{16\, p_L^2\,
B_1}{3 {\cal D}}\, {\cal D}_{\chi \chi}\, \left( 2 h - H + 2\,
\frac{\dot{B}_1}{B_1} \right)\, \left( H - 2 h +
\frac{\dot{B}_1}{2 B_1} - \frac{\dot{\cal D}_{\chi \chi}}{2 {\cal
D}_{\chi \chi}} + \frac{\dot{\cal D}}{2 {\cal D}} \right)
\nonumber\\
&& + \frac{8 p_L^2\, B_1}{3 {\cal D}}\, {\cal D}_{\chi \chi}\,
\left( 2 \frac{\dot{B}_1^2}{B_1^2} + 2 \dot{h} - \dot{H} + 2
\left( 2 h - H \right)\, \frac{\dot{B}_1}{B_1} + 2
\frac{\ddot{B}_1}{B_1} \right) \nonumber\\
{\cal A}_{65} &=& 2 \dot{h} - \dot{H} + \frac{16\, p_L^2}{3 {\cal
D}}\, {\cal D}_{\chi \chi}\, \left( 2 h - H +
\frac{\dot{B}_1}{B_1} \right)\, \left( H - 2 h +
\frac{\dot{B}_1}{2 B_1} - \frac{\dot{\cal D}_{\chi \chi}}{2 {\cal
D}_{\chi \chi}} + \frac{\dot{\cal D}}{2 {\cal D}} \right)
\nonumber\\
&& - \frac{8 p_L^2}{3 {\cal D}}\, {\cal D}_{\chi \chi}\, \left( 2
\dot{h} - \dot{H} + \left( 2 h - H \right)\, \frac{\dot{B}_1}{B_1}
+ \frac{\ddot{B}_1}{B_1} \right) \nonumber\\
{\cal A}_{66} &=& - \frac{p_T^2}{p_L^2}\, \left( \dot{H} - 2
\dot{h} - 6 h\, H + 12 h^2 \right) \nonumber\\
{\cal A}_{67} &=& -\ddot{B}_1 + \left( 2 h - H \right)\, \dot{B}_1
+ \left( 2 \dot{h} - \dot{H} \right)\, B_1 + \frac{16 p_L^2\,
\left( 6 + B_1^2 \right)}{3 B_1\, {\cal D}}\, {\cal D}_{\chi
\chi}\, {\cal H}\, \left[ H - 2 h - \frac{B_1\, \dot{B}_1}{6 +
B_1^2} + \frac{\dot{B}_1}{2 B} - \frac{\dot{\cal D}_{\chi \chi}}{2
{\cal D}_{\chi \chi}} + \frac{\dot{\cal D}}{2 {\cal D}} -
\frac{\dot{\cal H}}{2 {\cal H}} \right] \nonumber\\
{\cal A}_{68} &=& -\frac{2 p_L^2\, p_T^2}{3 B_1^2\, {\cal D}^2}\,
 {\cal D}_{\chi \chi}\, \Bigg\{ \left[ B_1^4 + 4\, \left( 3 + 2
 \frac{p_L^2}{p_T^2}\, {\cal D}_{\chi \chi} \right)\, B_1^2 + 12\,
 \left( 3 - 4 \frac{p_L^2}{p_T^2}\, {\cal D}_{\chi \chi} \right)
 \right]\, \dot{B}_1 - B_1\, \left( 6 + B_1^2 \right)^2\, \left( 6
 h + \frac{\dot{\cal D}}{{\cal D}} \right) \Bigg\} \nonumber\\
{\cal A}_{69} &=& \frac{2 p_T^6}{3 p_L^2\, {\cal D}^2}\, \Bigg\{
3\, \left[ 3 B_1^4 + 36\, \left( 3 + B_1^2 \right) - 16
\frac{p_L^2}{p_T^2}\, \left( 6 + B_1^2 \right)\, {\cal D}_{\chi
\chi}\, \left( 3 - \frac{p_L^2}{p_T^2}\, {\cal D}_{\alpha_0
\alpha_0} \right) + 192 \frac{p_L^4}{p_T^4}\, {\cal D}_{\chi
\chi}^2 \right]\, h \nonumber\\
&& + \frac{p_L^2}{p_T^2}\, {\cal D}_{\alpha_0 \alpha_0}\, \Bigg[
-16 \frac{p_L^2}{p_T^2}\, {\cal D}_{\chi \chi}\, B_1\, \dot{B}_1 +
\left( 8 \frac{p_L^2}{p_T^2}\, {\cal D}_{\chi \chi} + B_1^2
\right)\, \frac{\dot{\cal D}_{\alpha_0 \alpha_0}}{{\cal
D}_{\alpha_0 \alpha_0}}\, B_1^2 + 4 \left( 3 + B_1^2 \right)\,
\left( 3 - 4 \frac{p_L^2}{p_T^2}\, {\cal D}_{\chi \chi} \right)\,
\frac{\dot{\cal D}_{\alpha_0 \alpha_0}}{{\cal D}_{\alpha_0
\alpha_0}} \nonumber\\
&& + 8 \left( 6 + B_1^2 \right) \frac{p_L^2}{p_T^2}\, \dot{\cal
D}_{\chi \chi} \Bigg] \Bigg\} \label{A6}
\end{eqnarray}
Thus, we have the full form of the equation (\ref{system}), which we
integrate numerically.

\section{Early time canonical action and initial conditions}

\label{appB}

We discuss here how we set the initial conditions for the perturbations entering in the linearized system
 (\ref{system}). As in the standard case \cite{mfb}, 
the initial conditions follow from the quantization of the quadratic action for the dynamical 
modes. As we show in Section \ref{sec:ghost}, the quadratic action - formally written in eq. (\ref{action-integrated}) - is obtained by integrating the nondynamical fields out of the quadratic action for the perturbations - formally written in eq. (\ref{formal-act}).

The quadratic action of the $2$d scalar perturbations of  the model (\ref{act-V0}) is given in eq. 
(\ref{act-2dS-cc}). The corresponding action for the dynamical modes is extremely lengthy, and we do not explicitly write it here. Fortunately, to set the initial conditions for the perturbations we only need the 
leading terms in an early time expansion of this action. We are interested in the phenomenologically relevant case of moderate anisotropy ($B_1 \ll 1$), for which $H_a \simeq H_b$ are nearly constant. Therefore, the two scale factors $a$ and $b$ grow nearly exponentially with time, and $p/H$ (where $p$ is either the longitudinal or the transverse component, and $H$ is either $H_a$ or $H_b$) is exponentially large in the asymptotic past. Therefore, the early time expansion of the action coincides with the sub-horizon $p / H \gg 1$ expansion, exactly as in the standard case.

Specifically, we first compute the exact matrices $K ,\, \Lambda ,\, \Omega^2$ that form the action for the dynamical modes (cf. eq. (\ref{klo})), and then expand them for $p / H \gg 1$. Since the resulting expressions are still quite involved, we further expand them for $B_1 \ll1$. This procedure is legitimate provided that
\begin{equation}
\frac{H}{p} \ll B_1 \ll 1
\end{equation}
which, as we remarked, is always true in the case of small anisotropy, and for sufficiently early times. At leading order, we obtain the expression
\begin{equation}
S_{\rm can} \simeq \frac{1}{2}\, \int\, d^3k\, dt\, \left\{ \vert
\dot{H}_+ \vert^2 - p^2\, \vert H_+ \vert^2 + \vert \dot{\Delta}_+
\vert^2 - p^2\, \vert \Delta_+ \vert^2 + \vert \dot{\Delta}_{-}
\vert^2 - p^2\, \vert \dot{\Delta}_{-} \vert^2 \right\}
\end{equation}
where the modes $H_+ ,\, \Delta_+ ,\,$ and $\Delta_-$  are related to the original perturbations by

\begin{eqnarray}
\hat\Psi &=& \frac{1}{\sqrt{a\, b^2}}\, \Bigg\{
\left(\frac{\sqrt{2}\, p^2}{p_T^2} - \frac{3 H_0^4\, \left( p_L^2
+ 10 p_T^2 \right) - m^2\, p_T^2\, \left( 6 H_0^2 - m^2
\right)}{18\, \sqrt{2}\, H_0^4\, p_T^2}\, B_1^2\right)\, H_{+} - \left(1 - \frac{m^2}{3 H_0^2} \right)\, \frac{p}{p_T}\, B_1\, \Delta_{+} \nonumber\\
&& \qquad\qquad\qquad + \frac{3 H_0^2\, \left( 4 p_L^4 + 7 p_L^2\,
p_T^2 + p_T^4 \right) - m^2\, p_T^2\, p^2}{6\, \sqrt{6}\,
H_0^2\, p_T^4}\, B_1^2 \, \Delta_{-} \Bigg\} \nonumber\\
\hat\alpha &=& \frac{1}{\sqrt{a\, b^2}}\, \Bigg\{ \left(
\frac{p}{2 p_T} - \frac{3 H_0^4\, \left( p_L^2 - 5 p_T^2 \right)\,
\left( p_L^2 + 3 p_T^2 \right) + 4 H_0^2\, m^2\, p_T^2\, \left(
p_L^2 + 7 p_T^2 \right) - 4 m^4\, p_T^4}{144 H_0^4\, p_T^3\, p}\,
B_1^2 \right)\, \Delta_{+} \nonumber\\
&& \qquad\qquad\qquad + \left( \frac{\sqrt{6}}{B_1} - \frac{p_T^2
- 3 p_L^2}{8\, \sqrt{6}\, p_T^2}\, B_1 \right)\,
\Delta_{-} \Bigg\} \nonumber\\
\hat\alpha_1 &=& \frac{1}{\sqrt{a\, b^2}}\, \Bigg\{
-\frac{p}{\sqrt{2}\, p_T}\, \Delta_{+} + \left(
\frac{\sqrt{6}}{B_1} - \frac{3 H_0^2\, \left( p_L^2 - 3 p_T^2
\right) + 8 m^2\, p_T^2}{24\, \sqrt{6}\, H_0^2\, p_T^2}\, B_1
\right)\, \Delta_{-} \Bigg\} \label{cc-canonical}
\end{eqnarray}

The modes $H_+ ,\, \Delta_+ ,\,$ and $\Delta_-$ are the canonical variables of the system (they are the analogous of the Mukhanov-Sasaki~\cite{musa} variable $v$ in the standard case of scalar field isotropic inflation). As in the standard case, their early times frequency is given by the momentum $p$, up to  ${\rm O } \left( H / p \right)$ subdominant corrections. Since the momentum changes adiabatically at early times ($\dot{p} / p^2 = {\rm O } \left( H / p \right)$), we can set the initial conditions for the canonical modes according to the adiabatic vacuum prescription, precisely as done in the standard case \cite{mfb}:
\begin{equation}
H_{+, \, in} = \Delta_{+,\, in} = \Delta_{-\, in} =
\frac{1}{\sqrt{2 p}} \,\,\,\, , \,\,\,\,\, \dot{H}_{+, \, in} =
\dot{\Delta}_{+,\, in} = \dot{\Delta}_{-\, in} = -i\,
\sqrt{\frac{p}{2}} \label{adiabatic}
\end{equation}
which are ${\rm O}\left( H/p\right)$ accurate. From eqs. (\ref{cc-canonical}) and (\ref{adiabatic}) we thus obtain the initial conditions for $\{ \hat\Psi, \, \hat\alpha, \, \hat\alpha_1\}$ and their time derivatives. Finally, the first, third, and fourth of eqs. (\ref{einstein-cc}) provide the initial conditions for the nondynamical modes ${\hat \Phi} ,\, {\hat \chi} ,\,$ and ${\hat \alpha}_0 \,$. In this way, we have the initial conditions for all the modes of the system (\ref{system}).

\end{document}